\documentclass[12pt]{article}
\usepackage{amsmath,amssymb,bm,epsf,epsfig,graphicx,hyperref,physics, cleveref, subfig, slashed, stmaryrd, tensor, float, booktabs}
\usepackage[backend=bibtex,natbib,style=numeric-comp,sorting=none, doi=false, isbn=false,url=false]{biblatex}
\makeatletter
\def\input@path{{./}{draft/}}
\makeatother
\graphicspath{{./}{draft/}}
\IfFileExists{ref.bib}{\addbibresource{ref}}{\addbibresource{draft/ref}}
\input epsf.sty
\usepackage[]{todonotes}
\usepackage{tikz}
\usetikzlibrary{arrows.meta, decorations.markings}

\usepackage[weather, alpine]{ifsym}

\definecolor{bisque}{rgb}{1.0, 0.89, 0.77}
\definecolor{forestgreen(web)}{rgb}{0.13, 0.55, 0.13}

\def\be{\begin{equation}}
\def\ee{\end{equation}}
\def\bea{\begin{eqnarray}}
\def\eea{\end{eqnarray}}
\def\ie{\begin{equation}\begin{aligned}}
\def\fe{\end{aligned}\end{equation}}
\numberwithin{equation}{section}

\renewcommand{\title}[1]{\vbox{\center\LARGE{#1}}\vspace{5mm}}
\renewcommand{\author}[1]{\vbox{\center#1}\vspace{5mm}}
\newcommand{\address}[1]{\vbox{\center\em#1}}
\newcommand{\email}[1]{\vbox{\center\tt#1}\vspace{5mm}}

\begin{document}

\unitlength = .8mm

\hypersetup{pageanchor=false}
\begin{titlepage}

\begin{center}

\hfill \\
\hfill \\
\vskip 1cm

\title{A Sojourn Through $c=1$ String Theory}

\author{
Sam Christian\textsuperscript{1},
Yuchen Wang\textsuperscript{1},
Xi Yin\textsuperscript{1,2},
Yutai Zhang\textsuperscript{1}
}

\address{
\textsuperscript{1}Jefferson Physical Laboratory, Harvard University,
Cambridge, MA 02138 USA\\
\textsuperscript{2}OpenAI
}

\email{
    samchristian@fas.harvard.edu, yuchen\_wang@fas.harvard.edu, xiyin@fas.harvard.edu, yutaizhang@fas.harvard.edu}

\end{center}

\abstract{We revisit the $S$-matrix, the thermal $S$-matrix and the thermal free energy of $c=1$ string theory from the worldsheet perspective. By numerically integrating Virasoro conformal blocks with DOZZ structure constants, we compute the scattering amplitudes and thermal amplitudes up to 6-point at tree level and up to 3-point at one loop, in addition to the genus two thermal free energy. All results are in agreement with the predicted answers from the $c=1$ matrix model.}

\vfill

\end{titlepage}
\hypersetup{pageanchor=true}

\eject

\begingroup
\hypersetup{linkcolor=black}

\tableofcontents

\endgroup

\section{Introduction}

Two-dimensional $c=1$ string theory has long served as a useful laboratory for understanding string duality and string perturbation theory \cite{Klebanov:1991qa,Ginsparg:1993is}. The worldsheet formulation for the $c=1$ string theory is based on a time-like free boson CFT, the $c=25$ Liouville CFT and the $bc$ ghost system. The perturbative excitations in $c=1$ string theory consist of massless scalar bosons in $1+1$ dimensions that interact and scatter off a Liouville wall in the spatial direction. A particularly attractive feature of the theory is that it admits both an exactly solvable worldsheet CFT description and a conjecturally dual formulation in terms of a particular $U(N)$ matrix quantum mechanics model, known as the $c=1$ matrix model \cite{GrossMiljkovic:1990,BrezinKazakovZamolodchikov:1990,
GinspargZinnJustin:1990,Parisi:1990DiscretizedString, Moore:1991zv, GrossKlebanov:1990Circle}.

The matrix model description provides exact predictions for string observables such as the $S$-matrix, the thermal $S$-matrix and the perturbative thermal free energy. In contrast, direct worldsheet computations of these observables remain less developed. Existing analytic computations of the tree level $S$-matrix, based on analytic continuation from special ``resonance momenta'' \cite{DiFrancescoKutasov:1991,DiFrancescoKutasov:1992}, do not straightforwardly extend to generic kinematics or genus $g\geq 1$.

A complementary numerical approach to worldsheet amplitudes is to evaluate the moduli-space integrand via conformal block decomposition, using the exact Liouville structure constants \cite{Dorn:1994xn,Zamolodchikov:1995aa,Teschner:1995,FateevZamolodchikovZamolodchikov:2000,Hosomichi:2001,PonsotTeschner:2001}. This approach has enabled the worldsheet evaluation of perturbative string amplitudes at generic momenta and beyond tree level, including the tree level four-point amplitudes and the one loop $1\rightarrow 1$ amplitude \cite{BalthazarRodriguezYin:2019}. This approach has also been extended beyond perturbation theory through the study of ZZ-instanton contributions to closed string scattering amplitudes \cite{BalthazarRodriguezYin:2023ZZ,Sen:2020FixingAmbiguity, BalthazarRodriguezYin:2023MultiInstanton,Sen:2021Divergent,Sen:2020DInstantonPerturbation,Sen:2021Cutkosky,Sen:2021DInstantonsTwoDimensional,Sen:2021DInstantonNormalization,EniceicuEtAl:2022ZZAnnulus} and to long string scattering \cite{BalthazarRodriguezYin:2018LongString}. A similar numerical approach has also been applied to two dimensional non-critical Type 0B string theory \cite{BalthazarRodriguezYin:2022Type0B,BalthazarRodriguezYin:2022Type0BInstanton}.

In this paper, we extend this numerical program in two directions: to higher point scattering amplitudes and to the thermal free energy at higher genus. We compute both the ordinary and the thermal $1\to n$ amplitudes up to 6-point at tree level and up to 3-point at genus one. We also compute the thermal free energy at genus two, the first genus at which interaction corrections to the free energy appear.

The common ingredient in these computations is the integration of the worldsheet correlators or partition functions over the moduli space of Riemann surface with punctures. We evaluate the Liouville correlators and partition functions using conformal block decompositions, computing the higher point and higher genus blocks numerically using the $c$-recursion relation of \cite{Cho-Collier-Yin}.  The moduli space integration is evaluated using Monte Carlo method.

The $1\to n$ $S$-matrix of $c=1$ string theory takes the form \cite{Ginsparg:1993is}
\ie{}
&S_{1\to n}(\omega; \omega_1,\cdots, \omega_n) = \delta\left(\omega - \sum_{i=1}^n \omega_i\right) {\cal A}_{1\to n}(\omega_1,\cdots,\omega_n),
\fe
where $\omega$ is the energy of the incoming massless particle, and $\omega_1,\cdots,\omega_n$ are the energies of the outgoing particles (after reflection from the Liouville wall). The amplitude ${\cal A}_{1\to n}$ is defined to be
\ie{}
&{\cal A}_{1\to n}(\omega_1,\cdots,\omega_n)=
\sum_{L=0}^\infty g_s^{n-1+2L} {\cal A}_{1\to n}^{(L)}(\omega_1,\cdots,\omega_n) + \mathcal{O}\left(e^{-1/g_s}\right),
\fe
where $g_s$ is the string coupling constant, ${\cal A}^{(L)}_{1\to n}$ is the genus $L$ string scattering amplitude defined in string perturbation theory, and the non-perturbative corrections in $\mathcal{O}\left(e^{-1/g_s}\right)$ are defined through D-instanton perturbation theory on ZZ-instantons.

From the matrix model, one anticipates the answer
\ie\label{matrixanswer}
{\cal A}_{1\to2}^{(0)}
&=2\pi i\,\omega_1\omega_2\omega,
\quad
{\cal A}_{1\to3}^{(0)}
=4\pi^2 i\,\omega_1\omega_2\omega_3\omega\,
(1+i\omega), \quad {\cal A}_{1\to1}^{(1)}
=\frac{\pi^2}{6}
\left(i\omega^2+2i\omega^4-\omega^5\right),
\\
{\cal A}_{1\to4}^{(0)}
&=8\pi^3 i\,\omega_1\omega_2\omega_3\omega_4\omega\,
(1+i\omega)(2+i\omega),
\\
{\cal A}_{1\to5}^{(0)}
&=16\pi^4 i\,\omega_1\omega_2\omega_3\omega_4\omega_5\omega\,
(1+i\omega)(2+i\omega)(3+i\omega),
\\
{\cal A}_{1\to2}^{(1)}
&=-\frac{i\pi^3}{3}\,
\omega_1\omega_2\omega\,
(\omega-i)(\omega-2i)
\left(\omega_1^2+\omega_2^2-i\omega+1\right),
\fe
where we have written $\omega = \sum_{i=1}^n \omega_i$. In previous work \cite{BalthazarRodriguezYin:2019}, the tree level $1\to 3$ amplitude $\mathcal{A}^{(0)}_{1\to3}$ and the one loop $1\to 1$ amplitude $\mathcal{A}^{(1)}_{1\to 1}$ were computed from the string worldsheet by numerically integrating conformal blocks. In this paper, we extend the comparison to the tree level $1\to 4$ amplitude $\mathcal{A}^{(0)}_{1\to 4}$, the tree level $1\to 5$ amplitude $\mathcal{A}^{(0)}_{1\to 5}$ and the one loop $1\to2$ amplitude $\mathcal{A}^{(1)}_{1\to2}$. All of these results agree with the predictions of the matrix model, providing an extensive check of the conjectured equivalence between the perturbative $S$-matrices of $c=1$ string theory and the dual matrix model.

The duality between the $c=1$ string theory and the matrix model extends to observables at finite temperature, such as the thermal free energy and the thermal $S$-matrix. In particular, the perturbative thermal free energy $F(R)$ of $c=1$ string theory is expected to agree, up to local counterterms, with the grand canonical free energy $F_{\mathrm{MM}}(R)$ of the matrix model. The agreement between the genus one thermal free energy and the matrix model result has been checked analytically \cite{Bershadsky:1990xb,Klebanov:1991qa}. In this work, we extend the worldsheet computation of the perturbative thermal free energy to genus two. The matrix model predicts that the genus two thermal free energy is
\ie\label{eq:Genus2ThermalFreeEnergy}
 F^{(2)}(R)
&=-\frac{\pi}{2880}\left(7+10R^{-2}+7R^{-4}\right).
\fe
Our numerical worldsheet result agrees with this prediction, providing the first non-trivial check of the duality at genus two. We also compute the tree level $1\to 4$ and $1\to 5$, as well as the one loop $1\to 1$ and $1\to 2$ thermal $S$-matrices. All of these results agree with the matrix model predictions.

The rest of this paper is organized as follows. In section~\ref{sec:worldsheet-formulation}, we review the worldsheet formulation of $c=1$ string theory, the $c=1$ matrix model, and the duality between the two theories. Matrix model results for the $S$-matrix, the thermal $S$-matrix and the grand canonical free energy are also reviewed in section~\ref{sec:worldsheet-formulation}. Our worldsheet computations of the $S$-matrix, the thermal $S$-matrix and the genus two free energy are presented in sections~\ref{sec:higher-point-amplitudes}, \ref{sec:thermal_S_matrix} and \ref{sec:genus2-free-energy}, respectively. We conclude in section~\ref{sec:discussion} with a discussion of future directions. The code for our numerical computations is available in a GitHub repository.\footnote{\url{https://github.com/YutaiZhang2003/c1-computation}}

\section[Worldsheet formulation of \texorpdfstring{$c=1$}{c=1} string theory and the dual matrix model]{Worldsheet formulation of \texorpdfstring{$c=1$}{c=1} string theory \\and the dual matrix model}\label{sec:worldsheet-formulation}
In this section, we recall the basic setup of $c=1$ string theory as well as its duality with the matrix model. The worldsheet convention follows \cite{BalthazarRodriguezYin:2019}.

\subsection{The worldsheet theory}

The worldsheet formulation for the $c=1$ string theory is based on a time-like free boson CFT, the $c=25$ Liouville CFT and the $bc$ ghost system. Throughout the paper, we will take $\alpha'=1$. The Liouville CFT is governed by the action
\ie
S_L = {1\over 4\pi} \int d^2z \sqrt{g} \left( g^{mn} \partial_m \phi \partial_n \phi + Q R \phi + 4\pi \mu_L e^{2b\phi} \right),
\fe
whose central charge is related to the background charge $Q=b+b^{-1}$ by $c=1+6Q^2$. The case of interest $c=25$ corresponds to $b=1$. The Virasoro primaries of Liouville CFT are scalar operators $V_P$ labeled by the ``Liouville momentum'' $P\in\mathbb{R}_{\geq 0}$. $V_P$ has scaling dimension $\Delta = h+\tilde h = {Q^2\over 2} + 2P^2$. Our normalization convention for the Liouville vertex operator follows \cite{BalthazarRodriguezYin:2019} to be
\ie\label{Liouville2pt}
\left\langle V_{P_1}(z,\bar z) V_{P_2}(0) \right\rangle = \pi{\delta(P_1-P_2)\over |z|^{\Delta_1+\Delta_2}}.
\fe
At $b=1$, with an appropriate rescaling of the Liouville cosmological constant $\mu_L$, the three point coefficients of the Liouville primaries are given by the expression  \cite{Dorn:1994xn, Zamolodchikov:1995aa}
\ie\label{cppp}
{\cal C}(P_1, P_2, P_3) = {1\over \Upsilon_1(1+i(P_1+P_2+P_3)) } \left[ { 2P_1 \Upsilon_1(1+2iP_1)\over  \Upsilon_1(1 + i(P_2+P_3-P_1)) } \times (2~{\rm permutations}) \right],
\fe
where $\Upsilon_b(x)$ is the Upsilon function, defined by the analytic continuation of
\ie
\log\Upsilon_b(x) = \int_0^\infty {dt\over t} \left[ \left( {Q\over 2}-x \right)^2 e^{-t} - {\sinh^2\left[\left({Q\over 2}-x\right){t\over 2} \right]\over \sinh{tb\over 2}\sinh{t\over 2b} } \right],~~~ 0<{\rm Re} (x) < Q.
\fe
We will denote the time-like free boson by $X^0$, and normalize the sphere one point function of free boson vertex operator by
\ie\label{eq:free-boson-sphere-normalization}
\left\langle e^{i \omega X^0} \right\rangle = \frac{2\pi i}{g_s^2}\delta(\omega).
\fe
The weight $(h,\tilde{h})=(1,1)$ vertex operators in the worldsheet matter CFT are given by
\ie\label{eq:MatterVertex}
\mathcal{V}^{\pm}_{\omega} = g_s e^{ \pm i \omega X^0} V_{\frac{\omega}{2}}\qquad \omega>0,
\fe
where $g_s$ is the string coupling constant. The BRST cohomology classes corresponding to 1-particle asymptotic states can be represented by the vertex operators $c\tilde{c}\, \mathcal{V}^{\pm}_{\omega}$.
Here, ${\cal V}_\omega^+$ and ${\cal V}_\omega^-$ represent incoming and outgoing modes, respectively, of energy $\omega>0$. 

To compare with the result in the matrix model side, we normalize the asymptotic string state as
\ie\label{eq:AsymptoticState}
{}_{\mathrm{in}}\langle\omega'|\omega\rangle_{\mathrm{in}}
={}_{\mathrm{out}}\langle\omega'|\omega\rangle_{\mathrm{out}}
=\omega\,\delta(\omega-\omega'),
\qquad \omega,\omega'>0.
\fe
The normalization of the moduli integrand for genus $g$ $n$-point scattering amplitude can be pinned down by requiring factorization of the amplitude, given the conventions for the normalization of the vertex operators in equation \eqref{eq:free-boson-sphere-normalization} and equation \eqref{eq:MatterVertex} as well as the asymptotic string state (equation \ref{eq:AsymptoticState}).

\subsection{The dual matrix model}\label{sec:MQM}

The $c=1$ matrix model is defined as a suitable $N\to \infty$ limit of the $U(N)$ gauged matrix quantum mechanics (MQM) with the Hamiltonian $H = {1\over 2} {\rm Tr} (P^2-X^2)$, where $X$ is a Hermitian $N\times N$ matrix on which the $U(N)$ gauge symmetry acts by the adjoint representation, and $P$ is the canonically conjugate momentum matrix \cite{GrossMiljkovic:1990,BrezinKazakovZamolodchikov:1990,GinspargZinnJustin:1990,Parisi:1990DiscretizedString}. The singlet sector of MQM admits an equivalent formulation as a system of infinitely many non-interacting fermions in the inverted harmonic oscillator potential $V(x)=-x^2/2$, subject to the Hamiltonian $H={1\over 2}(p^2-x^2) $\cite{BrezinItzyksonParisiZuber:1978}.

In the free fermion description of the $c=1$ MQM, the incoming scattering states $\bigl\{\ket{E}^{\mathrm{in}}_{L},\ket{E}^{\mathrm{in}}_{R}\mid E\in\mathbb{R}\bigr\}$ form an energy eigenbasis of the single fermion Hilbert space.
Here $L$ and $R$ indicate that the incoming wave approaches the inverted harmonic oscillator potential from the left or right. An alternative energy eigenbasis consists of outgoing scattering states, denoted as $\bigl\{\ket{E}^{\mathrm{out}}_{L},\ket{E}^{\mathrm{out}}_{R}\mid E\in\mathbb{R}\bigr\}$. The two bases are related by the single fermion $S$-matrix, obtained by solving the scattering problem in the inverted harmonic oscillator potential.

The in- and out-vacuum states \(\lvert\Omega_a\rangle\), with \(a\in \{\mathrm{in},\mathrm{out}\}\), relevant to the scattering process in \(c=1\) string theory are defined by filling the single-fermion states \(\lvert E\rangle_R^a\) with \(E\leq-\mu<0\). In the semi-classical description, these vacua correspond to a Fermi sea occupying the \(x>\sqrt{p^2+2\mu}\) region of the phase space.

The quasi-particles of the $c=1$ matrix model that are conjecturally dual to the asymptotic states of the $c=1$ string theory are collective excitations of the Fermi surface, namely the particle-hole pairs. Detailed descriptions of the collective excitations are given in \cite{Sengupta:1990bt, Das:1990kaa, Polchinski:1991uq, Moore:1991zv, Moore:1992gb, Polchinski:1994mb, Gross:1990st}. 
Working in the second-quantized formalism, we introduce the operators \(b_{\mathrm{in},\sigma}^{\dagger}(E)\) and \(b_{\mathrm{out},\sigma}^{\dagger}(E)\), which are the creation operators for the single-fermion states \(\ket{E}^{\mathrm{in}}_\sigma\) and \(\ket{E}^{\mathrm{out}}_\sigma\), respectively, with \(\sigma=L,R\). They satisfy the canonical anticommutation relations
\ie
\{b_{\mathrm{in},\sigma}(E),b_{\mathrm{in},\sigma'}^\dagger(E')\}
=\delta_{\sigma\sigma'}\delta(E-E'), \quad \{b_{\mathrm{out},\sigma}(E),b_{\mathrm{out},\sigma'}^\dagger(E')\}
=\delta_{\sigma\sigma'}\delta(E-E').
\fe
By solving the scattering problem for a single fermion in the inverted harmonic oscillator potential, we obtain the following relations between the in and out creation operators:
\ie\label{eq:in/out_oscillator}
b_{\mathrm{in},R}^\dagger(E)
&=
\mathcal R(E)\,b_{\mathrm{out},R}^\dagger(E)
+
\mathcal T(E)\,b_{\mathrm{out},L}^\dagger(E),\\ 
b_{\mathrm{in},L}^{\dagger}(E)
&=
\mathcal R(E)\,b_{\mathrm{out},L}^{\dagger}(E)
+
\mathcal T(E)\,b_{\mathrm{out},R}^{\dagger}(E),
\fe
where $\mathcal R(E)$ is the reflection factor and $\mathcal T(E)$ is the transmission factor, given by\footnote{We choose the branch for the reflection factor and transmission factor such that $\mathcal R(0)=\frac{i}{\sqrt2}, \mathcal T(0)=\frac1{\sqrt2}$.}
\ie\label{eq:reflectionphase}
\mathcal{R}(E) = i \mu^{iE} \left[ \frac{1}{1+e^{2\pi E}}\frac{\Gamma({1\over 2} - i E)}{\Gamma({1\over 2} + i E)} \right]^{1\over 2}, 
\mathcal T(E)
=
\mu^{iE}
\left[
\frac{1}{1+e^{-2\pi E}}\,
\frac{\Gamma\!\left(\frac12-iE\right)}
{\Gamma\!\left(\frac12+iE\right)}
\right]^{1/2}.
\fe
The creation operator for an incoming/outgoing particle-hole pair is defined as
\ie\label{Eq:particle-hole pair}
a_{\mathrm{in}}^\dagger(\omega)=\int_{-\infty}^{\infty}dE \,b_{\mathrm{in},R}^{\dagger}(E+\omega)b_{\mathrm{in},R}(E),\quad a_{\mathrm{out}}^\dagger(\omega)=\int_{-\infty}^{\infty}dE \,b_{\mathrm{out},R}^{\dagger}(E+\omega)b_{\mathrm{out},R}(E).
\fe
The $1\to n$ $S$-matrix of particle-hole pairs in the matrix model is defined as
\ie\label{eq:MMSmatrix}
S_{1\to n,\mathrm{MM}}(\omega|\omega_1,\dots \omega_n):= \frac{\bra{\Omega_{\mathrm{out}}}\left( \prod_{i=1}^n a_{\mathrm{out}}(\omega_i) \right)   a_{\mathrm{in}}^\dagger(\omega)\ket{\Omega_{\mathrm{in}}}}{\bra{\Omega_{\mathrm{out}}}\ket{\Omega_{\mathrm{in}}}}.
\fe
The $1\to n$ scattering amplitude of the collective excitations \(\mathcal{A}_{1\to n,\mathrm{MM}}\) is defined by factoring out the energy-conserving delta function in the $1\to n$ $S$-matrix:
\ie
S_{1\to n,\mathrm{MM}}(\omega; \omega_1,\cdots, \omega_n) = \delta\left(\omega - \sum_{i=1}^n \omega_i\right) {\cal A}_{1\to n,\mathrm{MM}}(\omega_1,\cdots,\omega_n).
\fe
Using equations \eqref{eq:in/out_oscillator}, \eqref{Eq:particle-hole pair}, and \eqref{eq:MMSmatrix}, the exact $1\to n$ scattering amplitude of particle-hole pairs is given by\footnote{Due to the different choice of Fermi-sea vacuum, the reflection of the hole contributes $\mathcal R(E)^{-1}$ rather than $\mathcal R(E)^*$ as in the type~0B matrix model \cite{BalthazarRodriguezYin:2022Type0BInstanton}.}
\ie\label{eq:mqm1tonamplitdue}
\mathcal{A}_{1\to n,\mathrm{MM}}(\omega_1,\ldots,\omega_n)
=
-\sum_{S_1\sqcup S_2=S}
(-)^{|S_2|}
\int_{0}^{\omega(S_2)} dx\,
\mathcal{R}(-\mu+\omega-x)\,
\bigl(\mathcal{R}(-\mu-x)\bigr)^{-1},
\fe
where $\omega=\sum_{i=1}^n \omega_i$,  \(S_1\) and \(S_2\) are disjoint subsets of $S=\{\omega_1,\ldots,\omega_n\}$ such that \(S_1\sqcup S_2=S\), \(|S_2|\) denotes the number of elements of \(S_2\) and \(\omega(S_2)\) is the sum of all elements of \(S_2\). For instance, the exact $1\to 2$ scattering amplitude is given by
\ie\label{aam}
& {\cal A}_{1\to 2,\mathrm{MM}} (\omega_1, \omega_2) = \int_0^{\omega_2} dx \frac{\mathcal{R}(-\mu + \omega- x)}{\mathcal{R}(-\mu - x) } - \int_{\omega_1}^\omega dx \,  \frac{\mathcal{R}(-\mu + \omega - x)}{\mathcal{R}(-\mu - x)},~~~\omega=\omega_1+\omega_2.
\fe
The asymptotic expansion of Eq. \eqref{eq:mqm1tonamplitdue} admits the structure of a genus expansion in $\mu^{-1}$ as
\ie\label{eq:Genus_expansion}
\mathcal A_{1\to n,\mathrm{MM}}(\omega_1,\ldots,\omega_n)
\sim
\sum_{L=0}^{\infty}
\mu^{-(n-1+2L)}
\mathcal A_{1\to n,\mathrm{MM}}^{(L)}(\omega_1,\ldots,\omega_n),
\fe
which suggests the identification between $\mu^{-1}$ and the string coupling constant $g_s$. The precise statement of the duality is explained in section \ref{sec:c1MQMdual}.

The duality between $c=1$ string theory and matrix model extends to thermodynamic quantities, including the thermal partition function and thermal $S$-matrix. The thermal partition function of the $c=1$ string is expected to agree, up to local counterterms, with the grand canonical partition function of the matrix model. The latter is defined by a suitable regularization and counterterm prescription for the formal expression\footnote{We use the same parameter \(\mu\) introduced above to specify the vacuum state, with the understanding that in the zero temperature limit, the fermion occupation reproduces the one-sided Fermi sea vacuum defined above.} 
\ie\label{eq:grandpartitionfunc}
\Xi(\mu,R)
=\Tr_{\mathcal H_R} e^{-\beta(H_{\mathrm{MM}}+\mu N)},
\qquad \beta=2\pi R,
\fe
where $\mathcal H_R$ is the Fock space generated by $\ket{E}^{\mathrm{out}}_R$, $H_{\mathrm{MM}}$ is the matrix model Hamiltonian, and $N$ is the fermion number operator. The grand canonical free energy of $c=1$ MQM $F_{\mathrm{MM}}(\mu,R)$ is given by\footnote{We use the standard convention for the grand canonical free energy $F_{\mathrm{MM}}(\mu,R):= -\frac{1}{2\pi R}\log \Xi(\mu,R)$.} \cite{GrossKlebanov:1990Circle,AlexandrovKaushik:2026MultiInstantons,AlexandrovMahajanSen:2024SineLiouville}
\ie\label{grandcanonical_free_energy}
F_{\mathrm{MM}}(\mu,R)
=
\lim_{\Lambda\to\infty}
\left[
\frac{1}{8\pi R}
\int_{\Lambda^{-1}}^\infty\frac{dt}{t}\,
\frac{e^{-i\mu t}}
{\sinh(t/2)\sinh\!\bigl(t/(2R)\bigr)}
+F_{\mathrm{ct}}(\mu,R;\Lambda)
\right],
\fe
where $\Lambda$ is a regulator.  $F_{\mathrm{ct}}(\mu,R;\Lambda)$ denotes the counterterm, which is a polynomial in $\mu$ chosen to ensure a finite $\Lambda\to\infty$ limit. For concreteness, we take the counterterm prescription
\ie
F_{\mathrm{ct}}(\mu,R;\Lambda)
={}-\frac{\Lambda^2}{4\pi}
+\frac{i\mu\Lambda}{2\pi}+\left[
\frac{\mu^2}{4\pi}
+\frac{1+R^{-2}}{48\pi}
\right]
\left(\log\Lambda-\gamma_E-\frac{i\pi}{2}\right)
+\frac{3\mu^2}{8\pi}.
\fe
where $\gamma_E$ is the Euler–Mascheroni constant. From equation \eqref{grandcanonical_free_energy}, one can extract the large $\mu$ expansion of the matrix model free energy:
\ie
F_{\mathrm{MM}}(\mu,R)
=\frac{\mu^2\log\mu}{4\pi}
+\frac{1+R^{-2}}{48\pi}\log \mu
-\frac{7+10R^{-2}+7R^{-4}}{11520\pi\,\mu^2}
+O(\mu^{-4}).
\fe
We compare this expression with the genus expansion of the free energy of $c=1$ string theory in section \ref{sec:c1MQMdual}.

The thermal $S$-matrix of $c=1$ MQM is the finite temperature analog of the usual $S$-matrix, and has a lot of similar properties. To define it, we first introduce notation for the fermion occupation states. Let \(\nu\) denote a set of occupied energies in the continuous single-fermion spectrum that labels the fermion occupation configuration. We introduce $\ket{\nu_{\mathrm{in}}}$ and $\ket{\nu_{\mathrm{out}}}$ as 
\ie
\ket{\nu_{\mathrm{in}}}
:=\prod_{E\in \nu}
b^\dagger_{\mathrm{in},R}(E)\bigr|0\rangle,
\quad \ket{\nu_{\mathrm{out}}}
:=\prod_{E\in \nu}
b^\dagger_{\mathrm{out},R}(E)|0\rangle.
\fe
where the product is in the increasing energy order for the fermion creation operators. The energy of the state is given by $\mathcal E_\nu= \sum_{E\in\nu} E$ and the particle number of the state is given by $N_{\nu}= \left|\nu\right|$. The overlap between $|\nu_{\text{in}}\rangle$ and $|\nu_{\text{out}}\rangle$ is given by $\prod_{E\in \nu}\mathcal{R}(E)$.

The $1\to n$ thermal $S$-matrix element is defined as\footnote{In the Type 0B matrix model, the in/out vacua in the scattering procedure are related by a phase, which allows for expressing the thermal $S$-matrix as a grand canonical thermal trace of creation and annihilation operators \cite{BalthazarRodriguezYin:2022Type0BInstanton}. In the $c=1$ matrix model, the definition is more intricate due to the distinction between in and out vacua in the scattering procedure.}
\ie\label{eq:thermalSmatrix}
S^{(T)}_{1\to n,\mathrm{MM}}\left(\omega\mid\omega_1,\dots \omega_n\right)
:= \sum_\nu \frac{e^{-\beta(\mathcal E_\nu+\mu N_\nu)}}{\Xi} \,S_{1\to n,\mathrm{MM}}^\nu(\omega|\omega_1,...,\omega_n),
\fe
where the sum over $\nu$ runs over the distinct subsets of single fermion spectrum and is assumed to be properly regularized. $S_{1\to n,\text{MM}}^\nu(\omega|\omega_1,...,\omega_n)$ is defined as
\ie
S_{1\to n,\mathrm{MM}}^\nu(\omega|\omega_1,...,\omega_n)
:=
\frac{\bra{\nu_{\mathrm{out}}}
\left(\prod_{i=1}^n a_{\mathrm{out}}(\omega_i)\right)
a_{\mathrm{in}}^\dagger(\omega)
\ket{\nu_{\mathrm{in}}}}
{ \bra{\nu_{\mathrm{out}}}\ket{\nu_{\mathrm{in}}}}.
\fe
The thermal S-matrix has the structure
\ie
S^{(T)}_{1\to n,\mathrm{MM}}(\omega\mid \omega_1,\ldots,\omega_n)
=\delta\!\left(\omega-\sum_{i=1}^n\omega_i\right)
\mathcal A^{(T)}_{1\to n,\mathrm{MM}}(\omega_1,\ldots,\omega_n).
\fe
$\mathcal{A}^{(T)}_{1\to n,\mathrm{MM}}$ can be solved in a similar way as the standard zero temperature scattering amplitude, resulting in the expression
\ie
\mathcal A^{(T)}_{1\to n,\mathrm{MM}}(\omega_1,\ldots,\omega_n)
=
\frac{1}{\prod_{a=1}^{n}
(1-e^{-\beta\omega_a})}
\int_{-\infty}^{\infty}dE\,
\frac{\mathcal{R}(E+\omega)}{\mathcal{R}(E)}
\sum_{J\subseteq\{1,\ldots,n\}}
(-1)^{|J|}\,n_F(E+\omega_J),
\fe
where $\mathcal{R}$ is the reflection amplitude in Eq.\eqref{eq:reflectionphase} and 
\ie
\omega_J:= \sum_{a\in J}\omega_a, \quad n_F(E):=\frac{1}{1+e^{\beta(E+\mu)}}, \quad \beta=2\pi R.
\fe
In the limit $\beta \rightarrow \infty$, the only state contributing in the thermal sum is the one-sided Fermi sea vacua, and the thermal $S$-matrix in MQM reduces to the usual $S$-matrix.

To make contact with Euclidean correlation functions, we analytically continue $\mathcal{A}^{(T)}_{1\to n,\mathrm{MM}}$ to the upper half complex $\omega_j$ plane towards the Matsubara frequency $\frac{i m_j}{R}$, where $m_j\in \mathbb{Z}_{>0}$. In this procedure, poles of $n_F(E+\omega_J)$ cross the $E$ integration contour, giving rise to residue contributions that do not vanish as $\omega_j\rightarrow \frac{i m_j}{R}$. The remaining integral over $E$ gives a finite contribution to $\mathcal A^{(T)}_{1\to n,\mathrm{MM}}$ in this limit. In the limit $\omega_j \rightarrow \frac{i m_j}{R}$, $\mathcal A^{(T)}_{1\to n,\mathrm{MM}}$ has the behavior
\ie\label{eq:thermalLSZ}
\mathcal A^{(T)}_{1\to n,\mathrm{MM}}(\omega_1,\ldots,\omega_n) \longrightarrow \frac{1}{\prod_{a=1}^{n}(1-e^{-\beta\omega_a})} \widetilde{\mathcal{A}}^{(T)}_{1\to n,\mathrm{MM}}\left(\omega_j= \frac{i m_j}{R}\right),
\fe
where
\ie\label{eq:thermalamplitude}
\widetilde{\mathcal A}^{(T)}_{1\to n,\mathrm{MM}}
&=
-\frac{i}{R}\sum_{J\subseteq\{1,\ldots,n\}}(-1)^{|J|}
\sum_{k=0}^{m_J-1}
\frac{
\mathcal{R}\!\left(-\mu+\frac{i}{R}(M-k-\tfrac12)\right)}
{\mathcal{R}\!\left(-\mu-\frac{i}{R}(k+\tfrac12)\right)},
\fe
and $m_J =\sum_{j\in J} m_j$, $M= \sum_{j=1}^n m_j$. This relation can be viewed as the thermal analog of the LSZ reduction formula, and so we refer to $\tilde{\mathcal{A}}^{(T)}_{1\to n}$ as the thermal amplitude. It also has the structure of a genus expansion in $\mu^{-1}$ as in the standard $S$-matrix.

It is instructive to look at the perturbative expansion of thermal amplitudes in equation \eqref{eq:thermalamplitude}. The perturbative expansion of thermal $1\to1$ and $1\to2$ amplitude up to one loop is given by the expressions
\ie
\widetilde{\mathcal A}^{(T)}_{1\to1,\mathrm{MM}}
&=
\omega
-\frac{\omega^2(\omega-i)}{24\mu^2}
\left(1-i\omega+\omega^2+\frac1{R^2}\right)
+O(\mu^{-4}),\\
\widetilde{\mathcal A}^{(T)}_{1\to2,\mathrm{MM}}
&={}\frac{i\omega\omega_1\omega_2}{\mu}-\frac{i\omega\omega_1\omega_2}{24\mu^3}
(\omega-i)(\omega-2i)
\left(1-i\omega+\omega_1^2+\omega_2^2+\frac1{R^2}\right)
+O(\mu^{-5}),
\fe
where $\omega_j=i m_j/R$ with $m_j\in\mathbb Z_{>0}$, and $\omega=\sum_j\omega_j$. At tree level, the thermal amplitudes coincide with the zero temperature amplitudes analytically continued to Matsubara frequencies. At genus one, they receive additional contributions proportional to $R^{-2}$.

\subsection{Duality between \texorpdfstring{$c=1$}{c=1} string theory and \texorpdfstring{$c=1$}{c=1} MQM}\label{sec:c1MQMdual}

In this subsection, we will explain the conjectured duality between $c=1$ string theory and $c=1$ MQM, focusing on the correspondence between their $S$-matrix, thermal free energy, and thermal $S$-matrix. 

For the $S$-matrix, the duality predicts that scattering amplitudes of $c=1$ string theory agree with the scattering amplitudes of collective excitations of the matrix model, with $\mu^{-1}$ proportional to the string coupling $g_s$:
\ie
\mathcal{A}_{m\to n}(\omega_i ,g_s) = \mathcal{A}_{m\to n,\mathrm{MM}}(\omega_i, \mu^{-1}=c\,  g_s).
\fe
The proportionality constant $c$ can be determined by comparing the tree level $1\to 2$ and  $1\to 3$ scattering amplitudes with the matrix model results \cite{BalthazarRodriguezYin:2019}. This comparison pins down the dictionary to further be
\ie
\mu^{-1}=2\pi g_s.
\fe
Specializing to $1\to n$ scattering amplitude, the duality predicts that
\ie
\mathcal{A}_{1\to n}(\omega_i ,g_s) = \mathcal{A}_{1\to n,\mathrm{MM}}(\omega_i, \mu^{-1}=2\pi g_s),
\fe
where the matrix model amplitude on the RHS is given by equation~\eqref{eq:mqm1tonamplitdue}.

An analogous conjecture relates the thermal free energy of $c=1$ string theory to the grand canonical free energy of MQM. Its precise formulation requires specifying both the terms included in the comparison and the regularization prescription.  The genus $g$ perturbative free energy of $c=1$ string theory $F^{(g)}(R)$ can be computed by the genus $g$ vacuum amplitude in the thermal background at $g\geq1$. The genus 0 free energy, on the other hand, is not well-defined within this prescription. Therefore, we restrict attention to
\ie
F(g_s,R):=\sum_{g\geq1}g_s^{2g-2}F^{(g)}(R) + \mathcal{O}(e^{-1/g_s}),
\fe
where $\mathcal{O}(e^{-1/g_s})$ represents the non-perturbative corrections to the thermal free energy defined via $D$-instanton perturbation theory. 

A further subtlety arises at genus one. The worldsheet vacuum amplitude can be evaluated analytically using the unfolding trick \cite{Bershadsky:1990xb,Klebanov:1991qa}, which produces
\ie\label{eq:worldsheetgenus1}
F^{(1)}(R)=-\frac{V_{\phi}}{24\pi}
\left(1+\frac{1}{R^2}\right),
\fe
where $V_{\phi}$ is the regularized length of the Liouville direction. This result agrees with the genus 1 matrix model free energy if one makes a further identification
\ie
V_{\phi}=-\frac12\log\mu.
\fe
Along with the identification of $\mu^{-1}=2\pi g_s$, the duality predicts
\ie
F(g_s,R)
=
\left[
F_{\mathrm{MM}}(\mu,R)
- \frac{\mu^2\log\mu}{4\pi}
\right]_{\mu^{-1}=2\pi g_s},
\fe
where the matrix model free energy on the RHS is given exactly by equation \eqref{grandcanonical_free_energy} and we exclude the genus $0$ free energy of the matrix model from this comparison.\footnote{A priori, the counterterm prescription for the matrix model free energy is not unique. Choosing a different prescription shifts $F_{\mathrm{MM}}(\mu,R)$ by a finite polynomial in $\mu$, whose coefficients may depend on $R$. Such a shift leaves the coefficients of $\log\mu$ and $\mu^{-2L}$ ($L\geq1$), as well as the nonperturbative contributions, unchanged. Comparing these terms with the corresponding terms in the $c=1$ string free energy is therefore independent of this counterterm ambiguity.}

Apart from the thermal free energy, another thermal observable one can access from the worldsheet is the thermal $S$-matrix. The worldsheet formulation of  $1\to n$ thermal $S$-matrix and thermal amplitude $\widetilde{\mathcal{A}}^{(T)}_{1\to n}(\omega_i ,g_s) $ will be discussed in detail in section \ref{sec:thermal_S_matrix}.  The duality predicts the thermal amplitudes to be
\ie
\widetilde{\mathcal{A}}^{(T)}_{1\to n}(\omega_i ,g_s) = \widetilde{\mathcal{A}}^{(T)}_{1\to n,\mathrm{MM}}(\omega_i, \mu^{-1}=2\pi g_s),
\fe
where the matrix model thermal amplitude on the RHS is given by equation \eqref{eq:thermalamplitude}.

Lastly, we emphasize that the identification $\mu^{-1}=2\pi g_s$ as well as $V_{\phi}= -\frac{1}{2}\log \mu$ is part of the conjectured duality, and neither follows from the structure of the genus expansion alone. We may make a weaker version of the conjecture that allows the dictionary to be of the form
\ie
\mu^{-1}=m_1(R)g_s+m_3(R)g_s^3+\mathcal{O}(g_s^5), \quad V_{\phi}= -\frac{1}{2}\log \mu+a_3(R) \mu^{-2}+\mathcal{O}\left(\mu^{-4}\right).
\fe
This more general dictionary preserves the structure of genus expansion for string theory and matrix model observables, while mixing the coefficients in the $g_s$ expansion and $\mu^{-1}$ expansion. A comparison at tree level alone cannot rule out the higher order corrections in $g_s$ in the dictionary, as well as the $R$ dependent corrections. 

As we will see in sections ~\ref{sec:numerical_amplitude}, \ref{sec:numerical_thermal_amplitude}, and \ref{sec:genus2numerics}, the worldsheet tree level and one loop thermal amplitudes, as well as the genus two free energy, suggest that
\ie
m_1(R)=2\pi,\qquad m_3(R)=0, \qquad a_3(R) =0.
\fe
Therefore, our computations provide support for the exact identification of  $\mu^{-1}=2\pi g_s$ as well as $V_{\phi}= -\frac{1}{2}\log \mu$.

\section{Higher point amplitude from worldsheet}\label{sec:higher-point-amplitudes}

In this section, we compute the tree level $1\to 4$ amplitude $\mathcal{A}^{(0)}_{1\to 4}$, tree level $1\to 5$ amplitude $\mathcal{A}^{(0)}_{1\to 5}$ and one loop $1\to 2$ amplitude $\mathcal{A}^{(1)}_{1\to 2}$ in $c=1$ string theory from the worldsheet moduli integration numerically, extending the analysis of \cite{BalthazarRodriguezYin:2019}. 

In each case, we compute the worldsheet amplitude by integrating a product of the $bc$ ghost, time-like free boson, and Liouville correlators over moduli space. We evaluate the Liouville correlators via conformal block
decomposition and compute the conformal block numerically using the $c$ recursion relation in~\cite{Cho-Collier-Yin}. The precise moduli integrand is introduced in section \ref{sec:1to4amp}, for the tree level $1\to4$, $1\to 5$ amplitude, and section \ref{sec:g1_1to2amp} for genus one $1\to 2$ amplitudes. We perform the moduli integration using Monte Carlo methods, which will be elaborated in Sec.\ref{sec:numerical_amplitude}.

At physical real frequencies, the moduli integrand has power law divergences due to OPE singularities, and the integral must be regularized. In principle, this can be achieved by explicitly subtracting divergent terms, as explained in \cite{BalthazarRodriguezYin:2019}. Such subtractions become increasingly cumbersome for higher point amplitudes because of the multiple degeneration regions and potentially large numerical cancellations. We instead analytically continue the scattering amplitude from real tachyon energy to imaginary tachyon energy $\omega_j=it_j$, and evaluate the scattering amplitudes at energies where the moduli integral converges. This procedure would require a proper prescription for the analytic continuation of the Liouville correlators.

We evaluate the worldsheet amplitude numerically in the imaginary frequency regime and compare with the matrix model predictions analytically continued to the same imaginary frequencies. As illustrated in section \ref{sec:numerical_amplitude}, the worldsheet numerics agree with the matrix model answer, providing a direct test of the proposed duality.

\subsection{Tree level \texorpdfstring{$1\to 4$}{1 to 4} and \texorpdfstring{$1\to 5$}{1 to 5} scattering amplitudes}\label{sec:1to4amp}
The general $1\to n$ tree level amplitude can be computed by the expression
\ie\label{eq:tree1tonamplitude}
&\delta\left(\omega-\sum_{j=1}^n\omega_j\right)
g_s^{n-1}\mathcal{A}_{1\to n}^{(0)}
\\
&=\int d^2z_0\prod_{i=4}^n d^2z_i\,
\left\langle
\mathcal{V}_\omega^+(z_0,\bar z_0)
c\tilde c\,\mathcal{V}_{\omega_1}^-(0)
c\tilde c\,\mathcal{V}_{\omega_2}^-(1)
c\tilde c\,\mathcal{V}_{\omega_3}^-(\infty)
\prod_{i=4}^n\mathcal{V}_{\omega_i}^-(z_i,\bar z_i)
\right\rangle
\\
&=2\pi i\,g_s^{n-1}
\delta\left(\omega-\sum_{j=1}^n\omega_j\right)
\int d^2z_0\prod_{i=4}^n d^2z_i\,
\prod_{\substack{0\leq i<j\leq n\\ i,j\neq3}}
|z_{ij}|^{-k_i k_j}
\\
&\qquad\times
\left\langle
V_{\frac{\omega}{2}}(z_0,\bar z_0)
V_{\frac{\omega_1}{2}}(0)
V_{\frac{\omega_2}{2}}(1)
V_{\frac{\omega_3}{2}}(\infty)
\prod_{i=4}^n V_{\frac{\omega_i}{2}}(z_i,\bar z_i)
\right\rangle_{\rm Liouville}.
\fe
where
\[
k_0=\omega,\qquad
k_i=-\omega_i\quad(i=1,\ldots,n),\qquad
z_{ij}=z_i-z_j,\qquad
(z_1,z_2,z_3)=(0,1,\infty).
\]
In this paper, we specifically consider the $1\to 4$ and $1\to 5$ amplitudes.

\begin{figure}[htbp]
\centering
\tikzset{every picture/.style={line width=0.75pt}}

\begin{tikzpicture}[x=0.75pt,y=0.75pt,yscale=-1,xscale=1]
\draw (80.29,201.16) -- (303.96,200.96);
\draw (129.99,126.94) -- (131,201.96);
\draw (191.11,126.04) -- (192.12,201.06);
\draw (252.63,125.43) -- (253.63,200.44);

\draw (373.93,201.16) -- (597.6,200.96);
\draw (423.63,126.94) -- (424.64,201.96);
\draw (479.65,124.01) -- (506.5,159.71);
\draw (506.5,159.71) -- (533.65,124.12);
\draw (506.5,159.71) -- (506.5,200.98);

\draw (82.32,177.36) node [anchor=north west,inner sep=0.75pt,align=left]
  {$\displaystyle h_{\frac{\omega_2}{2}}$};
\draw (121.81,103.35) node [anchor=north west,inner sep=0.75pt,align=left]
  {$\displaystyle h_{\frac{\omega}{2}}$};
\draw (183.5,103.35) node [anchor=north west,inner sep=0.75pt,align=left]
  {$\displaystyle h_{\frac{\omega_1}{2}}$};
\draw (245.19,103.35) node [anchor=north west,inner sep=0.75pt,align=left]
  {$\displaystyle h_{\frac{\omega_3}{2}}$};
\draw (278.5,177.36) node [anchor=north west,inner sep=0.75pt,align=left]
  {$\displaystyle h_{\frac{\omega_4}{2}}$};
\draw (149.83,179.19) node [anchor=north west,inner sep=0.75pt,align=left]
  {$\displaystyle h_{P_1}$};
\draw (211.52,179.19) node [anchor=north west,inner sep=0.75pt,align=left]
  {$\displaystyle h_{P_2}$};

\draw (375.97,177.36) node [anchor=north west,inner sep=0.75pt,align=left]
  {$\displaystyle h_{\frac{\omega_2}{2}}$};
\draw (415.45,103.35) node [anchor=north west,inner sep=0.75pt,align=left]
  {$\displaystyle h_{\frac{\omega}{2}}$};
\draw (526.49,106.2) node [anchor=north west,inner sep=0.75pt,align=left]
  {$\displaystyle h_{\frac{\omega_3}{2}}$};
\draw (472.14,106.2) node [anchor=north west,inner sep=0.75pt,align=left]
  {$\displaystyle h_{\frac{\omega_1}{2}}$};
\draw (572.14,177.36) node [anchor=north west,inner sep=0.75pt,align=left]
  {$\displaystyle h_{\frac{\omega_4}{2}}$};
\draw (454.58,179.19) node [anchor=north west,inner sep=0.75pt,align=left]
  {$\displaystyle h_{P_1}$};
\draw (508.8,168.65) node [anchor=north west,inner sep=0.75pt,align=left]
  {$\displaystyle h_{P_2}$};
\end{tikzpicture}
\caption{Two examples of the sphere 5-point Virasoro conformal block in linear
channel. The left diagram is the channel used in
\eqref{eq:5ptblock_decomposition}.}
\label{fig:sphere-5-point-linear-block}
\end{figure}
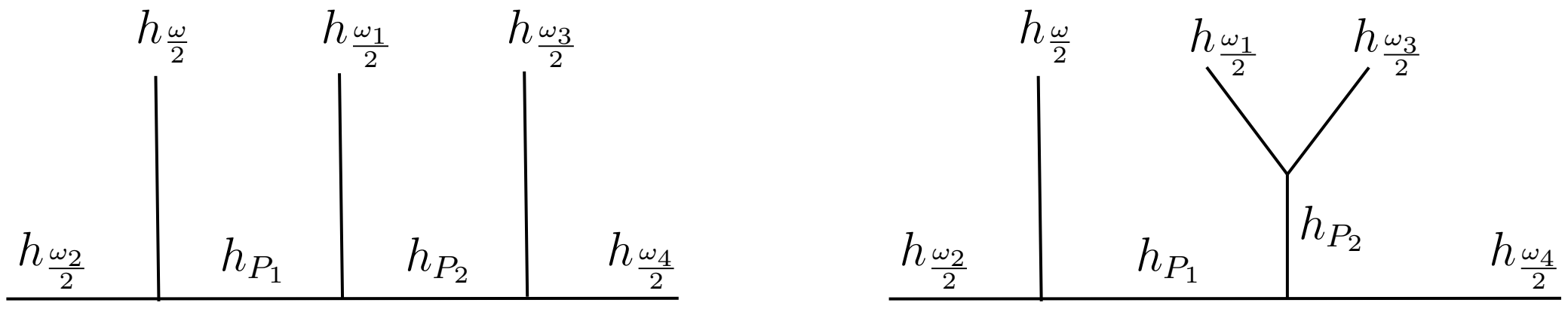

We compute the Liouville 5- and 6-point functions by conformal block decomposition.  As an example, consider the domain of the 5-point function where $0<|z_1|<|z_2|<1$. The Liouville correlator can be decomposed into Virasoro 5-point block as (see LHS of Fig. \ref{fig:sphere-5-point-linear-block})
\ie\label{eq:5ptblock_decomposition}
{}&\left\langle V_{\omega\over 2}(z_1,\bar z_1) V_{\omega_1\over 2}(z_2,\bar z_2) V_{\omega_2\over 2}(0) V_{\omega_3\over 2}(1) V_{\omega_4\over 2}(\infty) \right\rangle_{\rm Liouville}=\\
& \int_{0}^{\infty} \frac{dP_1}{\pi}\int_0^{\infty}\frac{dP_2}{\pi}   \, {\cal C}\left({\omega_2\over 2}, {\omega\over 2}, P_1\right) {\cal C}\left(P_1,{\omega_1\over 2}, P_2\right) {\cal C}\left(P_2, {\omega_3\over 2}, {\omega_4\over 2} \right)
\left|G(P_i,\omega_i\mid z_1,z_2)\right|^2,
\fe
where
\ie
G(P_i, \omega_i\mid  z_1,z_2) = z_1^{P_1^2-1- \frac{\omega^2+\omega_2^2}{4}} z_2^{P_2^2-P_1^2-1- \frac{\omega_1^2}{4}} F_5\left(P_i,\omega_i \mid \frac{z_1}{z_2},z_2\right)
\fe
and $F_5$ is the 5-point holomorphic Virasoro block in the linear channel. The 5-point block is computed numerically using the c-recursion relation derived in \cite{Cho-Collier-Yin}, and the Liouville momentum integrals are also evaluated numerically.  %

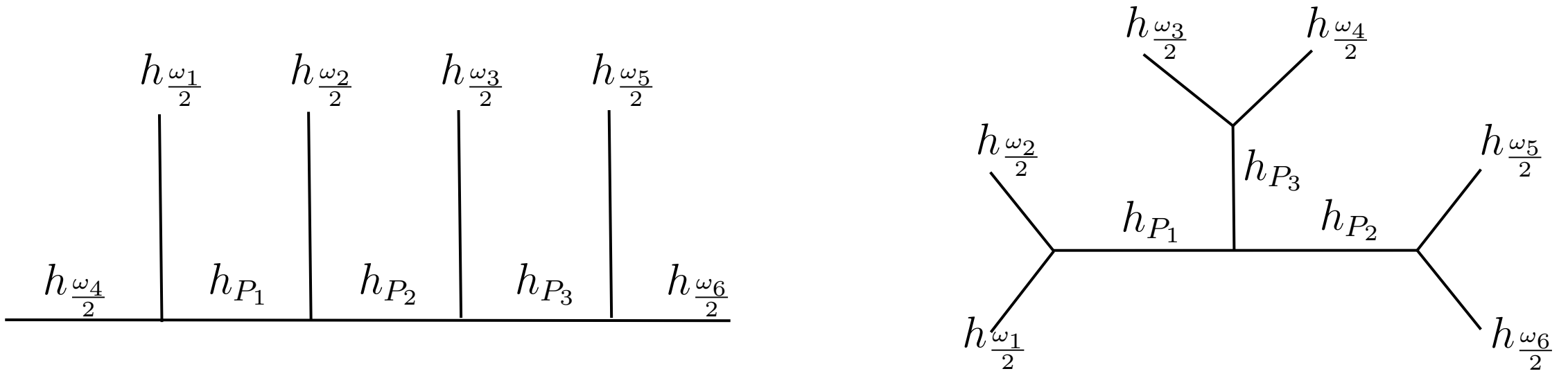
\begin{figure}[htbp]
\centering
\tikzset{every picture/.style={line width=0.75pt}}

\begin{tikzpicture}[x=0.75pt,y=0.75pt,yscale=-1,xscale=1]
\draw (31.5,231.52) -- (294.08,231.84);
\draw (87.38,157.03) -- (88.27,232.32);
\draw (141.36,156.13) -- (142.25,231.41);
\draw (195.68,155.52) -- (196.57,230.8);
\draw (250.16,155.52) -- (251.04,230.8);

\draw (410.77,206.37) -- (542.7,206.34);
\draw (565.76,177) -- (542.7,206.34);
\draw (565.76,235) -- (542.7,206.34);
\draw (443.63,135.31) -- (475.99,161.21);
\draw (504.63,133.91) -- (475.99,161.21);
\draw (475.99,161.21) -- (476.43,206.03);
\draw (388.36,236.03) -- (411.16,206.63);
\draw (388.17,178.03) -- (411.16,206.63);

\draw (43.65,209.7) node [anchor=north west,inner sep=0.75pt,align=left]
  {$\displaystyle h_{\frac{\omega_4}{2}}$};
\draw (78.51,133.42) node [anchor=north west,inner sep=0.75pt,align=left]
  {$\displaystyle h_{\frac{\omega_1}{2}}$};
\draw (132.99,133.42) node [anchor=north west,inner sep=0.75pt,align=left]
  {$\displaystyle h_{\frac{\omega_2}{2}}$};
\draw (187.47,133.42) node [anchor=north west,inner sep=0.75pt,align=left]
  {$\displaystyle h_{\frac{\omega_3}{2}}$};
\draw (103.44,209.48) node [anchor=north west,inner sep=0.75pt,align=left]
  {$\displaystyle h_{P_1}$};
\draw (157.92,209.48) node [anchor=north west,inner sep=0.75pt,align=left]
  {$\displaystyle h_{P_2}$};
\draw (241.95,133.42) node [anchor=north west,inner sep=0.75pt,align=left]
  {$\displaystyle h_{\frac{\omega_5}{2}}$};
\draw (214.57,209.48) node [anchor=north west,inner sep=0.75pt,align=left]
  {$\displaystyle h_{P_3}$};
\draw (269.27,209.7) node [anchor=north west,inner sep=0.75pt,align=left]
  {$\displaystyle h_{\frac{\omega_6}{2}}$};

\draw (500.62,116.46) node [anchor=north west,inner sep=0.75pt,align=left]
  {$\displaystyle h_{\frac{\omega_4}{2}}$};
\draw (376.49,228.85) node [anchor=north west,inner sep=0.75pt,align=left]
  {$\displaystyle h_{\frac{\omega_1}{2}}$};
\draw (381.35,158.85) node [anchor=north west,inner sep=0.75pt,align=left]
  {$\displaystyle h_{\frac{\omega_2}{2}}$};
\draw (435.23,116.35) node [anchor=north west,inner sep=0.75pt,align=left]
  {$\displaystyle h_{\frac{\omega_3}{2}}$};
\draw (433.99,186.91) node [anchor=north west,inner sep=0.75pt,align=left]
  {$\displaystyle h_{P_1}$};
\draw (505.86,186.25) node [anchor=north west,inner sep=0.75pt,align=left]
  {$\displaystyle h_{P_2}$};
\draw (563.71,158.99) node [anchor=north west,inner sep=0.75pt,align=left]
  {$\displaystyle h_{\frac{\omega_5}{2}}$};
\draw (477.99,168.21) node [anchor=north west,inner sep=0.75pt,align=left]
  {$\displaystyle h_{P_3}$};
\draw (567.58,228.98) node [anchor=north west,inner sep=0.75pt,align=left]
  {$\displaystyle h_{\frac{\omega_6}{2}}$};
\end{tikzpicture}
\caption{Two examples of the sphere 6-point Virasoro conformal block
channels. The left diagram is the linear channel. The right diagram is the star channel.}
\label{fig:sphere-6-point-linear-block}
\end{figure}

For the sphere 5-point function, there are 120 different channels one can choose for the conformal block decomposition. For the sphere 6-point function, there are 1440 distinct conformal block decomposition channels, including 720 linear channels and 720 star channels. The two topologies of the conformal block decomposition channels for sphere 6-point function are shown in Fig.~\ref{fig:sphere-6-point-linear-block}.  For each point on the moduli space, there are multiple choices of channels. We choose one based on the convergence of the power series expansion of the conformal block and numerical efficiency considerations. The details of our choice of the conformal decomposition channel are explained in Appendix~\ref{App:Genus01to4Numerics} for tree level $1\to 4$ amplitude and  Appendix~\ref{App:Genus01to5Numerics} for tree level $1\to 5$ amplitude.

The integrand in equation \eqref{eq:tree1tonamplitude} suffers from power law divergences near $z_i\rightarrow z_j$ due to OPE singularities, and thus the moduli integral must be regularized. Analyticity of the scattering amplitude suggests the following regularization scheme: evaluate the amplitude for complex tachyon energies $\{\omega_i\}$ subject to energy conservation, in a domain where the integral in equation \eqref{eq:tree1tonamplitude} converges, and then analytically continue the result to the physical values of $\omega_i$.\footnote{This analytic regularization scheme is in the same spirit as the one used in the textbook computation of the Virasoro--Shapiro amplitude, where divergences at the boundaries of moduli space are treated by analytic continuation in the Mandelstam variables.} In our computation, we choose
\ie
\omega_j = i t_j \quad t_j >0,
\fe
with $t_j$ sufficiently small so that the moduli integrand is integrable.

In this prescription, the Liouville correlator also needs to be analytically continued carefully.  For real external Liouville momenta $\{\frac{\omega_i}{2}\}$, Liouville correlators are computed by conformal block decomposition with the internal momenta integrated along $P_i\in[0,\infty)$.  As the external momenta are analytically continued to complex values, the poles of the DOZZ structure constants move in the internal momentum planes.  If a pole crosses an integration contour, the contour must be deformed to preserve analyticity.  For example, for the Virasoro 5-point function  in the linear channel \eqref{eq:5ptblock_decomposition}, the poles are located at
\ie
P_1=\frac{\epsilon_0\omega+\epsilon_1\omega_2}{2}+in,
\quad
P_2=\epsilon P_1+\frac{\epsilon_2\omega_1}{2}+in,
\quad
P_2=\frac{\epsilon_3\omega_3+\epsilon_4\omega_4}{2}+in,
\fe
where $ \epsilon, \epsilon_i= \pm 1$ and $n\in\mathbb Z\setminus\{0\}$. When we analytic continuate in the locus where a pole in $P_1,P_2$ passes across the Liouville momentum integration contour $P_i\geq 0$, an appropriate contour deformation will be required. We will follow the prescription of \cite{BalthazarRodriguezYin:2019}. Whenever a pole crosses the Liouville momentum integration contour, we include the corresponding residue contribution in the correlation function.

With the analytic continuation prescription, the moduli integrand is unambiguously specified and is accessible numerically. The strategy we use for numerical integration over the moduli space is explained further in section~\ref{sec:numerical_amplitude}. We explain the precise numerical details in Appendix \ref{App:Genus01to4Numerics}.

\subsection{Genus one \texorpdfstring{$1\to 2$}{1 to 2} Scattering amplitude}\label{sec:g1_1to2amp}

The genus one $1\to 2$ scattering amplitude in $c=1$ string theory can be computed by the formula
\ie\label{eq:genus1threepoint}
{}&\delta\left(\omega-\sum_{i=1}^2\omega_i\right)g_s^3 \mathcal{A}_{1\to 2}^{(1)} \\
{}&=2\pi^2  \int_{\mathcal{F}_1} d^2\tau \int_{T^2(\tau)} d^2z_1 d^2z_2 \left\langle \tilde{b}b \mathcal{V}^+_{\omega}(z_1,\bar{z}_1)\mathcal{V}^-_{\omega_1}(z_2,\bar{z}_2)c\tilde{c}\mathcal{V}^-_{\omega_2}(0) \right\rangle_{T^2(\tau)} \\
&=  2\pi^2 i g_s^3 ~\delta\left(\omega-\sum_{i=1}^2\omega_i\right) \\
&\times \int_{\mathcal{F}_1} d^2\tau \int_{T^2(\tau)} d^2z_1 d^2z_2~ \frac{|\eta(\tau)|^2}{\sqrt{\tau_2}} \prod_{1\leq i<j\leq3} E\left(z_{ij}\mid\tau\right)^{-k_i k_j} \left\langle V_{\frac{\omega}{2}}(z_1,\bar{z}_1)V_{\frac{\omega_1}{2}}(z_2,\bar{z}_2)V_{\frac{\omega_2}{2}}(0)\right\rangle_{\mathrm{Liouville}, T^2(\tau)},
\fe
where
\ie\label{eq:genus1defs}
{}&k_1= \omega,\quad k_2= -\omega_1,\quad k_3= -\omega_2,\quad z_{ij}=z_i-z_j, \quad z_3=0,\\
&E\left(z_{ij}\mid\tau\right):= \left|
\frac{2\pi}{\partial_z\theta_1(0\mid\tau)}
\theta_1\!\left(\frac{z_{ij}}{2\pi}\,\middle|\,\tau\right)
\right|
\exp\!\left[
-\frac{\left(\operatorname{Im}z_{ij}\right)^2}{4\pi\tau_2}
\right],\\
{}&\theta_1(w\mid\tau)
:=
2\sum_{n=0}^{\infty}(-1)^n
e^{\pi i\tau(n+\frac12)^2}
\sin\!\bigl[(2n+1)\pi w\bigr],
\fe
and $\mathcal{F}_1$ is the fundamental domain for the action of
$\operatorname{PSL}(2,\mathbb{Z})$ on the upper half plane 
\ie\label{eq:genus1funddomain}
\mathcal{F}_1
=
\left\{
\tau\in\mathbb{C}
\;\middle|\;
\operatorname{Im}\tau> 0,\quad -\frac{1}{2}\leq \operatorname{Re}\tau\leq\frac{1}{2},
\quad |\tau|\geq1
\right\}.
\fe
The torus 3-point function of Liouville primaries can be computed by conformal block decomposition. We compute the conformal block numerically from the c-recursion relation. 
\begin{figure}[htbp]
\centering
\tikzset{every picture/.style={line width=0.75pt}}

\begin{tikzpicture}[x=0.75pt,y=0.75pt,yscale=-1,xscale=1]
\draw (67.53,153.25) .. controls (67.53,134.34) and (82.19,119.01) .. (100.27,119.01)
  .. controls (118.35,119.01) and (133,134.34) .. (133,153.25)
  .. controls (133,172.16) and (118.35,187.49) .. (100.27,187.49)
  .. controls (82.19,187.49) and (67.53,172.16) .. (67.53,153.25) -- cycle;
\draw (100.27,118.8) -- (100.56,83.63);
\draw (73.22,172.93) -- (43.62,200.61);
\draw (157.7,200.61) -- (127.32,172.93);

\draw (312.46,120.57) .. controls (330.54,120.54) and (345.22,135.85) .. (345.26,154.76)
  .. controls (345.29,173.67) and (330.66,189.03) .. (312.58,189.06)
  .. controls (294.5,189.09) and (279.82,173.78) .. (279.78,154.87)
  .. controls (279.75,135.96) and (294.38,120.6) .. (312.46,120.57) -- cycle;
\draw (345.45,154.76) -- (379.07,155.01);
\draw (246.16,154.62) -- (219.65,123.71);
\draw (219.76,186.45) -- (246.16,154.62);
\draw (246.16,154.62) -- (279.79,154.88);

\draw (572.17,120.57) .. controls (590.25,120.54) and (604.93,135.85) .. (604.96,154.76)
  .. controls (605,173.67) and (590.37,189.03) .. (572.28,189.06)
  .. controls (554.2,189.09) and (539.52,173.78) .. (539.49,154.87)
  .. controls (539.46,135.96) and (554.09,120.6) .. (572.17,120.57) -- cycle;
\draw (460.75,195.13) -- (505.31,153.96);
\draw (505.31,153.96) -- (460.73,101.37);
\draw (456.76,145.35) -- (478.82,123.87);
\draw (505.31,153.96) -- (538.93,154.21);

\draw (45.04,122.42) node [anchor=north west,inner sep=0.75pt,align=left]
  {$\displaystyle h_{P_3}$};
\draw (131.42,124.88) node [anchor=north west,inner sep=0.75pt,align=left]
  {$\displaystyle h_{P_1}$};
\draw (92.08,187.08) node [anchor=north west,inner sep=0.75pt,align=left]
  {$\displaystyle h_{P_2}$};
\draw (96.63,68.17) node [anchor=north west,inner sep=0.75pt,align=left]
  {$\displaystyle h_{\frac{\omega}{2}}$};
\draw (156.79,197.35) node [anchor=north west,inner sep=0.75pt,align=left]
  {$\displaystyle h_{\frac{\omega_1}{2}}$};
\draw (36.54,198.66) node [anchor=north west,inner sep=0.75pt,align=left]
  {$\displaystyle h_{\frac{\omega_2}{2}}$};

\draw (305.15,97.82) node [anchor=north west,inner sep=0.75pt,align=left]
  {$\displaystyle h_{P_3}$};
\draw (252.22,133.8) node [anchor=north west,inner sep=0.75pt,align=left]
  {$\displaystyle h_{P_1}$};
\draw (304.15,188.08) node [anchor=north west,inner sep=0.75pt,align=left]
  {$\displaystyle h_{P_2}$};
\draw (213.38,104.47) node [anchor=north west,inner sep=0.75pt,align=left]
  {$\displaystyle h_{\frac{\omega}{2}}$};
\draw (210.98,181.82) node [anchor=north west,inner sep=0.75pt,align=left]
  {$\displaystyle h_{\frac{\omega_1}{2}}$};
\draw (379.74,144.99) node [anchor=north west,inner sep=0.75pt,align=left]
  {$\displaystyle h_{\frac{\omega_2}{2}}$};

\draw (563.85,101.28) node [anchor=north west,inner sep=0.75pt,align=left]
  {$\displaystyle h_{P_3}$};
\draw (487.41,118.04) node [anchor=north west,inner sep=0.75pt,align=left]
  {$\displaystyle h_{P_1}$};
\draw (512.28,133.49) node [anchor=north west,inner sep=0.75pt,align=left]
  {$\displaystyle h_{P_2}$};
\draw (450.57,83.32) node [anchor=north west,inner sep=0.75pt,align=left]
  {$\displaystyle h_{\frac{\omega}{2}}$};
\draw (445.57,140.07) node [anchor=north west,inner sep=0.75pt,align=left]
  {$\displaystyle h_{\frac{\omega_1}{2}}$};
\draw (448.57,188.66) node [anchor=north west,inner sep=0.75pt,align=left]
  {$\displaystyle h_{\frac{\omega_2}{2}}$};
\end{tikzpicture}
\caption{Three examples of Virasoro conformal block decomposition channels for the torus 3-point function. Left to right, they are the necklace, pair-OPE, and comb channels.}
\label{fig:torus-3-point-necklace-block}
\end{figure}
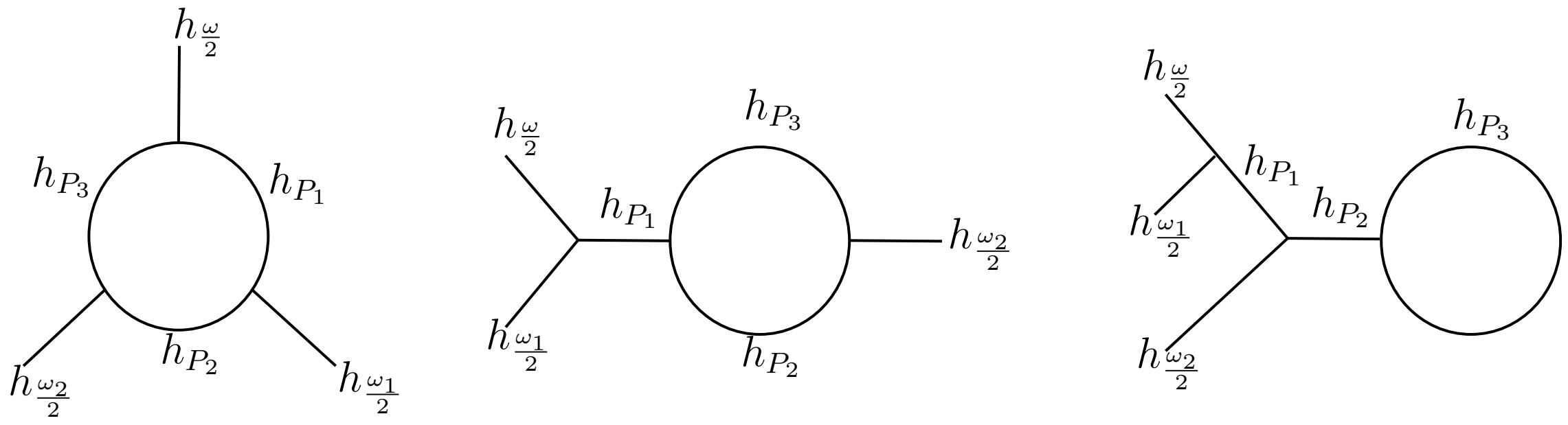

As in the case of $1\to 4$ scattering amplitude, the moduli integrand suffers power law divergences, thus the integral needs to be regularized. We regularize the integral by analytic continuation of the tachyon energy, as elaborated in section \ref{sec:1to4amp}. In our computation, we choose
\ie
\omega_j = i t_j \quad t_j >0 \quad j=1,2 ,
\fe
with $t_j$ sufficiently small so that the moduli integrand is integrable. The Liouville correlator is also treated using the same prescription as in section \ref{sec:1to4amp}. The details of the numerics are presented in Appendix \ref{App:Genus11to2Numerics}.

\subsection{Numerical Results}\label{sec:numerical_amplitude}

Based on the complete prescription for the moduli integrands for $\mathcal{A}^{(0)}_{1\to 4}$, $\mathcal{A}^{(0)}_{1\to 5}$ and $\mathcal{A}^{(1)}_{1\to 2}$ explained in the previous sections, we carry out the moduli integration using Monte Carlo methods. 

\begin{figure}[p]
\centering
\captionsetup{font=small,skip=6pt}
\subfloat[tree level $1\to4$]{%
  \includegraphics[width=0.48\textwidth]{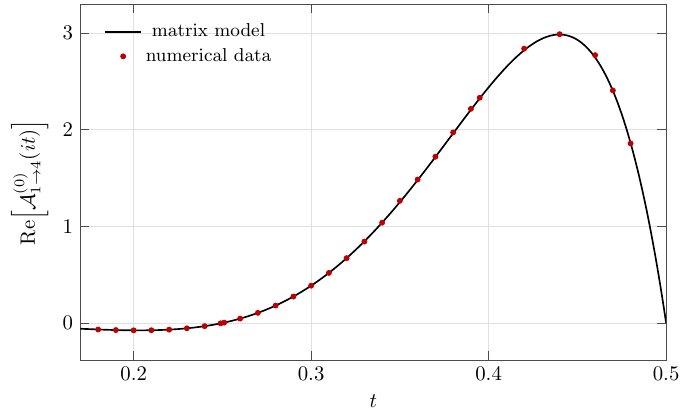}%
  \label{fig:tree-one-to-four-separate}}
\hfill
\subfloat[tree level $1\to5$]{%
  \includegraphics[width=0.48\textwidth]{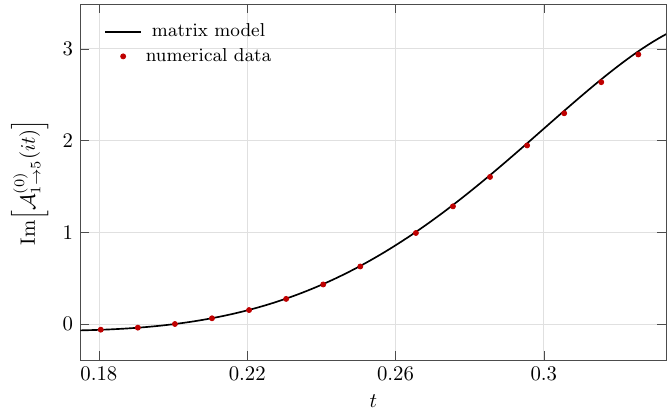}%
  \label{fig:tree-one-to-five-separate}}
\par\medskip
\subfloat[Genus one $1\to2$]{%
  \includegraphics[width=0.48\textwidth]{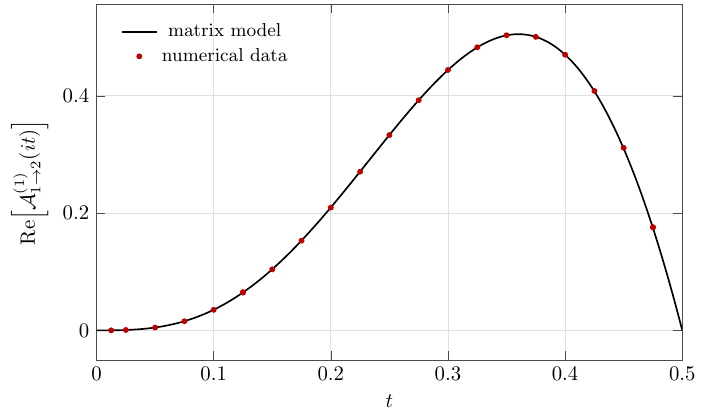}%
  \label{fig:one loop-one-to-two-separate}}
\caption{Numerical worldsheet results (red points) compared with the
matrix model predictions (black curves) using the dictionary $\mu^{-1}=2\pi g_s$.
Panels (a) and (b) show the real part of the tree level $1\to4$ amplitude
and the imaginary part of the tree level $1\to5$ amplitude, respectively,
at equal outgoing energies $\omega_j=it$. Panel (c) shows the real part of
the genus one $1\to2$ amplitude at equal outgoing energies
$\omega_1=\omega_2=it$, with incoming energy $\omega=2it$. The components not shown are
at most of order $10^{-18}$ numerically and vanish in the matrix model
predictions. Panels (a) and (b) are computed via equation \eqref{eq:tree1tonamplitude} and panel (c) via equation \eqref{eq:genus1threepoint}.}
\label{fig:amplitude-numerical-comparisons}
\end{figure}

We consider the kinematic region where the outgoing energies of the tachyons take the same value, $\{\omega_j=it ,t > 0 \}$, study the $t$ dependence of the scattering amplitude and compare with the matrix model answer. The numerical amplitudes at the chosen values of $t$ are plotted alongside the matrix model predictions in Fig.~\ref{fig:amplitude-numerical-comparisons}. The results show agreement with the matrix model prediction, providing a further check of the duality between $c=1$ string theory and its dual matrix model.  

The method can also be applied to other choices of tachyon energies, provided the moduli integrals converge. The genus one $1\to2$ result also tests the relation between $\mu$ and $g_s$ beyond leading order. If the coupling dictionary takes the form
\ie
\mu^{-1}=2\pi g_s+m_3(\infty)g_s^3+\mathcal{O}(g_s^5),
\fe
matching the coefficients of $\mathcal{A}_{1\to 2}$ and  $\mathcal{A}_{1\to 2,\mathrm{MM}}$ at order $g_s^3$ gives the following relation between the genus-one amplitudes:
\ie
\mathcal A^{(1)}_{1\to2}
=(2\pi)^3\mathcal A^{(1)}_{1\to2,\mathrm{MM}}
+m_3(\infty)\mathcal A^{(0)}_{1\to2,\mathrm{MM}}.
\fe
The observed agreement therefore supports $m_3(\infty)=0$ within our numerical accuracy. In Sec.~\ref{sec:thermal_S_matrix}, we use thermal amplitudes to further investigate the possible temperature dependence of the coupling dictionary through order $g_s^3$.

\clearpage

\clearpage
\section{Thermal \texorpdfstring{$S$}{S}-matrix from the worldsheet}\label{sec:thermal_S_matrix}

In this section, we formulate the thermal amplitudes in $c=1$ string theory from the worldsheet and compare them with the results from the matrix model.

To formulate the worldsheet theory of the \(c=1\) string in a thermal background, we analytically continue the time like free boson $X^0$ to a Euclidean compact boson $X=i X^0$ with the identification $X \sim X+2\pi R$. Closed string modes therefore carry discrete momentum corresponding to the Matsubara frequencies \(\omega=\frac{i m}{R}\), with \(m\in\mathbb{Z}\), and can have integer winding numbers around the thermal circle. The closed string vertex operators carrying discrete momentum take the form
\ie
g_s c\tilde c\,e^{\pm i m X/R}\,V_{\frac{im}{2R}}.
\fe
Here $V_{\frac{im}{2R}}$ is the Liouville primary operator after analytic continuation, whose correlation functions follow from the same prescription introduced in section \ref{sec:1to4amp}. 
We define the thermal amplitudes of $c=1$ string theory $\tilde{\mathcal{A}}^{(T)}\left(\omega_j =\frac{i m_j}{R}\right)$ through the usual prescription for the worldsheet amplitudes with the insertion of vertex operators of the form $c\tilde c\,e^{\pm i m X/R}V_{\frac{im}{2R}}$, where a factor of $-i R\delta_{m,\sum m_i}$ is stripped off from the amplitude. We strip off a factor of $g_s$ from the definition of the vertex operators, instead including it as a prefactor to the amplitude as in equation \eqref{eq:ThermalAmplitude}.

For concreteness, we consider the $1\to n$ thermal amplitude in $c=1$ string theory. By construction, it admits the structure of a genus expansion:
\ie\label{eq:ThermalAmplitude}
\tilde{\mathcal{A}}_{1\to n}^{(T)}\left(\omega_j =\frac{i m_j}{R}, g_s\right)= \sum_{L=0}^{\infty} g_s^{2L-1+n} \tilde{\mathcal{A}}_{1\to n}^{(T,L)} \left(\omega_j =\frac{i m_j}{R}\right)+\mathcal{O}(e^{-1/g_s}).
\fe
Here $\tilde{\mathcal{A}}^{(T,L)}$ is the worldsheet genus $L$ $1\to n$ amplitude with vertex operators $g_s c\tilde c\,e^{\pm i m X/R}V_{\frac{im}{2R}}$ insertion, computed by integrating over the product of Liouville correlator, compact boson correlator and $b c$ ghost correlator over the moduli space of genus $L$ Riemann surface with $n+1$ punctures.  The omitted terms in $\mathcal{O}(e^{-1/g_s})$ are the non-perturbative corrections defined by $D$-instanton perturbation theory.

Usually, the thermal amplitude defined in this way does not have a clear physical interpretation, as the concept of asymptotic state is not usually well-defined in a system interacting with a heat bath. In $c=1$ string theory, however, the interaction dies off when the string states are far away from the Liouville wall. This makes it possible to connect the worldsheet thermal amplitude with the $S$-matrix like thermal observables.

In analogy with the ordinary string $S$-matrix, we define the $1\to n$ thermal $S$-matrix of $c=1$ string theory by analytically continuing the thermal amplitude $\widetilde{\mathcal A}^{(T)}_{1\to n}$ from Matsubara frequencies $\omega_j=im_j/R$ through the upper half complex $\omega_j$-plane to the positive real axis, $\omega_j\to E_j+i0^+$ with $E_j>0$, and applying the thermal LSZ prescription.\footnote{This analytic continuation is analogous to the $i\epsilon$ prescription in the worldsheet definition of scattering amplitudes. For a general $m\to n$ thermal $S$-matrix, one also needs to consider the decomposition of the S-matrix into connected contributions via cluster decomposition and a suitable thermal LSZ prescription to relate the thermal $S$-matrix to connected thermal amplitudes, as discussed in \cite{BalthazarRodriguezYin:2022Type0BInstanton}.}
Writing the real external energies as $\omega_j$, we define
\ie\label{eq:WSthermalSmatrix}
S^{(T)}_{1\to n}(\omega\mid\omega_1,\dots,\omega_n)
:=
\delta\left(\omega-\sum_{j=1}^n\omega_j\right)
\prod_{j=1}^n\frac{1}{1-e^{-\beta\omega_j}}\,
\widetilde{\mathcal A}^{(T)}_{1\to n}
(\omega_1,\dots,\omega_n),
\qquad \omega_j>0.
\fe
Its counterpart on the matrix model side of the duality is the thermal $S$-matrix of equation \eqref{eq:thermalSmatrix}.

At each order in the genus expansion, the analytic continuation of the thermal amplitude from Matsubara frequencies is unique under the assumption that \(\widetilde{\mathcal A}^{(T,L)}_{1\to n}\) is a polynomial in the external tachyon energies \(\omega_j\), as anticipated from the matrix model.\footnote{An analogous analytic continuation prescription may be proposed for D-instanton contributions to thermal amplitudes. Motivated by the matrix model, we tentatively assume that, at each order in the string coupling within a fixed instanton sector, the thermal amplitude analytically continued from Matsubara frequencies belongs to the class of finite sums of terms \(P(\omega_i;R)e^{2\pi\sum_i r_i\omega_i}\), where \(P\) is a polynomial in the external energies and \(r_i\) are real constants independent of \(R\). Independent worldsheet computations of thermal amplitudes at Matsubara frequencies would provide a test of this proposal and are left for future work.}
A suggestive interpretation of the Bose-Einstein factor in equation~\eqref{eq:WSthermalSmatrix} is that it arises from the analytic continuation of the time-like free boson on the complexified time cylinder, under which the path integral contour originally running along the spatial circle winds repeatedly around the thermal circle \cite{BalthazarRodriguezYin:2022Type0BInstanton, JafferisSchneider:2021StringyEREPR}.

We propose that the thermal $S$-matrix of $c=1$ string theory is dual to the thermal $S$-matrix of the collective excitations in MQM. To check this proposal, we compare the worldsheet thermal amplitudes with the matrix model thermal amplitudes \(\tilde{\mathcal{A}}_{\mathrm{MM}}^{(T)}\) at Matsubara frequencies. The analogous proposal has been made in the context of the 2d type 0B string theory in  \cite{BalthazarRodriguezYin:2022Type0BInstanton}.

\subsection{Tree level thermal amplitudes}

As a sanity check of the proposal, we compare the tree level $1\to2$ thermal amplitude with the matrix model result at Matsubara frequencies. The worldsheet thermal amplitude is computed by the expression
\ie
-iR\,\delta_{M,m_1+m_2} g_s \widetilde{\mathcal{A}}^{(T,0)}_{1\to 2}&= g_s \big\langle c\tilde c(\infty)\, c\tilde c(1)\, c\tilde c(0)\rangle\,\langle e^{ i M X/R}V_{\frac{iM}{2R}}(\infty)\,e^{-i m_1 X/R}V_{\frac{im_1}{2R}}(1)\,e^{- i m_2 X/R}V_{\frac{im_2}{2R}} (0)\big\rangle\\
 &=-i R\, 2\pi g_s \frac{\,M m_1m_2}{R^3}\,
\delta_{M,m_1+m_2},
\fe
which gives
\ie
g_s \widetilde{\mathcal{A}}^{(T,0)}_{1\to 2} = 2\pi g_s \frac{\,M m_1m_2}{R^3}.
\fe
This agrees with the matrix model prediction specified in section \ref{sec:c1MQMdual} under the dictionary $\mu^{-1}=2\pi g_s$.

The moduli integrands for higher point tree level thermal amplitudes, such as $\widetilde{\mathcal{A}}_{1\to 3}^{(T,0)}$, $\widetilde{\mathcal{A}}_{1\to 4}^{(T,0)}$, and $\widetilde{\mathcal{A}}_{1\to 5}^{(T,0)}$, can be constructed using an analogous procedure to that of section~\ref{sec:higher-point-amplitudes}.  In fact, sphere correlators of KK vertex operators in the compact boson CFT coincide with the corresponding correlators of the noncompact timelike boson analytically continued to Matsubara frequencies after stripping off the respective zero mode factors.  It is straightforward to show that the moduli integrand for the tree level $1\to n$ thermal amplitude agrees with that of the standard tree level scattering amplitude evaluated at Matsubara frequencies. We can therefore directly use the numerical results for sphere amplitudes obtained in the previous section to determine the tree level thermal amplitudes at the corresponding Matsubara frequencies. From the matrix model side, this is consistent with the fact that tree level thermal amplitudes agree with the standard tree level string amplitudes analytically continued to Matsubara frequencies. The numerical computation of $\widetilde{\mathcal{A}}_{1\to 4}^{(T,0)}$ and $\widetilde{\mathcal{A}}_{1\to 5}^{(T,0)}$ is presented in section \ref{sec:numerical_thermal_amplitude}.

The agreement between the tree level $1\to 2$ thermal amplitude and the matrix model result suggests that the dictionary $\mu^{-1}=2\pi g_s$ is temperature independent at order $g_s$. The matching of higher point tree level amplitudes including $\widetilde{\mathcal{A}}_{1\to 4}^{(T,0)}$ and $\widetilde{\mathcal{A}}_{1\to 5}^{(T,0)}$ provides independent support for this conclusion.

\subsection{Genus one thermal amplitude}

The one loop $1\to 1$ thermal amplitude can be computed by the expression
\begin{equation}\label{eq:thermal-genus1-two-point}
\begin{aligned}
g_s^2\widetilde{\mathcal{A}}^{(T,1)}_{1\to1}\!\left(n;R\right)
={}&2\pi^2 g_s^2 i\\
\times&
\int_{\mathcal F_1} d^2\tau
\int_{T^2(\tau)} d^2z\,
\frac{|\eta(\tau)|^2}{\sqrt{\tau_2}}\,
E(z\mid\tau)^{-n^2/R^2}
\Theta_R^{(1)}\!\left(\frac{nz}{2\pi}\,\middle|\,\tau\right)
\left\langle
V_{\frac{in}{2R}}(z,\bar z)
V_{\frac{in}{2R}}(0)
\right\rangle_{\mathrm{Liouville},\,T^2(\tau)} .
\end{aligned}
\end{equation}
where the genus 1 prime form $E(z\mid\tau)$ is defined in Eq.\eqref{eq:genus1defs}, the genus 1 fundamental domain $\mathcal{F}_1$ is defined in \eqref{eq:genus1funddomain} and the genus 1 non-holomorphic theta function is defined as
\ie
\Theta_R^{(1)}(u\mid\tau)
:=
\sum_{m,w\in\mathbb Z}
\exp\!\left[
-\frac{\pi R^2}{\tau_2}|w-m\tau|^2
+\frac{2\pi i}{\tau_2}
\operatorname{Im}\!\bigl((w-m\bar\tau)u\bigr)
\right].
\fe
The non-holomorphic theta function arises from the compact boson correlator on the torus. It introduces additional $R$ dependence into the moduli integrand compared with the standard string scattering amplitude with the external momenta analytically continued to Matsubara frequencies. The numerical details of the computation are discussed in Appendix \ref{App:Genus11to1Numerics}.

The one loop thermal $1\to 2$ amplitude can be computed by the expression
\ie\label{eq:thermal-genus1-3-point}
g_s^3\widetilde{\mathcal A}^{T\,(1)}_{1\to2}
={}2\pi^2 g_s^3 i
\int_{\mathcal F_1} d^2\tau&
\int_{T^2(\tau)} d^2z_1\,d^2z_2\,
\frac{|\eta(\tau)|^2}{\sqrt{\tau_2}}\,
\prod_{1\leq i<j\leq3}
E(z_{ij}\mid\tau)^{-k_i k_j}
\\
&\times
\Theta_R^{(1)}\!\left(
\frac{R}{2\pi i}\sum_{j=1}^3 k_jz_j
\,\middle|\,\tau
\right)
\left\langle
V_{\frac{in}{2R}}(z_1,\bar z_1)
V_{\frac{in_1}{2R}}(z_2,\bar z_2)
V_{\frac{in_2}{2R}}(0)
\right\rangle_{\mathrm{Liouville},\,T^2(\tau)} .
\fe
where
\ie\label{eq:thermal-genus1-3-point-defs}
k_1=\frac{in}{R},\quad
k_2=-\frac{in_1}{R},\quad
k_3=-\frac{in_2}{R},\quad
n=n_1+n_2,\quad
z_{ij}=z_i-z_j,\quad z_3=0.
\fe
We compute the Liouville three point function via a conformal block decomposition, as in section \ref{sec:g1_1to2amp}. The numerical details of the computation are discussed in appendix \ref{App:Genus11to2Numerics}.

\subsection{Numerical Results}\label{sec:numerical_thermal_amplitude}

\begin{figure}[H]
\centering
\subfloat[Genus zero $1\to4$]{%
  \includegraphics[width=0.48\textwidth]{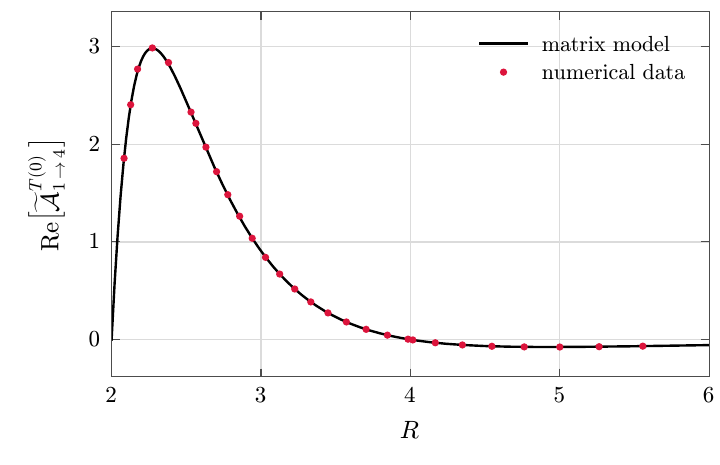}%
  \label{fig:thermal-tree-one-to-four}}
\hfill
\subfloat[Genus zero $1\to5$]{%
  \includegraphics[width=0.48\textwidth]{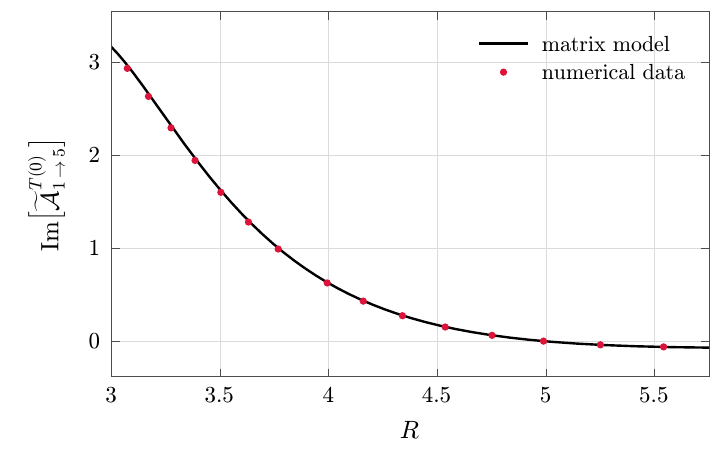}%
  \label{fig:thermal-tree-one-to-five}}

\par\vspace{1em}
\subfloat[Genus one $1\to1$]{%
  \includegraphics[width=0.48\textwidth]{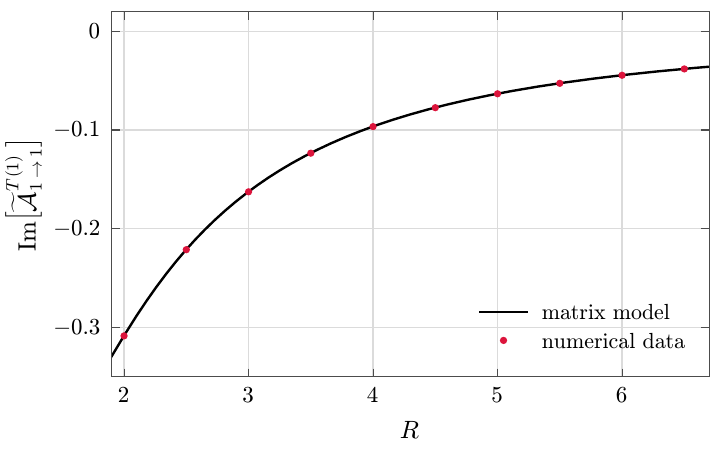}%
  \label{fig:thermal-one loop-one-to-one}}
\hfill
\subfloat[Genus one $1\to2$]{%
  \includegraphics[width=0.48\textwidth]{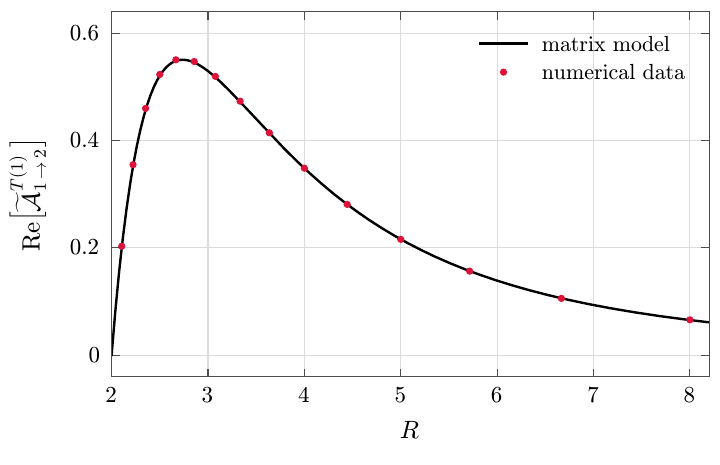}%
  \label{fig:thermal-one loop-one-to-two}}
\caption{Radius dependence of the thermal amplitudes. Panel (a) shows the real part of the genus zero $1\to 4$ thermal amplitudes.  Panel (b) shows the imaginary part of the genus zero $1\to 5$ thermal amplitudes. Panel (c) shows the imaginary part of the genus one  $1\to 1$ thermal amplitude. Panel (d) shows the real part of the genus one $1\to 2$ thermal amplitudes. Panels (a) and (b) are computed using equation~\eqref{eq:tree1tonamplitude}, while panels (c) and (d) are computed using
equations~\eqref{eq:thermal-genus1-two-point}
and~\eqref{eq:thermal-genus1-3-point}, respectively. All outgoing
Matsubara modes are set to $n_a=1$, with zero external winding. Red points
are the numerical worldsheet results and black curves are the matrix model
predictions with the dictionary $\mu^{-1}=2\pi g_s$.}
\label{fig:thermal-amplitude-radius-comparisons}
\end{figure}

In the computation of tree level and genus one thermal amplitudes, we numerically integrate the moduli integrand using Monte Carlo methods, as in the case of the standard scattering amplitude.  Numerical results of the $R$ dependence of tree level thermal amplitude $\widetilde{\mathcal{A}}^{(T,0)}_{1\to 4}$, $\widetilde{\mathcal{A}}^{(T,0)}_{1\to 5}$ as well as the one loop $1\to1$ thermal amplitude $\widetilde{\mathcal{A}}^{(T,1)}_{1\to 1}$ and one loop $1\to2$ thermal amplitude $\widetilde{\mathcal{A}}^{(T,1)}_{1\to 2}$ are shown in figure \ref{fig:thermal-amplitude-radius-comparisons}.  The results show agreement with the matrix model prediction, providing a further check of the duality between $c=1$ string theory and its dual matrix model.

The method can also be applied to other choices of tachyon energies, provided the moduli integrals converge. The genus-one $1\to2$ thermal amplitude result also tests the relation between $\mu$ and $g_s$ beyond leading order. If the coupling dictionary takes the form
\ie
\mu^{-1}=2\pi g_s+m_3(R)g_s^3+\mathcal{O}(g_s^5),
\fe
matching the coefficients of $\widetilde{\mathcal{A}}^{(T)}_{1\to 2}$ and  $\widetilde{\mathcal{A}}^{T}_{1\to 2,\mathrm{MM}}$ at order $g_s^3$ gives the following relation between the genus-one amplitudes:
\ie
\widetilde{\mathcal A}^{(T,1)}_{1\to2}
=(2\pi)^3\widetilde{\mathcal A}^{(T,1)}_{1\to2,\mathrm{MM}}
+m_3(R)\widetilde{\mathcal A}^{(T,0)}_{1\to2,\mathrm{MM}}.
\fe
The observed agreement therefore supports $m_3(R)=0$ within our numerical accuracy.

\section{Genus two thermal free energy}\label{sec:genus2-free-energy}

In this section, we numerically compute the genus two thermal free energy of $c=1$ string theory, providing the first nontrivial check of the duality between $c=1$ string theory and matrix model at genus two.

The worldsheet theory of $c=1$ string theory in the thermal background consists of a Euclidean compact boson $X\sim X+2\pi R$, the $c=25$ Liouville CFT, and the $bc$ ghost system. The genus two free energy is obtained by integrating the product of the genus two compact boson partition function, genus 2 Liouville partition function and the ghost correlator over the moduli space of genus 2 Riemann surfaces $\mathcal F_2$. We parameterize the moduli space using period matrices and perform the numerical integration over the Siegel--Gottschling fundamental domain. The period matrix parametrization is described in section~\ref{sec:genus2moduli}.

To compute the moduli integrand, we factor out the known genus two measure of the critical bosonic string and express the remaining contributions as ratios of partition functions that are Weyl invariant, as explained in section~\ref{sec:genus2integrand}. The compact boson contribution is expressed in terms of a genus two non-holomorphic theta function. The remaining
ratio involving the Liouville partition function and free boson partition function, which we call the Liouville factor $\mathcal Q_L$, is evaluated in the plumbing frame using conformal block decomposition,
as described in section~\ref{sec:genus2liouville}. 

Finally, in section~\ref{sec:genus2numerics}, we perform the moduli integral using Monte Carlo sampling. The resulting free energy agrees with the matrix model prediction within numerical uncertainties, providing strong evidence for the duality at two loops.

\subsection{The moduli space of genus two Riemann surfaces}\label{sec:genus2moduli}

An efficient parametrization of the moduli space of genus two Riemann surfaces is given by the period matrix $\Omega$, defined as follows. Let $\left\{\alpha_I,\beta_I \mid I=1,2\right\} \in H_1(\Sigma_2,\mathbb{Z})$ be a basis of homology cycles for a genus two Riemann surface $\Sigma_2$, with the following property under intersection pairing
\ie
\alpha_I \cdot \beta_{J} =\delta_{IJ} ,\qquad \alpha_I \cdot \alpha_{J} =0 ,\qquad \beta_I \cdot \beta_{J} =0.
\fe
There exists a basis of holomorphic one forms $\left\{\omega_I \mid I =1,2\right\}$ on $\Sigma_2$ that satisfies
\ie
\int_{\alpha_I} \omega_J =\delta_{IJ}.
\fe
The period matrix $\Omega$ is a $2\times 2$ matrix, with matrix elements
\ie
\Omega_{IJ} := \int_{\beta_I} \omega_J, \quad I,J=1,2.
\fe
It can be viewed as the higher genus analog of $\tau$ parameter for genus one Riemann surface. The period matrix $\Omega$ takes values in the genus two Siegel upper half plane $\mathbb{H}_2$
\ie
\mathbb H_2
:=
\left\{
\Omega\in\operatorname{Mat}_{2\times2}(\mathbb C)
\;\middle|\;
\Omega^{\mathsf T}=\Omega,\quad
\operatorname{Im}\Omega>0
\right\}.
\fe
A change of homology basis in $H_1\left(\Sigma_2,\mathbb{Z}\right)$ induces a $\operatorname{SP}\left(4,\mathbb{Z}\right)$ action on the Siegel upper half plane $\mathbb{H}_2$
\ie
\Omega \rightarrow (A\Omega+B)(C\Omega+D)^{-1}, \quad  \begin{pmatrix}
A & B\\
C & D
\end{pmatrix}
\in \operatorname{Sp}(4,\mathbb Z).
\fe
The moduli space of genus two Riemann surfaces $\mathcal F_2$ can be characterized by the quotient space $\operatorname{Sp}(4,\mathbb Z)\backslash\mathbb H_2$.
A particularly convenient choice of fundamental domain for the $\operatorname{Sp}(4,\mathbb Z)$ action, known as the Siegel–Gottschling fundamental domain $\mathcal{F}_2$, admits the explicit prescription \cite{Siegel:1943,Gottschling:1959}
\ie
\mathcal F_2
=
\left\{
\Omega=X+iY\in\mathbb H_2
\;\middle|\;
\begin{aligned}
&-\frac12\leq X_{IJ}<\frac12,
\qquad
0\leq2Y_{12}\leq Y_{11}\leq Y_{22},
\\
&|\det(C\Omega+D)|\geq1
\quad \forall
\begin{pmatrix}A&B\\ C&D\end{pmatrix}
\in\operatorname{Sp}(4,\mathbb Z)
\end{aligned}
\right\}.
\fe
The last condition on the determinants would truncate to 19 Gottschling inequalities, see Appendix~\ref{App:Genus2Numerics}.  This will be the domain we are integrating over.

\subsection{Genus two Moduli Integrand}\label{sec:genus2integrand}

The genus two thermal free energy can be computed by the expression
\ie\label{eq:genus2freeEnergy}
-2\pi R\,g_s^2 F^{(2)}(R) =8\pi^4 g_s^2
\int_{\mathcal F_2}
|d^3\Omega|^2\,
\left|
\left\langle
\prod_{a=1}^{3}(b,\mu_a)
\right\rangle
\right|^2
Z^{(2)}_c(R)~ Z^{(2)}_{\mathrm{Liouville}},
\fe
where $\mu_a$ are the Beltrami differentials on genus two Riemann surface, $Z_c^{(2)}(R)$ is the genus two compact boson partition function with compact boson radius $R$, $Z^{(2)}_{\mathrm{Liouville}}$ is the genus two Liouville partition function and $(b,\mu_a)$ is the pairing between $b$ ghost and Beltrami differentials
\ie
(b,\mu_a)
:=
\frac{1}{2\pi}
\int_{\Sigma_2}d^2z\,
b_{zz}\,(\mu_a)_{\bar z}{}^{z}.
\fe
We also use the convention $|d^3\Omega|^2= d^2\Omega_{11}d^2\Omega_{12}d^2\Omega_{22}$ for the integration measure.

In equation \eqref{eq:genus2freeEnergy}, the ghost correlator and the matter partition functions are computed in the same Weyl frame. The moduli integrand is independent of the Weyl frame and is only a function of period matrix $\Omega$.
The normalization constant $8\pi^4g_s^2$ is fixed by the factorization of the moduli integrand and the $\mathbb{Z}_2$ conformal killing group on a generic genus two Riemann surface. Further details of the determination of the normalization constant are presented in Appendix~\ref{App:Factorialization}.

To compute the moduli integrand, we rewrite the moduli integral as
\ie\label{eq:genus2integral}
-2\pi R\,g_s^2 F^{(2)}(R) =8\pi^4 g_s^2
\int_{\mathcal F_2}
|d^3\Omega|^2\,
\left|
\left\langle
\prod_{a=1}^{3}(b,\mu_a)
\right\rangle
\right|^2
\bigl(Z_{\mathrm{free}}^{(2)}\bigr)^{26}~
\frac{Z_c^{(2)}(R)}{Z_{\mathrm{free}}^{(2)}}~
\frac{Z^{(2)}_{\mathrm{Liouville}}}{\bigl(Z_{\mathrm{free}}^{(2)}\bigr)^{25}},
\fe
where $Z^{(2)}_{\mathrm{free}}$ is the genus two free boson partition function.\footnote{In defining the free boson partition function, we discard the target space volume factor coming from the zero mode integration in the path integral.}

The free boson times ghost factor can be identified with the moduli integrand of critical bosonic string, given by the expression \cite{BelavinKnizhnik:1986,Moore:1986TwoLoop}
\ie
\left|
\left\langle
\prod_{a=1}^{3}(b,\mu_a)
\right\rangle
\right|^2
\bigl(Z^{(2)}_{\mathrm{free}}\bigr)^{26}
=
\frac{1}
{(4\pi^2)^{25}
 \bigl(\det\operatorname{Im}\Omega\bigr)^{13}
 \left|\widehat{\chi}_{10}(\Omega)\right|^2},
\fe
where $\hat{\chi}_{10}(\Omega)$ is the Igusa cusp form, defined by \cite{Igusa:1962}
\ie
\widehat{\chi}_{10}(\Omega)
=
2^{-12}\prod_{\substack{
\boldsymbol a,\boldsymbol b\in\{0,1\}^2\\
\boldsymbol a\cdot\boldsymbol b=0\pmod 2
}}
\theta\!\begin{bmatrix}\boldsymbol a/2\\ \boldsymbol b/2\end{bmatrix}
(0|\Omega)^2,
\fe
\ie
\theta\!\begin{bmatrix}\boldsymbol a/2\\ \boldsymbol b/2\end{bmatrix}
(\boldsymbol z|\Omega)
:=
\sum_{\boldsymbol n\in\mathbb Z^2}
\exp\!\left[
\pi i\left(\boldsymbol n+\frac{\boldsymbol a}{2}\right)^{\!T}
\Omega
\left(\boldsymbol n+\frac{\boldsymbol a}{2}\right)
+2\pi i\left(\boldsymbol n+\frac{\boldsymbol a}{2}\right)^{\!T}
\left(\boldsymbol z+\frac{\boldsymbol b}{2}\right)
\right].
\fe
The ratio between compact boson and free boson partition function can be computed by the path integral, resulting in the expression
\ie
\frac{Z_c^{(2)}(R)}{Z_{\mathrm{free}}^{(2)}}
=
2\pi R\,\Theta_R^{(2)}(\Omega),
\fe
where $\Theta_R^{(2)}(\Omega)$ is the genus 2 non-holomorphic theta function defined by
\ie
\Theta_R^{(2)}(\Omega)
:=
\sum_{\boldsymbol m,\boldsymbol n\in\mathbb Z^2}
\exp\!\left[
-\pi R^2
(\boldsymbol m+\overline{\Omega}\boldsymbol n)^T
(\operatorname{Im}\Omega)^{-1}
(\boldsymbol m+\Omega\boldsymbol n)
\right].
\fe
The last recipe for the moduli integrand is the Weyl invariant ratio
\ie\label{eq:LiouvilleFactor}
\mathcal{Q}_L(\Omega) := \frac{Z^{(2)}_{\mathrm{Liouville}}}{\bigl(Z_{\mathrm{free}}^{(2)}\bigr)^{25}},
\fe
which we refer to as the Liouville factor. The computation of the Liouville factor is explained in the next subsection.

The overall formula we use to compute the genus two thermal free energy is then
\ie\label{eq:genus2-moduli-integral}
g_s^2 F^{(2)}(R)
=
-\frac{8\pi^4 g_s^2}{(4\pi^2)^{25}}
\int_{\mathcal F_2}
\frac{|d^3\Omega|^2}
{\bigl(\det\operatorname{Im}\Omega\bigr)^{13}
 \left|\widehat{\chi}_{10}(\Omega)\right|^2}
\,
\Theta_R^{(2)}(\Omega)\,
\mathcal Q_L(\Omega).
\fe

\subsection{Computation of the Liouville factor \texorpdfstring{$\mathcal{Q}_L$}{Q\_L}}\label{sec:genus2liouville}

In this subsection, we explain the computation of the Liouville factor $\mathcal{Q}_L(\Omega)$ defined in equation \eqref{eq:LiouvilleFactor}.

Due to Weyl anomaly cancellation, $\mathcal{Q}_L(\Omega)$ is a function of $\Omega$ and is independent of the Weyl frame. We choose a preferred Weyl frame to compute it. For our purposes, we compute $\mathcal{Q}_L(\Omega)$ using the plumbing frame, defined via the plumbing fixture of genus 2 Riemann surface.

\begin{figure}[htbp]
\centering
\tikzset{every picture/.style={line width=0.75pt}} %

\begin{tikzpicture}[x=0.75pt,y=0.75pt,yscale=-1,xscale=1]

\draw  [draw opacity=0] (154.31,229.53) .. controls (147.19,239) and (136.77,244.97) .. (125.53,244.78) .. controls (115.07,244.6) and (106.02,239.11) .. (100.11,230.53) -- (128.31,203.32) -- cycle ; \draw   (154.31,229.53) .. controls (147.19,239) and (136.77,244.97) .. (125.53,244.78) .. controls (115.07,244.6) and (106.02,239.11) .. (100.11,230.53) ;  
\draw  [draw opacity=0] (106.1,237.26) .. controls (112.17,234.62) and (119.03,233.14) .. (126.27,233.2) .. controls (133.37,233.25) and (140.03,234.76) .. (145.88,237.39) -- (125.3,270.17) -- cycle ; \draw   (106.1,237.26) .. controls (112.17,234.62) and (119.03,233.14) .. (126.27,233.2) .. controls (133.37,233.25) and (140.03,234.76) .. (145.88,237.39) ;  
\draw  [draw opacity=0] (179,251.42) .. controls (171.03,266.55) and (151.2,277.28) .. (128,277.29) .. controls (97.75,277.31) and (73.22,259.14) .. (73.2,236.69) .. controls (73.19,214.25) and (97.69,196.04) .. (127.94,196.02) .. controls (150.06,196.01) and (169.13,205.73) .. (177.78,219.73) -- (127.97,236.66) -- cycle ; \draw   (179,251.42) .. controls (171.03,266.55) and (151.2,277.28) .. (128,277.29) .. controls (97.75,277.31) and (73.22,259.14) .. (73.2,236.69) .. controls (73.19,214.25) and (97.69,196.04) .. (127.94,196.02) .. controls (150.06,196.01) and (169.13,205.73) .. (177.78,219.73) ;  
\draw  [draw opacity=0] (351.89,226.64) .. controls (344.77,236.11) and (334.35,242.08) .. (323.11,241.89) .. controls (312.64,241.71) and (303.6,236.22) .. (297.69,227.64) -- (325.89,200.43) -- cycle ; \draw   (351.89,226.64) .. controls (344.77,236.11) and (334.35,242.08) .. (323.11,241.89) .. controls (312.64,241.71) and (303.6,236.22) .. (297.69,227.64) ;  
\draw  [draw opacity=0] (303.59,234.41) .. controls (309.68,231.74) and (316.58,230.25) .. (323.85,230.31) .. controls (331.15,230.36) and (337.99,231.95) .. (343.94,234.72) -- (322.88,267.27) -- cycle ; \draw   (303.59,234.41) .. controls (309.68,231.74) and (316.58,230.25) .. (323.85,230.31) .. controls (331.15,230.36) and (337.99,231.95) .. (343.94,234.72) ;  
\draw  [draw opacity=0] (265.7,218.36) .. controls (273.84,202.1) and (296.3,190.16) .. (322.89,189.7) .. controls (356.58,189.11) and (384.15,207.18) .. (384.45,230.06) .. controls (384.75,252.95) and (357.68,271.97) .. (323.99,272.56) .. controls (297.32,273.03) and (274.49,261.8) .. (265.98,245.76) -- (323.44,231.13) -- cycle ; \draw   (265.7,218.36) .. controls (273.84,202.1) and (296.3,190.16) .. (322.89,189.7) .. controls (356.58,189.11) and (384.15,207.18) .. (384.45,230.06) .. controls (384.75,252.95) and (357.68,271.97) .. (323.99,272.56) .. controls (297.32,273.03) and (274.49,261.8) .. (265.98,245.76) ;  
\draw    (177.93,253.3) .. controls (184.47,244.01) and (189,247.14) .. (196.23,246.28) ;
\draw    (176.44,217.69) .. controls (181.13,224.73) and (193.04,225.25) .. (199.9,225.06) ;
\draw    (250.96,222.72) .. controls (256.65,221.94) and (263.66,222.34) .. (266.68,216.54) ;
\draw    (245.93,244.89) .. controls (261,244.67) and (262.62,241.46) .. (265.98,245.76) ;
\draw    (199.9,225.06) -- (250.96,222.72) ;
\draw    (196.23,246.28) -- (245.93,244.89) ;
\draw   (434,238.77) .. controls (434,220.67) and (449.67,206) .. (469,206) .. controls (488.33,206) and (504,220.67) .. (504,238.77) .. controls (504,256.87) and (488.33,271.54) .. (469,271.54) .. controls (449.67,271.54) and (434,256.87) .. (434,238.77) -- cycle ;
\draw    (504,238.77) -- (567,239) ;
\draw   (567,238.77) .. controls (567,220.67) and (582.67,206) .. (602,206) .. controls (621.33,206) and (637,220.67) .. (637,238.77) .. controls (637,256.87) and (621.33,271.54) .. (602,271.54) .. controls (582.67,271.54) and (567,256.87) .. (567,238.77) -- cycle ;
\draw   (85.72,78.23) .. controls (85.72,49.09) and (152.43,25.46) .. (234.72,25.46) .. controls (317.01,25.46) and (383.72,49.09) .. (383.72,78.23) .. controls (383.72,107.37) and (317.01,131) .. (234.72,131) .. controls (152.43,131) and (85.72,107.37) .. (85.72,78.23) -- cycle ;
\draw    (140.55,91.3) .. controls (167.55,111.3) and (285.55,118.3) .. (329.55,92.3) ;
\draw    (131.3,92.63) .. controls (185.3,83.63) and (305.55,88.3) .. (337.55,93.3) ;
\draw    (326.55,68.3) .. controls (304.13,53.62) and (178.13,47.62) .. (142.55,67.3) ;
\draw    (335.3,67.63) .. controls (299.3,73.63) and (177.3,72.63) .. (131.3,65.63) ;
\draw   (443,84) .. controls (443,56.94) and (483.74,35) .. (534,35) .. controls (584.26,35) and (625,56.94) .. (625,84) .. controls (625,111.06) and (584.26,133) .. (534,133) .. controls (483.74,133) and (443,111.06) .. (443,84) -- cycle ;
\draw    (443,84) -- (625,84) ;
\draw [color={rgb, 255:red, 208; green, 2; blue, 27 }  ,draw opacity=1 ]   (234.67,109.7) .. controls (244.33,114.03) and (242.33,128.03) .. (234.72,131) ;
\draw [color={rgb, 255:red, 208; green, 2; blue, 27 }  ,draw opacity=1 ] [dash pattern={on 0.84pt off 2.51pt}]  (234.67,109.7) .. controls (220.38,115.86) and (232.38,134.86) .. (234.72,131) ;
\draw [color={rgb, 255:red, 208; green, 2; blue, 27 }  ,draw opacity=1 ]   (235.94,71.21) .. controls (245.49,74.78) and (243.14,85.66) .. (235.46,87.83) ;
\draw [color={rgb, 255:red, 208; green, 2; blue, 27 }  ,draw opacity=1 ] [dash pattern={on 0.84pt off 2.51pt}]  (235.94,71.21) .. controls (221.51,75.74) and (233.03,90.8) .. (235.46,87.83) ;
\draw [color={rgb, 255:red, 52; green, 92; blue, 228 }  ,draw opacity=1 ]   (100.46,80.47) .. controls (111.34,140.93) and (363.79,136.6) .. (368.46,82.47) ;
\draw [color={rgb, 255:red, 52; green, 92; blue, 228 }  ,draw opacity=1 ]   (100.46,80.47) .. controls (76.88,24.02) and (370.88,3.02) .. (368.46,82.47) ;
\draw [color={rgb, 255:red, 52; green, 92; blue, 228 }  ,draw opacity=1 ]   (129.46,60.47) .. controls (102.33,91.33) and (384.04,88.8) .. (344.04,61.8) ;
\draw [color={rgb, 255:red, 52; green, 92; blue, 228 }  ,draw opacity=1 ]   (129.46,60.47) .. controls (146.04,40.57) and (316.04,39.57) .. (344.04,61.8) ;
\draw [color={rgb, 255:red, 208; green, 2; blue, 27 }  ,draw opacity=1 ]   (127.67,244.7) .. controls (137.33,249.03) and (135.49,274.05) .. (127.88,277.02) ;
\draw [color={rgb, 255:red, 52; green, 92; blue, 228 }  ,draw opacity=1 ]   (90.96,234.09) .. controls (90,272.67) and (171,271.67) .. (168.38,233.41) ;
\draw [color={rgb, 255:red, 208; green, 2; blue, 27 }  ,draw opacity=1 ] [dash pattern={on 0.84pt off 2.51pt}]  (127.67,244.7) .. controls (113.38,250.86) and (125.54,280.88) .. (127.88,277.02) ;
\draw [color={rgb, 255:red, 52; green, 92; blue, 228 }  ,draw opacity=1 ]   (90.96,234.09) .. controls (89,210.67) and (163,197.67) .. (168.38,233.41) ;
\draw [color={rgb, 255:red, 208; green, 2; blue, 27 }  ,draw opacity=1 ]   (327.67,241.69) .. controls (337.33,246.02) and (337.69,267.73) .. (327.69,272.73) ;
\draw [color={rgb, 255:red, 208; green, 2; blue, 27 }  ,draw opacity=1 ] [dash pattern={on 0.84pt off 2.51pt}]  (327.67,241.69) .. controls (313.38,247.84) and (320.38,273.8) .. (327.69,272.73) ;
\draw [color={rgb, 255:red, 52; green, 92; blue, 228 }  ,draw opacity=1 ]   (282.96,228.09) .. controls (282,266.67) and (363,265.67) .. (360.38,227.41) ;
\draw [color={rgb, 255:red, 52; green, 92; blue, 228 }  ,draw opacity=1 ]   (282.96,228.09) .. controls (281,204.67) and (355,191.67) .. (360.38,227.41) ;

\draw (462,182) node [anchor=north west][inner sep=0.75pt]   [align=left] {$\displaystyle q_{L}$};
\draw (595,183) node [anchor=north west][inner sep=0.75pt]   [align=left] {$\displaystyle q_{R}$};
\draw (529,214) node [anchor=north west][inner sep=0.75pt]   [align=left] {$\displaystyle q_{B}$};
\draw (573,104) node [anchor=north west][inner sep=0.75pt]   [align=left] {$\displaystyle q_{0}$};
\draw (561,62) node [anchor=north west][inner sep=0.75pt]   [align=left] {$\displaystyle q_{1}$};
\draw (571,16) node [anchor=north west][inner sep=0.75pt]   [align=left] {$\displaystyle q_{\infty }$};
\draw (19,8) node [anchor=north west][inner sep=0.75pt] [font=\large\rmfamily] [align=left] {Theta};
\draw (19,167) node [anchor=north west][inner sep=0.75pt] [font=\large\rmfamily] [align=left] {Glass};

\end{tikzpicture}
\caption{The theta and glasses plumbing decompositions of a genus 2 Riemann surface. The loop marked in red represents the A cycles we choose for each plumbing channel, and loops marked in blue represent the B cycles. We suppress the orientation of the loop in the figure, which should be chosen appropriately to satisfy the condition $\alpha_I\cdot\beta_J=\delta_{IJ}$. }
\label{fig:genus-two-plumbing-channels}
\end{figure}
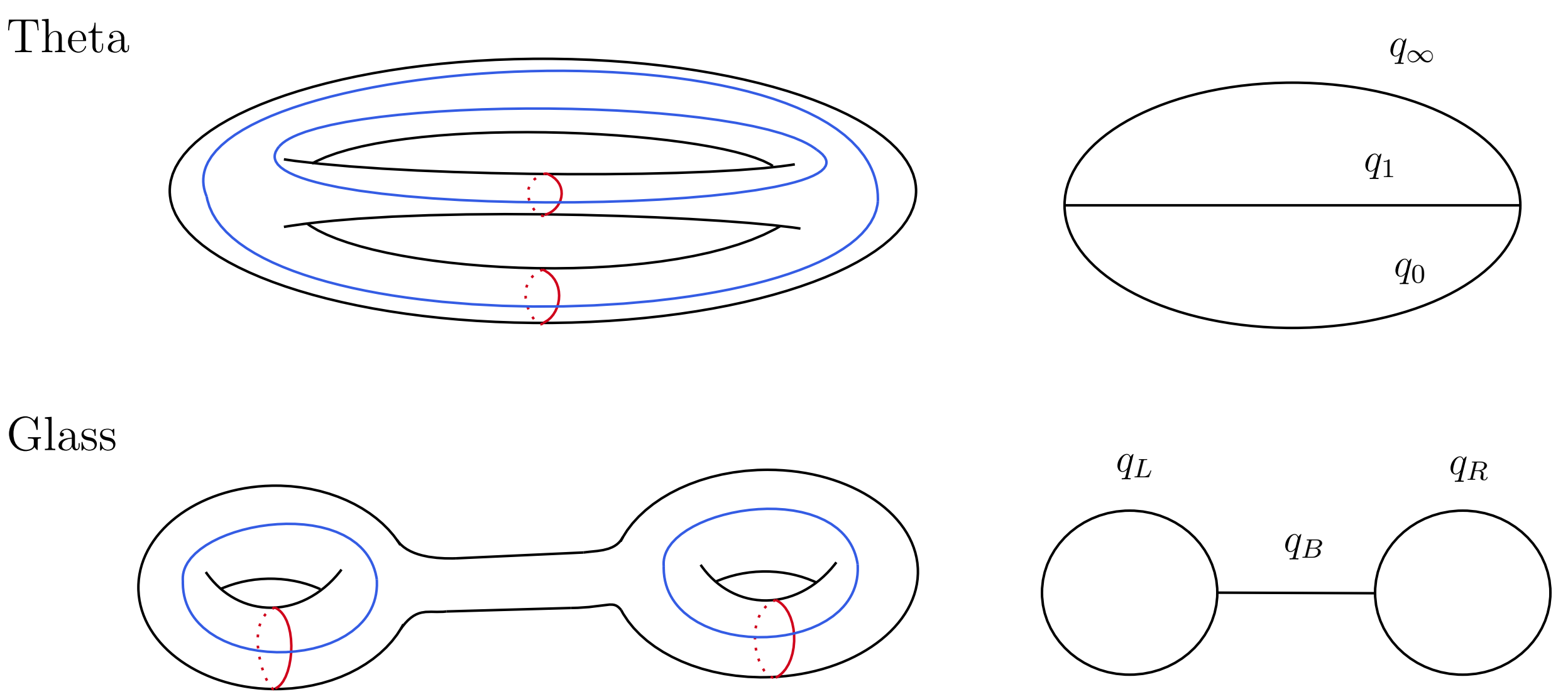

We first briefly review the plumbing fixture and the plumbing frame of genus 2 Riemann surface. Given two spheres with global coordinates \(z\) and \(w\), each having punctures at \(0,1,\infty\), there are two different ways of constructing a genus two Riemann surface from the plumbing fixture (see figure~\ref{fig:genus-two-plumbing-channels}). In the theta channel, the annular neighborhood of $(0,1,\infty)$ on each punctured sphere are sewn to each other:
\ie
\text{Theta:}\qquad
\begin{alignedat}{3}
zw&=q_0,
&\qquad&
\frac{|q_0|}{\epsilon_0}<|z|<\epsilon_0,
\\
(z-1)(w-1)&=q_1,
&&
\frac{|q_1|}{\epsilon_1}<|z-1|<\epsilon_1,
\\
\frac{1}{zw}&=q_\infty,
&&
\frac{|q_\infty|}{\epsilon_2}
<\left|\frac{1}{z}\right|<\epsilon_2 .
\end{alignedat}
\fe

In the glasses channel, the neighborhoods of 0 and \(\infty\) on each sphere are sewn, while the annular neighborhoods of $z=w=1$ on the two spheres are sewn to each other:
\ie
\text{Glasses:}\qquad
\begin{alignedat}{3}
z&\sim \frac{z}{q_L},
&\qquad&
\frac{|q_L|}{\epsilon_L}<|z|<\epsilon_L,
\\
w&\sim \frac{w}{q_R},
&&
\frac{|q_R|}{\epsilon_R}<|w|<\epsilon_R,
\\
(z-1)(w-1)&=q_B,
&&
\frac{|q_B|}{\epsilon_B}<|z-1|<\epsilon_B .
\end{alignedat}
\fe

The plumbing fixture is defined only when the three annular regions on the Riemann sphere do not overlap. For each plumbing channel $\Gamma$, the plumbing parameters \(\boldsymbol q\) determine the complex structure of the resulting genus two Riemann surface.  Each plumbing channel thus defines a local chart on the moduli space of genus two Riemann surfaces \(\mathcal F_2\), with the plumbing parameters \(\boldsymbol q\) serving as local coordinates on the chart. It is expected that the collection of the charts associated with all the plumbing channels forms an atlas covering the genus two moduli space $\mathcal{F}_2$~\cite{HubbardKoch:2014,HinichVaintrob:2010}. Namely, every \(\Omega\in\mathcal F_2\) can be expressed as \(\Omega=\Omega(\boldsymbol q)\) for some plumbing channel \(\Gamma\) and plumbing parameters \(\boldsymbol q\).\footnote{Strictly speaking, we are not aware of a proof in the mathematical literature that every genus-two surface can be obtained using our choice of plumbing coordinates. The standard plumbing coordinate theorem ensures the coverage of the moduli space by plumbing charts with plumbing relations of the form $f(z)g(w)=q$, where $f$ and $g$ are suitably chosen local coordinates near the punctures \cite{HinichVaintrob:2010}. In our construction, however, these local coordinates are fixed to $z,z-1,1/z$ and their $w$ counterparts in terms of the global sphere coordinates. Therefore, we check this statement numerically for every period matrix \(\Omega\) in our Monte Carlo sample of the genus two Siegel–Gottschling fundamental domain \(\mathcal F_2\). In each case, we found plumbing parameters \(\boldsymbol q\) in either the theta or the glasses channel such that the period matrix of the resulting sewn surface agrees with \(\Omega\) up to a numerical tolerance of $10^{-5}$.}

We now explain the explicit relation between the plumbing parameters \(\boldsymbol q\) and the period matrix \(\Omega\). We choose a basis of homology cycles as in figure~\ref{fig:genus-two-plumbing-channels} to define the period matrix $\Omega$. To state the relation, we need to introduce the Schottky groups \cite{Maskit:1988}. 
The Schottky groups $P_{\Gamma}$ associated with the two plumbing channels of genus two Riemann surface $\Gamma\in \{\mathrm{Theta},\mathrm{Glasses}\}$ are
rank two free subgroups of $\operatorname{PSL}(2,\mathbb C)$. The group elements are M\"obius transformations on the Riemann sphere. In the theta channel, 
\ie
P_{\mathrm{Theta}}
&=
\left\langle\gamma_0,\gamma_1\right\rangle,
\\
\gamma_0(z)
&=
q_0q_\infty z,
\\
\gamma_1(z)
&=
\frac{1+q_\infty(q_1-1)z}
     {1-q_\infty z}.
\fe
In the glasses channel, 
\ie
P_{\mathrm{Glasses}}
&=
\left\langle\gamma_L,\gamma_R\right\rangle,
\\
\gamma_L(z)
&=
q_Lz,
\\
\gamma_R(z)
&=
\frac{
(q_R+q_B-1)z+(q_R-1)(q_B-1)
}{
(q_R-1)z+q_R(q_B-1)+1
}.
\fe
Every nontrivial element $\gamma\in P_{\Gamma}$  has one attracting fixed point $a_\gamma$ and one repelling fixed point $r_\gamma$, satisfying
\ie
\gamma(a_\gamma)&=a_\gamma,
&
\left|\gamma'(a_\gamma)\right|&<1,
\\
\gamma(r_\gamma)&=r_\gamma,
&
\left|\gamma'(r_\gamma)\right|&>1.
\fe
The Schottky multiplier $k_\gamma$ is defined by
\ie
k_\gamma:=\gamma'(a_\gamma),
\qquad
\gamma'(r_\gamma)=k_\gamma^{-1},
\qquad
0<|k_\gamma|<1.
\fe
Equivalently, $k_\gamma$ is characterized by the solution of the equation
\ie
\frac{\gamma(z)-a_\gamma}
     {\gamma(z)-r_\gamma}
=
k_\gamma
\frac{z-a_\gamma}
     {z-r_\gamma}.
\fe
The period matrix is related with the plumbing parameters by
\ie\label{Period_Matrix_from_Schottsky}
\exp\!\left[2\pi i(\Omega_\Gamma)_{IJ}\right]
=
k_I^{\delta_{IJ}}
\prod_{\gamma\in P_{\Gamma}(IJ)}
\frac{
\bigl(a_I-\gamma(a_J)\bigr)
\bigl(r_I-\gamma(r_J)\bigr)}
{
\bigl(a_I-\gamma(r_J)\bigr)
\bigl(r_I-\gamma(a_J)\bigr)}
.
\fe
where \(I,J\in\{0,1\}\) in the theta channel and
\(I,J\in\{L,R\}\) in the glasses channel.  \(P_{\Gamma}(IJ)\) consists of the group elements of \(P_{\Gamma}\) that neither begin with \(\gamma_I^{\pm1}\) nor end with
\(\gamma_J^{\pm1}\). The identity element is excluded when \(I=J\)
and included when \(I\neq J\). This equation determines the
period matrix up to \(\Omega_{IJ}\rightarrow\Omega_{IJ}+1\). We fix this ambiguity by taking the solution with
\ie
-\frac12<\operatorname{Re}(\Omega_\Gamma)_{IJ}\leq\frac12.
\fe
In practice, when the plumbing parameters are large, it is more efficient to compute the period matrix by constructing holomorphic one forms in the plumbing fixture numerically. The numerical details of this are explained in appendix \ref{App:Genus2Numerics}.

For the plumbing fixture of the genus two Riemann surface, there exists a preferred Weyl frame called the plumbing frame. In this frame, the genus two partition function can be computed by sewing pair-of-pants amplitudes. The free boson partition function in the plumbing frame is given by \cite{McIntyreTakhtajan:2006}
\ie\label{eq:partition_free}
Z_{\mathrm{free},\Gamma}^{(2)}(\boldsymbol q)
=
\frac{1}{4\pi^2\sqrt{\det\operatorname{Im}\Omega_\Gamma(\boldsymbol q)}}
\prod_{[\gamma]\in \mathcal{P}_\Gamma}
\prod_{n=1}^{\infty}
\left|1-k_\gamma(\boldsymbol q)^n\right|^{-2},
\qquad
\Gamma\in\{\mathrm{Theta},\mathrm{Glasses}\}.
\fe
where $\mathcal{P}_{\Gamma}$ denotes the set of primitive conjugacy classes in the Schottky group for the plumbing channel $\Gamma$ modulo the identification\footnote{An element $\gamma$ in the Schottky group $P_{\Gamma}$ is called primitive if it cannot be written as $\gamma=\eta^m$ for a group element $\eta$ and an integer $m\geq 2$. Primitive conjugacy class refers to the conjugacy class of primitive elements. In $\mathcal{P}_{\Gamma}$, one further identifies the conjugacy classes $[\gamma]\sim[\gamma^{-1}]$. It can be shown that the Schottky multiplier $k_{\gamma}$ does not depend on the representative for the conjugacy class and satisfies $k_{\gamma} = k_{\gamma^{-1}}$. This fact ensures that equation \eqref{eq:partition_free} is well-defined. } \([\gamma]\sim[\gamma^{-1}]\), $k_{\gamma}$ denotes the Schottky multiplier for the Schottky group element $\gamma$, and $\Omega_\Gamma(\boldsymbol q)$ is the period matrix defined under the choice of homology cycle we specified in figure~\ref{fig:genus-two-plumbing-channels}.

The Liouville partition function in the plumbing frame can be computed by the conformal block decomposition.
In the theta channel, one has
\ie\label{eq:genus2-liouville-theta}
Z_{\mathrm{Liouville},\mathrm{Theta}}^{(2)}
=
\int_0^\infty
\frac{dP_0\,dP_1\,dP_\infty}{\pi^3}\,
\mathcal{C}(P_0,P_1,P_\infty)^2
\left|
q_0^{h_0}q_1^{h_1}q_\infty^{h_\infty}
F_{\mathrm{Theta}}
(\boldsymbol h;\boldsymbol q)
\right|^2 ,
\fe
where $(h_0,h_1,h_{\infty})=(1+P_0^2,1+P_1^2,1+P_\infty^2)$ and $F_{\mathrm{Theta}}(\boldsymbol h;\boldsymbol q)$ is the genus two conformal block in the theta channel.

In the glasses channel, one has
\ie
Z_{\mathrm{Liouville},\mathrm{Glasses}}^{(2)}
&=
\int_0^\infty
\frac{dP_L\,dP_R\,dP_B}{\pi^3}\,
\mathcal{C}(P_L,P_B,P_L)
\mathcal{C}(P_R,P_B,P_R)
\left|
q_L^{h_L}q_R^{h_R}q_B^{h_B}
F_{\mathrm{Glasses}}
(\boldsymbol h;\boldsymbol q)
\right|^2,
\fe
where $(h_L,h_R,h_B)=(1+P_L^2,1+P_R^2,1+P_B^2)$ and $F_{\mathrm{Glasses}}(\boldsymbol h;\boldsymbol q)$ is the genus 2 conformal block in the glasses channel. In practice, we compute  the conformal block numerically using the c-recursion in \cite{Cho-Collier-Yin}.

The Liouville factor $\mathcal{Q}_L$ can thus be computed explicitly in the plumbing frame by the expression
\ie\label{eq:genus2-liouville-factor-plumbing}
\mathcal{Q}_L(\Omega(\boldsymbol q)) = \frac{Z_{\mathrm{Liouville},\Gamma}^{(2)}(\boldsymbol q)}{\left(Z_{\mathrm{free},\Gamma}^{(2)}(\boldsymbol q)\right)^{25}}, \qquad \Gamma \in \left\{ \mathrm{Theta} ,\mathrm{Glasses}\right\}.
\fe
As a consistency check of our prescription, the Liouville factors computed in the theta and glasses channels must agree wherever the two plumbing charts overlap. We verify this numerically at five representative period matrices in the overlap, as shown in Table~\ref{tab:genus-two-liouville-channel-agreement}. The two evaluations agree to within \(10^{-7}\) relative accuracy.
\begin{table}[H]
\centering
\small
\renewcommand{\arraystretch}{1.35}
\begin{tabular}{c c c c}
\toprule
Period matrix $\Omega$
& $\mathcal Q_L^{\mathrm G}$
& $\mathcal Q_L^{\Theta}$
& $\mathcal Q_L^{\Theta}/\mathcal Q_L^{\mathrm G}$
\\
\midrule
$\left(\begin{smallmatrix}
-0.083+3.013i & 0.014+0.299i\\
0.014+0.299i & -0.353+3.095i
\end{smallmatrix}\right)$
& $5.576221\times10^{-27}$
& $5.576221\times10^{-27}$
& $0.9999999912$
\\[2pt]
$\left(\begin{smallmatrix}
-0.365+3.296i & 0.046+0.293i\\
0.046+0.293i & -0.268+3.694i
\end{smallmatrix}\right)$
& $1.455071\times10^{-30}$
& $1.455071\times10^{-30}$
& $1.000000017$
\\[2pt]
$\left(\begin{smallmatrix}
0.167+2.150i & -0.012+0.284i\\
-0.012+0.284i & -0.180+3.719i
\end{smallmatrix}\right)$
& $1.777813\times10^{-26}$
& $1.777813\times10^{-26}$
& $0.9999999689$
\\[2pt]
$\left(\begin{smallmatrix}
0.188+3.039i & -0.197+0.290i\\
-0.197+0.290i & -0.429+3.479i
\end{smallmatrix}\right)$
& $1.646699\times10^{-28}$
& $1.646698\times10^{-28}$
& $0.9999999304$
\\[2pt]
$\left(\begin{smallmatrix}
-0.499+3.095i & -0.168+0.266i\\
-0.168+0.266i & -0.455+3.532i
\end{smallmatrix}\right)$
& $4.216520\times10^{-29}$
& $4.216520\times10^{-29}$
& $0.9999999728$
\\
\bottomrule
\end{tabular}
\caption{Comparison of the Liouville factor computed in the glasses and theta
plumbing channels for five period matrices on the overlap of the two plumbing channels. We use $\mathcal{Q}_L^{G}$ for the Liouville factor computed in the glasses channel and $\mathcal{Q}_L^{\Theta}$ for the Liouville factor computed in the theta channel. }
\label{tab:genus-two-liouville-channel-agreement}
\end{table}

\subsection{Numerical Result}\label{sec:genus2numerics}

Based on the complete prescription for the moduli integrand discussed in sections \ref{sec:genus2integrand} and \ref{sec:genus2liouville}, we perform the moduli integral over the fundamental domain $\mathcal{F}_2$ using Monte Carlo sampling. At each sampled period matrix, we evaluate the integrand in an appropriate theta or glasses plumbing chart. To improve the numerical efficiency, we also treat separately the region where the genus two surface is represented by two genus one surfaces joined by a long plumbing tube. The specific numerical details are explained in appendix \ref{App:Genus2Numerics}.

Figure~\ref{fig:genus-two-free-energy} compares the numerical worldsheet results with the genus 2 matrix model free energy under the identification $\mu^{-1}=2\pi g_s$. The worldsheet numerics are in agreement with the matrix model answer. This provides the first non-trivial check for the duality between $c=1$ string theory and matrix model at two loops. The worldsheet numerics also suggest that the potential ambiguity in the identification of Liouville volume with chemical potential does not appear at order $\mu^{-2}$ at all temperatures
\ie
V_{\phi}= -\frac{1}{2}\log \mu+\mathcal{O}\left(\mu^{-4}\right).
\fe

\begin{figure}[htbp]
\centering
\includegraphics[width=0.72\textwidth]{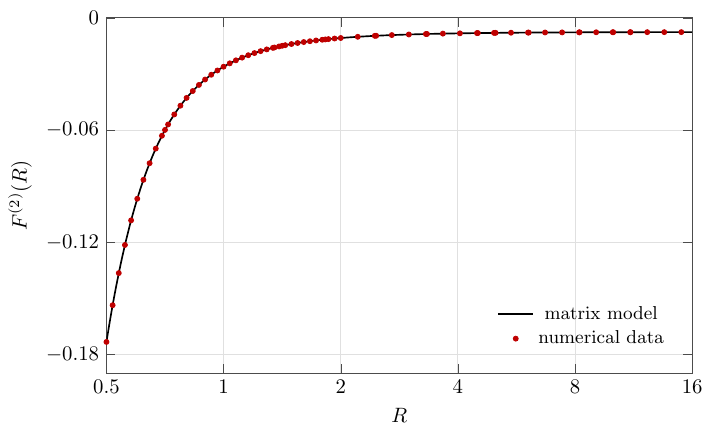}
\caption{The genus two free energy $F^{(2)}(R)$ obtained from the numerical worldsheet integration (red points), compared with the matrix model prediction (black curve).}
\label{fig:genus-two-free-energy}
\end{figure}

\section{Discussion}\label{sec:discussion}

In this paper, we have provided substantial evidence for the duality between $c=1$ string theory and $c=1$ matrix quantum mechanics through numerical worldsheet calculations of a variety of scattering amplitudes and the genus two thermal free energy.

To support the equivalence of perturbative $S$-matrix, we computed the tree level $1\to 4$ and $1\to 5$ scattering amplitudes, and the one loop $1\to 2$ scattering amplitudes by numerically integrating Virasoro conformal blocks with DOZZ structure constants on the sphere and torus for purely imaginary frequencies. For the thermal quantities in $c=1$ string theory, we formulated the $1\to n$ thermal $S$-matrix in $c=1$ string theory, clarified the relation between tree level thermal amplitudes and the standard string scattering amplitudes, and numerically computed the tree level $1\to 4$ and $1\to 5$ thermal amplitude. We also carried out the worldsheet computation of the one loop $1\to 1$ and $1\to 2$ thermal amplitudes, where the moduli integrand contains additional \(R\) dependence compared with the standard string scattering amplitude with the external momenta analytically continued to Matsubara frequency. Finally, we computed the genus 2 thermal free energy of $c=1$ string theory at finite temperature $F^{(2)}(R)$ by integrating over genus 2 Virasoro blocks with DOZZ structure constants. All the results find agreement with the prediction of $c=1$ MQM under the identification $\mu^{-1}=2\pi g_s$ and $V_{\phi}= -\frac{1}{2}\log \mu$.

The worldsheet computation of the thermal amplitude and thermal free energy supports the non-renormalization of the dictionary -- in particular the lack of dependence on temperature -- up to next-to-leading order in $\mu$:
\ie
\mu^{-1}=2\pi g_s+ \mathcal{O}(g_s^5),\quad V_{\phi} = -\frac{1}{2}\log \mu + \mathcal{O}(\mu^{-4}).
\fe
This suggests that the dictionary at leading order may be exact to all orders in perturbation theory. 

To evaluate $c=1$ string amplitudes, we developed an efficient method for computing higher-point and higher-genus correlation functions in Liouville CFT, combining $c$-recursion for conformal blocks with numerical integration over Liouville momenta. 
This approach can be extended to other irrational bosonic CFTs with known spectra and OPE coefficients. Degenerate representations, such as those appearing in Virasoro minimal models, require special treatment that is beyond the scope of the current approach. An alternative method would be needed in those cases in order to access higher point and higher genus correlation functions for operators in degenerate representations. Extensions to superconformal theories would require efficient evaluation of higher-point and higher-genus superconformal blocks, including channels with Ramond sector intermediate states. A method for dealing with genus $g$ $n$-point superconformal blocks with intermediate primaries in the R sector in $\mathcal{N}=1$ SCFTs will appear in an upcoming paper~\cite{ChristianEtAl:SuperconformalBlocks}.

In principle, the methods developed in this paper can be applied to compute the genus three free energy of $c=1$ string theory. In this case, the period matrix still provides coordinates on the moduli space of genus three Riemann surfaces. On the codimension one loci where the genus 3 Riemann surfaces can be represented as the hyperelliptic curves, the period matrix does not provide a set of local coordinates for the moduli space of genus three Riemann surfaces. That being said, one might expect it does not affect the moduli integration as long as the moduli integrand smoothly extends to the hyperelliptic locus. The main obstacle is the computational complexity, as the Liouville partition function on a genus three Riemann surface will involve six fold Liouville momentum integration, and the computation would be numerically expensive.

It would be interesting to apply the method developed in this paper to other 2d string theories, including non-critical Type 0A/B theory and Heterotic $\operatorname{Spin}(23)$/$\operatorname{E}_8\times \operatorname{Spin}(7)$ theory.  The current numerical method could provide a more extensive check on the duality between Type 0A/B and the proposed matrix model dual from the worldsheet perspective. The application to Heterotic $\operatorname{Spin}(23)$/$\operatorname{E}_8\times\operatorname{Spin}(7)$ theory is especially interesting, as there is no known matrix model description of the theories.

\section*{Acknowledgements}

We would like to thank Ben Mazel, Victor A. Rodriguez and Jaroslav Scheinpflug for discussions. YW and XY thank the RIKEN Interdisciplinary Theoretical and Mathematical Sciences Program (iTHEMS) and the Yukawa Institute for Theoretical Physics (YITP) at Kyoto University, where this work was completed during the iTHEMS-YITP workshop 'Resurgence, Matrix model, String field theory and Instantons' (YITP-W-26-11).  SC is supported in part by an NSF GRFP fellowship. This work is supported by DOE grant DE-SC0007870. ChatGPT 5.6 Sol and 6 Astra were used in this work for coding, algorithm optimization and assistance with writing and editing the manuscript.

\appendix

\section{Normalization of genus two moduli integrand}\label{App:Factorialization}

In this section, we determine the normalization constant for the genus two moduli integrand in Eq.~\eqref{eq:genus2integral} from a factorization argument. We denote the moduli integration for the genus 2 free energy as 
\ie\label{eq:integrandnormalization}
-2\pi R\,g_s^2 F^{(2)}(R) =\frac{1}{2}
\int_{\mathcal F_2}
\Omega^{(2,0)}_6,
\fe
where $\Omega^{(2,0)}_6$ is the six form on the moduli space of genus 2 Riemann surfaces 
\ie\label{eq:genus2integrand}
\Omega^{(2,0)}_6:=\mathcal{N}_2~ |d^3\Omega|^2\,
\left|
\left\langle
\prod_{a=1}^{3}(b,\mu_a)
\right\rangle
\right|^2
Z^{(2)}_c(R)~ Z^{(2)}_{\mathrm{Liouville}}.
\fe
The genus 2 compact boson partition function $Z_c^{(2)}(R)$, the genus 2 Liouville partition function $Z^{(2)}_{\mathrm{Liouville}}$ and $bc$ ghost correlators are normalized as in section \ref{sec:genus2integrand}. We use $\mathcal{N}_2$ to denote the normalization constant yet to be determined. 

In equation \eqref{eq:integrandnormalization}, we emphasize the factor $\frac{1}{2}$ coming from the $\mathbb{Z}_2$ conformal killing group on a generic genus 2 Riemann surface\footnote{The $\frac{1}{2}$ factor has the same origin as the familiar $\frac{1}{2}$ factor in the computation of genus one free energy.}. One way to demonstrate the $\mathbb{Z}_2$ action is to use the hyperelliptic curve representation of the genus 2 Riemann surface
\ie
y^2 = x(x-1)(x-e_1)(x-e_2)(x-e_3).
\fe
where the $\mathbb{Z}_2$ conformal killing group acts as $(x,y)\rightarrow (x,-y)$. We display the factor of $\frac{1}{2}$ separately as it does not follow from the argument based on the factorization of the string integrand.

We determine $\mathcal{N}_2$ by the factorization of the genus 2 moduli integrand $\Omega^{(2,0)}_6$ when the genus 2 Riemann surface degenerates into two genus 1 Riemann surfaces connected by a long tube. The precise statement of the factorization property requires an off-shell formalism, as elaborated in \cite{Sen:2014dca}. We review the statement briefly and apply it to our case.

Denote the compact boson CFT Hilbert space as $\mathcal{H}_X$, the Liouville CFT Hilbert space as $\mathcal{H}_{\mathrm{Liouville}}$ and $bc$ ghost Hilbert space as $\mathcal H_{bc}$. We define the ghost number two off-shell string Hilbert space $\mathcal{H}^{(2)}$ as\footnote{We do not include the level matching condition, as it is convenient to pin down the normalization before integrating over the twist angle in the factorization limit.}
\ie\mathcal H^{(2)}
=
\left\{
|\varphi\rangle\in
\mathcal H_X\otimes\mathcal H_{\mathrm{Liouville}}\otimes\mathcal H_{bc}
~\big |~
N_{\mathrm{gh}}|\varphi\rangle=2|\varphi \rangle, b_0|\varphi\rangle =\tilde{b}_0|\varphi\rangle=0
\right\}.
\fe
An example state in $\mathcal{H}^{(2)}$ is
\ie\label{eq:offshell_states}
|\varphi_P\rangle
=
g_s\,|\mathbf1\rangle_X
\otimes |P\rangle_{\mathrm{Liouville}}
\otimes c_1\widetilde c_1|0\rangle_{bc} \quad\in \mathcal H^{(2)}.
\fe
Under the state operator correspondence, this corresponds to the vertex operator $g_s c\tilde c V_P$, where $V_P$ is the Liouville primary with Liouville momentum $P$. The normalization of the state is\footnote{The normalization for the sphere 1-point function of identity operator in compact boson CFT is determined by taking the large-$R$ limit and comparing with the free-boson normalization in equation~\eqref{eq:free-boson-sphere-normalization}.}
\ie
\langle\!\langle\varphi_P|c_0\widetilde c_0|\varphi_{P'}\rangle
=
g_s^2RC_{S^2}\,\pi\delta(P-P')
=
2\pi R\,\pi\delta(P-P'),
\fe
where $C_{S^2}=\tfrac{2\pi}{g_s^2}$. We extend the family of states in equation \eqref{eq:offshell_states} to an orthogonal basis $\{\varphi_{a,P}\}$ of $\mathcal H^{(2)}$ with respect to the pairing
$\langle\!\langle\varphi_{a,P}|c_0\widetilde c_0|\varphi_{b,P'}\rangle$. The ghost number 4 dual basis $\ket{\varphi^c_{a,P}}$ is defined such that $\langle\!\langle\varphi^c_{a,P}|\varphi_{b,P'}\rangle
=\pi \delta_{ab}\,\delta(P-P')$. In particular, $\varphi^c$ and $\varphi$ are related as
\ie
|\varphi_P^c\rangle
=
\frac{c_0\widetilde c_0}{2\pi R}|\varphi_P\rangle,
\qquad
\langle\!\langle\varphi_P^c|\varphi_{P'}\rangle=\pi\delta(P-P').
\fe

The precise formulation of the genus one 1-point integration measure for off-shell string states requires the introduction of an auxiliary space $\mathcal P_{1,1}$, which denotes the space of one punctured tori equipped with a local coordinate $f$. The coordinate $f$ is defined in a neighborhood of the puncture and is zero at the puncture location. The data of the local coordinate remembers how the local operator associated with an off shell state is inserted on the torus. $\mathcal{P}_{1,1}\rightarrow\mathcal M_{1,1}$ defines a fiber bundle over the moduli space of one punctured tori $\mathcal{M}_{1,1}$, where the projection map forgets about the choice of the local coordinate $f$.

For an off shell string state $\ket{\varphi_{a,P}}\in\mathcal H^{(2)}$, the integration measure $\Omega^{(1,1)}_2(\varphi_{a,P})$ is a two form on $\mathcal P_{1,1}$. Choosing the local coordinate $f_{\tau}$ as a function of $\tau$ defines a section $S$ of the fiber bundle $\mathcal{P}_{1,1}\rightarrow\mathcal M_{1,1}$. The pullback $\Omega^{(1,1)}_2(\varphi_{a,P})\big|_S$ then gives the integration measure on $\mathcal{M}_{1,1}$, which admits the following expression:
\ie
\Omega^{(1,1)}_2(\varphi_{a,P})\big|_S
=
\mathcal{N}_{1,1}d^2\tau
\left\langle
\mathcal{B}_{\bar{\tau}}\mathcal{B}_{\tau}
[f_\tau^{-1}\circ\varphi_{a,P}](0)
\right\rangle_{\tau}.
\fe
$\mathcal B_\tau$ and $\mathcal B_{\bar{\tau}}$ are given by
\ie
\mathcal B_\tau
&=-i\oint_A du\,b(u)
-\oint_C\frac{du}{2\pi i}\,
\partial_\tau f_\tau^{-1}(w)\,b(u)
+\oint_C\frac{d\bar u}{2\pi i}\,
\partial_\tau\bar f_\tau^{-1}(\bar w)\,\widetilde b(\bar u),\\
\mathcal B_{\bar\tau}
&= i\oint_A d\bar u\,\widetilde b(\bar u)
-\oint_C\frac{du}{2\pi i}\,
\partial_{\bar\tau}f_\tau^{-1}(w)\,b(u)
+\oint_C\frac{d\bar u}{2\pi i}\,
\partial_{\bar\tau}\bar f_\tau^{-1}(\bar w)\,\widetilde b(\bar u).
\fe
Here $[f_\tau^{-1}\circ\varphi_{a,P}]$ denotes the vertex operator transformed from the local coordinate $w$ to the torus coordinate $u=f_\tau^{-1}(w)$, and $C$ is the anti-clockwise integration contour encircling the puncture. For example, for the local coordinate $f_\tau$ defined by the plumbing fixture of one punctured torus from a three punctured sphere, the integration measure admits the expression
\ie\label{eq:genus11offshell}
\Omega^{(1,1)}_2(\varphi_{P})\big|_S
=
\mathcal{N}_{1,1} d^2\tau~
 4\pi^2 g_s R
\frac{\Theta_R^{(1)}(0\mid\tau)|\eta(\tau)|^2}
{\sqrt{\operatorname{Im}\tau}}
\bigl\langle V_{P}(0)\bigr\rangle_{\mathrm{Liouville},\tau}.
\fe
The Liouville one point function can be computed from conformal block decomposition as
\ie
\bigl\langle V_{P}\bigr\rangle_{\mathrm{Liouville},\tau}
=
\int_0^\infty\frac{dP'}{\pi}\,
\mathcal C(P',P,P')
\left|
q^{h_{P'}-25/24}
F_{\mathrm{tor}}(h_{P'},h_P;q)
\right|^2,
\fe
where $F_{\mathrm{tor}}$ denotes the torus 1-point block and $q=e^{2\pi i\tau}$.

The statement of the factorization is that at the boundary of the moduli space where genus two Riemann surfaces degenerate into two genus one Riemann surfaces, the genus two moduli integrand factorizes as~\cite{StringNotes}
\ie\label{eq:fullfactorization}
\Omega^{(2,0)}_6\big|_S
={}&
\frac{d^2q_B}{|q_B|^2}
\sum_{a,b}\int_0^{\infty}\frac{dP}{\pi}\int_0^{\infty}\frac{dP'}{\pi}
\Omega^{(1,1)}_2(\varphi_{a,P})\big|_{S_L}
\wedge
\Omega^{(1,1)}_2(\varphi_{b,P'})\big|_{S_R}
\left\langle\!\left\langle\varphi_{a,P}^c\left|
\tilde{b}_0b_0\,
q_B^{L_0}\bar q_B^{\widetilde L_0}
\right|\varphi_{b,P'}^c\right\rangle\right.,
\fe
where $S_L$ and $S_R$ correspond to the sections induced by the section $S$ on two copies of the moduli space of one punctured tori. As equation \eqref{eq:fullfactorization} holds pointwise in the neighborhood of the boundary of genus 2 moduli space, we only need to compare the $q_B\rightarrow 0$ asymptotics of both sides to determine normalization between $\Omega^{(2,0)}_6$ and $\Omega^{(1,1)}_2(\varphi_{a,P})$ . Equation~\eqref{eq:fullfactorization} implies that the leading $q_B\rightarrow 0$ asymptotics of the moduli integrand behave as
\ie\label{eq:factorization}
\Omega^{(2,0)}_6\big|_S \sim \frac{1}{2\pi R}\frac{d^2q_B}{|q_B|^2}\int_0^{\epsilon}\frac{dP_B}{\pi} |q_B|^{2P_B^2}~\Omega^{(1,1)}_2(\varphi_{P_B})\big|_{S_L}\wedge
\Omega^{(1,1)}_2(\varphi_{P_B})\big|_{S_R},
\fe
where $\epsilon>0$ is a small real number.

We then use equation~\eqref{eq:factorization} to derive the relation between $\mathcal{N}_2$ and $\mathcal{N}_{1,1}$ from the explicit formula for $\Omega_{6}^{(2,0)}$ in equation \eqref{eq:genus2integrand}. Choosing a homology basis adapted to the degeneration limit, the moduli of two separate tori are given to leading order by $\tau_L=\Omega_{11}$ and $\tau_R=\Omega_{22}$. The leading relation between the separating coordinate and the plumbing parameter is
\ie
q_B\simeq-2\pi i\Omega_{12},
\qquad
d^2\Omega_{12}\simeq\frac{d^2q_B}{4\pi^2}.
\fe
In this region, the Liouville factor $\mathcal{Q}_L$ can be computed by the glasses channel conformal block decomposition. We approximate the Liouville partition
function by retaining only primary states propagating through
the separating tube. As $q_B\to0$, the glasses channel conformal block
factorizes as
\ie
F_{\mathrm{Glasses}}(\boldsymbol h;\boldsymbol q)
=
F_{\mathrm{tor}}(h_L,h_B;q_L)\,
F_{\mathrm{tor}}(h_R,h_B;q_R)
+\mathcal{O}(q_B),
\fe
where $F_{\mathrm{tor}}(h_a,h_B;q_a)$ is the torus 1-point
descendant block with internal weight $h_a$ and external weight
$h_B$. Therefore, the Liouville partition function asymptotes to
\ie
Z_{\mathrm{Liouville},\mathrm{Glasses}}^{(2)}
\simeq{}
\int_0^\infty
\frac{dP_L\,dP_R\,dP_B}{\pi^3}\,
\mathcal C(P_L,P_B,P_L)\,
\mathcal C(P_R,P_B,P_R)
\times
\left|
q_B^{h_B}
\prod_{a=L,R}
q_a^{h_a}F_{\mathrm{tor}}(h_a,h_B;q_a)
\right|^2,
\fe
where $h_a=1+P_a^2$ and $h_B=1+P_B^2$. When $q_B\rightarrow 0$, the rest of the moduli integrand also factorizes as
\ie
\frac{Z_c^{(2)}(R)}{Z_{\mathrm{free}}^{(2)}}
\frac{\bigl(Z_{\mathrm{free},\mathrm{Glasses}}^{(2)}\bigr)^{-25}}
{(4\pi^2)^{25}
 (\det\operatorname{Im}\Omega)^{13}
 |\widehat\chi_{10}(\Omega)|^2}
\simeq
\frac{2\pi R}{|q_B|^4}
\prod_{a=L,R}
\frac{
\Theta_R^{(1)}(0\mid\tau_a)\,
|q_a|^{-25/12}|\eta(\tau_a)|^2}
{\sqrt{\operatorname{Im}\tau_a}}.
\fe
Combining these factors with the Jacobian, we obtain the leading separating
form of the moduli measure:
\ie\label{eq:genus2-separating-moduli-integrand}
&\mathcal{N}_2 \frac{2\pi R}{(4\pi^2)^{25}}
\frac{|d^3\Omega|^2\,\Theta_R^{(2)}(\Omega)\,\mathcal Q_L(\Omega)}
{(\det\operatorname{Im}\Omega)^{13}|\widehat\chi_{10}(\Omega)|^2}
\\
&\qquad\simeq
\frac{1}{2\pi R}
\frac{d^2q_B}{|q_B|^2}
\int_0^\infty\frac{dP_B}{\pi}\,
|q_B|^{2P_B^2}
\prod_{a=L,R} 
\left[ d^2\tau_a\, \sqrt{\mathcal{N}_2} R
\frac{\Theta_R^{(1)}(0\mid \tau_a)|\eta(\tau_a)|^2}
{\sqrt{\operatorname{Im}\tau_a}}
\bigl\langle V_{P_B}\bigr\rangle_{\mathrm{Liouville},\tau_a}
\right],
\fe
By comparing equation \eqref{eq:genus2-separating-moduli-integrand} with equation \eqref{eq:factorization}, we determine the relation
\ie
\mathcal{N}_2 = 16 \pi^4 g_s^2 (\mathcal N_{1,1})^2.
\fe
The normalization constant $\mathcal N_{1,1}$ can be determined by repeating the same exercise for the factorization of the genus one 1-point integration measure into tree level 3-point integration measure, which determines
\ie
\mathcal N_{1,1} =1.
\fe
We thus pin down the genus 2 normalization constant to be
\ie\label{eq:NormalizationConstant}
\mathcal{N}_2=16\pi^4 g_s^2,
\fe
which is the normalization constant used in equation \eqref{eq:genus2integral}.

The same analysis has been used to determine the normalization of the genus one 2-point and 3-point moduli space integrand in  equation \eqref{eq:genus1threepoint} and equation $(4.1)$ of \cite{BalthazarRodriguezYin:2019}. In the work \cite{BalthazarRodriguezYin:2019}, the normalization was fixed by matching the numerical worldsheet computation to the matrix model prediction, as opposed to being determined intrinsically from the worldsheet theory. That being said, one can show that the normalization obtained in \cite{BalthazarRodriguezYin:2019} is precisely consistent with that required by factorization.

\section{Numerical Details}\label{App:Numerics}

\subsection{Genus zero \texorpdfstring{$1\to 4$}{1 to 4} amplitude}\label{App:Genus01to4Numerics}

\subsubsection*{Conformal Block}

For the Liouville 5-point function on the sphere, we choose the conformal block decomposition channel adaptively across the moduli space to ensure convergence and improve numerical efficiency. For each configuration of the five punctures, we enumerate the $120$ decomposition channels and use M\"obius transformations to place the
ordered punctures at $(0,x,y,1,\infty)$, with plumbing coordinates
$(q_1,q_2)=(x/y,y)$. Among the frames satisfying $|q_1|,|q_2|<1$, we select the one minimizing $\mathfrak r=\max\{|q_1|,|q_2|\}$.

In the selected channel, we compute the 5-point Virasoro block at
$c=25$ using the $c$-recursion relation
of~\cite{Cho-Collier-Yin}. The conformal block is truncated at total
level $N$ as
\[
F_5^{(N)}(q_1,q_2)
=\sum_{\substack{m,n\geq0\\m+n\leq N}}
F_{mn}\,q_1^m q_2^n,
\]
with $N=6$ or $8$ in the production. In section \ref{sec:numericalstability}, we analyze the difference between the $N=6$ and $N=8$ results, giving an estimate of the error introduced by level truncation.

\subsubsection*{Liouville Momentum Integration}

For the sphere 5-point function, we truncate each Liouville momentum integral at $P_{\max}=6$ and change the integration variable to $u_a$, with $P_a=P_{\max}u_a^{5/4}$, $a=1,2$. We evaluate each momentum integral in the variable $u_a$ using Gauss-Jacobi quadrature with weight $u_a^{1/4}$. 

We use 20 integration points in the $P_1$ direction and 21 in the $P_2$ direction, then interchange the integration point counts and average the two results. The reason we use an unequal number of integration points is that there is a term in the $c$-recursion relation has a simple pole when $P_1=P_2$, even though the full conformal block is regular.  At each external energy, we precompute the DOZZ factors and conformal block coefficients, and reuse them throughout the moduli integration.

The analytic continuation of the Liouville correlator requires additional residue contributions in certain ranges of the external energies. For equal outgoing energies $\omega_j=it$, there is no need to do contour deformation for $0<t<2/5$. For $2/5<t<1/2$, the pole in the DOZZ structure constant passes across the Liouville momentum integration contour. In these cases, we supplement the conformal block decomposition with
$-2i$ times the residue of the corresponding momentum integrand at
\ie
P_*=\frac52 it -i.
\fe
This is analogous to the procedure performed in appendix B of \cite{BalthazarRodriguezYin:2019}. Evaluating the residue contribution requires fixing one internal momentum at $P_*$. The remaining one dimensional integral is evaluated at the same integration points as the original integral.

\subsubsection*{Moduli Integration}

Once the genus zero $1\to 4$ string integrand is evaluated, we use Monte Carlo integration with scrambled Sobol
sequences. For each choice of tachyon energy $\omega_i$, we generate four independent sets of Monte Carlo sampling points for the integration and use the mean and variation of the four integral estimates to determine the result and its statistical error.

For each generation of the sampling point, we randomly select one of the $120$ coordinate frames and sample its plumbing coordinates with the weight function
\ie
p_\alpha(q_1,q_2)
=\left(\frac{\alpha}{2\pi}\right)^2
 |q_1|^{\alpha-2}|q_2|^{\alpha-2},
\qquad 0<|q_1|,|q_2|<1,
\qquad \alpha=0.15,
\fe
with respect to $d^2q_1\,d^2q_2$, and zero outside this domain. This distribution concentrates samples near degeneration limits. The choice of the weight function is for better numerical control.

For each sampled point, we evaluate the weighted integrand
\ie
W_\Gamma(q_1,q_2)
&=
\frac{\mathcal I(z_1,z_2)}{\rho_\Gamma(q_1,q_2)}
\left|
\det\frac{\partial(z_1,z_2)}{\partial(q_1,q_2)}
\right|^2,
\fe
where $\mathcal I(z_1,z_2)$ is the moduli integrand appearing in equation \eqref{eq:tree1tonamplitude}. The Jacobian factor between $z_1,z_2$ and plumbing parameters $q_1,q_2$ depends on the plumbing channel $\Gamma$. The factor $\rho_\Gamma$ in the denominator is defined as
\ie
\rho_\Gamma(q_1,q_2)
&:=
\frac1{120}
\sum_{\Gamma'}
p_\alpha(q'_1,q'_2)
\left|
\det\frac{\partial(q'_1,q'_2)}{\partial(q_1,q_2)}
\right|^2.
\fe
The sum includes only charts $\Gamma'$ containing the sampled point. $(q'_1,q'_2)$ are its coordinates in chart $\Gamma'$, and the
corresponding Jacobian is evaluated on the chart overlap. The factor $1/120$ is the probability of selecting each chart. Dividing by $\rho_\Gamma$ compensates for the nonuniform sampling distribution and avoids double counting on chart overlaps.

In practice, we compute the amplitude for each external energy by averaging the weighted integrand within
each of four independent sets of Monte Carlo sampling points, using $2^9$--$2^{11}$ points per set. The mean of the four integral estimates gives the moduli integral. We estimate the sampling standard error of this mean from the scatter among the four estimates.

\subsubsection*{Numerical Stability}
\label{sec:numericalstability}

We assess numerical stability by varying the parameters of the Liouville momentum integration and moduli integration, changing the total level of the conformal block, and comparing different conformal block decomposition channels.

 For the Liouville momentum integration, increasing the number of integration points $(20,21)$ to $(24,25)$ at $t=0.34$ changes the amplitude by $0.00055\%$. Varying the momentum cutoff over $P_{\max}=5,6,7$ changes the amplitude by $0.011\%$. 
 
For the moduli integration, calculations using $2^9$, $2^{10}$, and $2^{11}$ Monte Carlo sampling points in each of four independent sets find agreement within their estimated sampling standard errors.

Truncation effects are assessed by comparing the relative difference of the full $1\to 4$ amplitude at level 6 and 8 at $t=0.31,0.35,$ and $0.44$. Each calculation uses four independent sets of Monte Carlo sampling points, with \(2^{10}\) points per set for \(N=6\) and \(2^9\) for \(N=8\). The relative differences in the amplitudes, measured against the plotted values, are \(0.0354\%\), \(1.65\%\), and \(0.892\%\), respectively. Meanwhile, the sampling error is estimated at $0.168\%,0.177\%$, and $0.238\%$ for $N=6$ and $0.482\%,0.578\%,$ and $0.367\%$, respectively, for $N=8$.

We also check consistency between different conformal block decomposition channels in their overlap regions. At \(t=0.34\), the relative discrepancy between evaluations in two channels with comparable values of \(\mathfrak r\) decreases from \(0.44\%\) at total level \(N=6\) to \(0.069\%\) at \(N=8\).

\subsection{Genus zero \texorpdfstring{$1\to 5$}{1 to 5} amplitude}\label{App:Genus01to5Numerics}

\subsubsection*{Conformal Block}

For the Liouville 6-point function on the sphere, we choose the conformal block decomposition channel adaptively among 720 linear channels and 720 star channels. Using Möbius transformations, we map the punctures to
\[
\begin{aligned}
\text{linear}:&\quad(0,q_1q_2q_3,q_2q_3,q_3,1,\infty),\\
\text{star}:&\quad(0,q_1,1,1+q_2,1/q_3,\infty).
\end{aligned}
\]
In a selected conformal block decomposition channel, we compute the 6-point Virasoro block at \(c=25\) using the same \(c\)-recursion as for the 5-point block. The expansion in plumbing parameters is truncated at total level \(N\),
\ie\label{eq:truncationsixpoint}
F_6^{(N)}(q_1,q_2,q_3)
=
\sum_{\substack{m,n,\ell\geq0\\m+n+\ell\leq N}}
F_{mn\ell}\,q_1^m q_2^n q_3^\ell,
\fe
starting at \(N=12\) and increasing up to \(N=18\) at selected sampling points to assess truncation effects. 

The channel selection in the sphere 6-point case is more complicated than the 5-point case, due to the additional star channel. We use a decomposition channel only where the conformal block expansion in its plumbing parameters converges. To analyze the convergence, we introduce the following power series
\ie
F(s)=\sum_{N=0}^{\infty} s^N\sum_{\substack{m,n,\ell\geq0\\m+n+\ell= N}}
F_{mn\ell}\,q_1^m q_2^n q_3^\ell
\fe
where the conformal block corresponds to $s=1$. Therefore, in order for the truncation in Eq.\eqref{eq:truncationsixpoint} to be a good approximation of the conformal block, we ensure that the convergence radius of the series in $s$ is larger than one, using the following method.

In the star channel, the analysis of the convergence radius can be achieved by identifying $F(s)$ with the sphere 6-point block with punctures located as 
\(\left(0,sq_1,1,1+sq_2,\frac1{sq_3},\infty\right).\) The singularities on the complex $s$ plane are located at the loci where two of the punctures collide:
\ie
sq_1=1,\qquad 1+sq_2=0,\qquad sq_3=1,\\
s(q_1-q_2)=1,\qquad s^2q_1q_3=1,\qquad
sq_3(1+sq_2)=1.
\fe
Requiring the singularities to be outside of the unit disk in $s$ imposes $r_{\mathrm{star}}<1$, where
\ie
r_{\mathrm{star}}
=
\max\left\{
|q_1|,\ |q_2|,\ |q_3|,\ |q_1-q_2|,\
\sqrt{|q_1q_3|},\
|q_+|,\ |q_-|
\right\},
\qquad
q_\pm=\frac{q_3\pm\sqrt{q_3^2+4q_2q_3}}{2}.
\fe
We conduct a similar analysis on the convergence in the linear channel, which requires $ r_{\mathrm{linear}}=\max_i|q_i|<1$. For each conformal decomposition channel $\Gamma$, we retain channels with \(r_\Gamma<1\) and select the one with the smallest \(r_\Gamma\) to compute conformal block.

\subsubsection*{Liouville Momentum Integration}

The sphere 6-point function involves the integration over three internal Liouville momenta \(P_1,P_2,P_3\). Compared to the 5-point case, a more elaborate sampling grid must be used to perform the Liouville momentum integration to ensure numerical efficiency.

We treat the small $P$ region and large $P$ region in the Liouville momentum integration separately. We split each momentum integration into \(P_a\in(0,s_a)\) and \(P_a\in(s_a,\infty)\), and then change variables to $u$ in both intervals using
\ie
P_a=s_a u,
\qquad
P_a=s_a-2w_a\log(1-u),
\qquad 0<u<1,
\fe
where \(w_a=[\max(1,-2\log|q_a|)]^{-1/2}\). $s_a$ are parameters that we adjust to ensure the numerical efficiency. In the computation of the amplitude with $\omega_1=\dots=\omega_5=it$, we choose
\ie
s_a=
\begin{cases}
\min\!\left\{w_a,\,4(1-3t)\right\},
&\text{OPE of outgoing tachyon and incoming tachyon vertex operators},\\[2pt]
\min\!\left\{w_a,\,4(1-t)\right\},
&\text{OPE of two outgoing tachyon vertex operators},\\[2pt]
w_a,
&\text{middle line of the linear channel from OPE of three external operators}.
\end{cases}
\fe
This choice of splitting helps resolve the small momentum region. When a pole in imaginary momentum is close to the real axis, the contribution to the integrand from momentum close to the pole is large, so we change the splitting $s_a$ to be four times the distance of the poles at $\pm i(1-3t),\pm i(1-t)$ from real axis, respectively. For the middle line of the linear channel, the pole in a given momentum moves in position as the other two momenta vary. Therefore, there is not as simple of a rule as for the other integration variables.

On each interval in $u$, we use Gauss–Legendre quadrature to conduct the numerical integration, which is a special case of Gauss-Jacobi with a constant weight function.
21 out of 24 choices of channel plus incoming vertex operator location use \((6,8,10)\) integration points on each of the three finite intervals, while the other three use $(8,9,6),(8,7,8),$ and $(6,10,12)$ integration points. All 24 choices use \((8,9,10)\) integration points on their respective tails.

For equal outgoing energies \(\omega_j=it\), the first contour crossing occurs at \(t=1/3\). We work in the region \(0<t<1/3\) for simplicity, so all three momenta are integrated along the positive real axis without extra residue contributions.

\subsubsection*{Moduli Integration}

For the sphere $1\to5$ amplitude, we use Monte Carlo integration as in the sphere $1\to4$ calculation.
For each external energy, we generate eight independent sets of Monte Carlo sampling points, with $2^8$ points per set.

For each sampling point, we select one of the $1440$ coordinate
frames with equal probability, including $720$ linear channels and
$720$ star channels. In the computation of the amplitude with equal outgoing energies \(\omega_1=\dots=\omega_5=it\), we sample its plumbing coordinates with the
weight function
\ie
p(q_1,q_2,q_3)
&=
\sum_{\boldsymbol\alpha\in\{\alpha_{\mathrm E},\alpha_{\mathrm B}\}^3}
w_{\boldsymbol\alpha}
\prod_{a=1}^3
\frac{\alpha_a}{2\pi}|q_a|^{\alpha_a-2},
\qquad 0<|q_a|<1
\fe
with respect to $d^2q_1\,d^2q_2\,d^2q_3$, and zero outside this
domain. Here $\alpha_{\mathrm E}=2t^2,
\,
\alpha_{\mathrm B}=\frac34, $ and the coefficients $w_{\boldsymbol\alpha}$ are given by
\ie
w_{\boldsymbol\alpha}
=
\begin{cases}
\frac18,
& \boldsymbol\alpha=(\alpha_{\mathrm E},\alpha_{\mathrm E},\alpha_{\mathrm E}),\\[2pt]
\frac12,
& \boldsymbol\alpha=(\alpha_{\mathrm B},\alpha_{\mathrm B},\alpha_{\mathrm B}),\\[2pt]
\frac1{16},
& \text{otherwise},
\end{cases}
\qquad
\boldsymbol\alpha\in\{\alpha_{\mathrm E},\alpha_{\mathrm B}\}^3.
\fe
The smaller exponent terms in weight function from $\alpha_E$ concentrate samples near degeneration
limits, while the larger exponent from $\alpha_B$ improves sampling away from
these limits. Within each set, we allocate $32$, $128$, and
$16$ sampling points to these choices, respectively.

For each sampled point, we evaluate the weighted integrand
\ie\label{eq:weighted_integrand_sixpt}
W_\Gamma(q_1,q_2,q_3)
=
\frac{\mathcal I(z_1,z_2,z_3)}
{\rho_\Gamma(q_1,q_2,q_3)}
\left|
\det\frac{\partial(z_1,z_2,z_3)}
{\partial(q_1,q_2,q_3)}
\right|^2,
\fe
where $\mathcal I(z_1,z_2,z_3)$ is the moduli integrand in
equation~\eqref{eq:tree1tonamplitude}. The factor $\rho_\Gamma$ in the denominator
is defined as
\ie
\rho_\Gamma(q_1,q_2,q_3)
:=
\frac1{1440}
\sum_{\Gamma'}
p(q'_1,q'_2,q'_3)
\left|
\det\frac{\partial(q'_1,q'_2,q'_3)}
{\partial(q_1,q_2,q_3)}
\right|^2.
\fe
The sum includes all coordinate frames $\Gamma'$ containing the
sampled point, with $(q'_1,q'_2,q'_3)$ its coordinates in that
frame. The factor $1/1440$ is the probability of selecting each
frame. Dividing by $\rho_\Gamma$ compensates for the nonuniform
sampling distribution and avoids double counting on frame overlaps.
The conformal block is evaluated in the channel selected by the
convergence criterion above.

We treat separately the domain of the moduli space where the Riemann sphere with puncture is degenerate to have improved numerical control. In this computation, it suffices to treat the case where the correlator is computed by conformal block decomposition in the linear channel and $q_1, q_3$ are small. 
We define the region we treat as
\ie
D_{\mathrm{sep}}
=\left\{
|q_1|<0.02,\quad q_2\in D_2,\quad
|q_3|<0.01
\right\},
\fe
where
\ie
D_2=\left\{q_2:
-0.435\leq\operatorname{Re}q_2<-0.405,\quad
-0.898\leq\operatorname{Im}q_2<-0.878
\right\}.
\fe
In this region, we approximate the Liouville correlator by keeping only the propagation of conformal primaries through the two long tubes in the conformal block decomposition. In this limit,
\ie\label{eq:approximate_sixpt_block}
F_{6,\mathrm{linear}}(q_1,q_2,q_3)
=
F_4\!\left(
h_{P_2};
h_{P_1},h_{\omega_2/2},h_{\omega_3/2},h_{P_3}
\,\middle|\,q_2
\right)
+O(q_1)+O(q_3),
\qquad q_1,q_3\to0.
\fe
Here $F_4$ is the four-point descendant block with external
weights $(h_{P_1},h_{\omega_2/2},h_{\omega_3/2},h_{P_3})$
at $(0,q_2,1,\infty)$ and internal weight $h_{P_2}$. We evaluate the sphere 4-point block using Zamolodchikov's elliptic recursion relation, taking the $b\to1$ limit to obtain the result at $c=25$. The remaining integration over the moduli parameters $q_1$ and $q_3$ can be performed analytically: 
\ie
\label{eq:Isep}
I_{\mathrm{sep}}={}&
\int_{D_2}d^2q_2
\int_0^\infty\frac{dP_1\,dP_2\,dP_3}{\pi^3}\,
\mathcal C\!\left(P_2,\frac{it}{2},P_3\right)
\mathcal C\!\left(P_3,\frac{it}{2},\frac{it}{2}\right) \mathcal C\!\left(\frac{it}{2},\frac{5it}{2},P_1\right)
\mathcal C\!\left(P_1,\frac{it}{2},P_2\right)
\\
&\times
\frac{\pi e^{-2L_{1,0}(P_1^2+4t^2)}}{P_1^2+4t^2}
\frac{\pi e^{-2L_{3,0}(P_3^2+t^2)}}{P_3^2+t^2}
|q_2|^{2P_2^2+\frac92t^2-2}|1-q_2|^{t^2}
\left|
F_4\!\left(
h_{P_2};
h_{P_1},h_{it/2},h_{it/2},h_{P_3}
\,\middle|\,q_2
\right)
\right|^2,
\fe
where $L_{a,0}=-\log\rho_{a,0},~a=1,3.$ For Monte Carlo sampling points that are in $D_{\mathrm{sep}}$, we subtract from the weighted integrand in equation \eqref{eq:weighted_integrand_sixpt} the leading long tube approximation. We then add the analytically evaluated $I_{\text{sep}}$ given by equation \eqref{eq:Isep} separately to avoid double counting.

The eventual moduli integral is obtained by averaging the eight independent Monte Carlo estimates. We also compute the standard error of eight Monte Carlo estimates to obtain the uncertainty of the numerical result.

\subsubsection*{Numerical Stability}

We assess numerical stability by varying the Liouville momentum integration parameters and conformal block truncation, and comparing independent sets of moduli samples. For the momentum integration, we compare the result of the integration with two additional integration points on each finite or tail interval in turn, and with the width parameters $w_a$ halved, while keeping the block order fixed at $N=10$. At $t=0.1805,0.2505,0.3155$, the largest relative differences of the $1\to 5$ amplitude are approximately $0.034\%$, $0.027\%$, and $0.046\%$, respectively.

To assess the effect of conformal block truncation, we compare $N=10$ and $N=12$ on identical momentum nodes and moduli samples. Across the fifteen energies, the relative changes in the amplitudes range from $0.062\%$ to $0.187\%$. For the moduli integration, the eight independent Monte Carlo sampling sets of $2^8$ points give relative sampling standard errors of $1.46\%$ to $1.67\%$ over the fifteen energies.

\subsection{Genus one \texorpdfstring{$1\to 1$}{1 to 1} amplitude}\label{App:Genus11to1Numerics}

\subsubsection*{Conformal Block}

We fix the two vertex operator insertions at $0$ and $z$ on the torus. There are two conformal decomposition channels:
the necklace channel and the OPE channel in which the two external operators fuse before entering the handle.

In the necklace channel,
the plumbing parameters are given by
\ie\label{eq:g1_two_point_necklace_parameters}
q_1=e^{iz},\qquad q_2=e^{i(2\pi\tau-z)},\qquad
q_1q_2=e^{2\pi i\tau}.
\fe
We truncate the genus 1 two point block as 
\ie\label{eq:g1_two_point_sewing_series}
F_{\mathrm N}(q_1,q_2)
=\sum_{n_1,n_2\geq0}a_{n_1n_2}\,q_1^{n_1}q_2^{n_2},
\fe
with orders $(n_1,n_2)=(6,3)$. The coefficients $a_{n_1n_2}$ are computed using the
$c$-recursion of~\cite{Cho-Collier-Yin}.

In the OPE channel,  the plumbing parameters for the
handle and the OPE tube are related to $(\tau,z)$ by
\ie
q=e^{2\pi i\tau},\qquad v=e^{-iz}-1.
\fe
After factoring out the contributions from weights of primary operators, the genus one 2-point block in the OPE channel can be expanded in plumbing parameters as
\ie\label{eq:g1_two_point_ope_series}
F_{\mathrm O}(q,v)=\sum_{n,m\geq0}b_{nm}\,q^n v^m,
\fe
The coefficients $b_{nm}$ are computed
using the $c$-recursion
of~\cite{Cho-Collier-Yin}. In practice, we truncate the plumbing parameters $(q,v)$ at order $(3,8)$.

Defining
\ie
\delta=\min\left\{
|z|,\ |z-2\pi|,\ |z-2\pi\tau|,\ |z-2\pi(1+\tau)|
\right\},
\fe
we use the OPE channel for $|\delta|<0.3\pi$ and the
necklace channel otherwise.

\subsubsection*{Liouville Momentum Integration}

For the integration of Liouville momentum, we truncate each momentum integral at \(P_{\max}=6\) and change the variables as \ie\label{eq:g1_two_point_momentum_rule}
P=P_{\max}u^2,\qquad dP=2P_{\max}u\,du,\qquad 0<u<1.
\fe
We then use Gauss–Legendre quadrature to conduct numerical integration in $u$, with 32 integration points for each of the momenta. Integration in $u$ places more integration points where the Liouville momentum is small in order to improve numerical efficiency.

\subsubsection*{Moduli Integration}

As usual, we use Monte Carlo sampling to integrate over the moduli space. For each external energy, we generate eight independent sets of
$2^{16}$ sampling points. Their average gives the moduli integral,
and the deviation among the eight estimates determines its
sampling error.

For the moduli integration, we change the variable from $(\tau,z)$ to four real variables $(a,b,c,d) \in \left(0,1\right)^4$ as
\ie
\tau_1&=a-\frac12,\qquad &
\tau_2&=\sqrt{1-\tau_1^2}+(1-b)^{-2}-1,\\
c&=\frac{1}{2\pi}
\left(z_1-\frac{\tau_1}{\tau_2}z_2\right),\qquad &
d&=\frac{\kappa+\pi\tau_2}{\pi\tau_2}
\frac{z_2}{\kappa+z_2},
\qquad \kappa>0.
\fe
We choose $\kappa=4R^2$ in the computation of thermal amplitude for numerical efficiency. The choice for the zero temperature amplitude corresponds to taking $\kappa\to \infty$. The domain for the parameter $d$ only covers half of the torus, and the integration over the full torus is the double of it. We then perform the integration by sampling uniformly over $(a,b,c,d) \in \left(0,1\right)^4$. 

We treat the region in the moduli integration where two vertex operators collide separately. Consider the disk $D= \{|z|<1/2\}$. In that region, we approximate the OPE block by retaining only primary operators propagating through the OPE tube
\ie\label{eq:primaryapprox}
\bigl\langle V_{\omega/2}(z)V_{\omega/2}(0)\bigr\rangle_\tau
&\simeq
\int_0^\infty\frac{dP_{\mathrm O}}{\pi}\,
\mathcal C\!\left(\frac{\omega}{2},\frac{\omega}{2},P_{\mathrm O}\right)
|z|^{2h_{\mathrm O}-4h_{\omega/2}}
\bigl\langle V_{P_{\mathrm O}}(0)\bigr\rangle_\tau , \quad  h_{\omega/2}=1+\frac{\omega^2}{4}.
\fe
Under this approximation, the moduli parameter $z$ can be integrated analytically in the disk. When a sampling point is in the disk $D$, we subtract off the contribution from the primary propagation in order to avoid double counting once the analytic contribution is added.

\subsubsection*{Numerical Stability}

We check the numerical stability by varying the truncation level of conformal block, changing the number of Liouville momentum integration points and varying the region covered by pair-OPE channel. At $\omega=i/2$ and $R=2$,
increasing the necklace truncation and pair-OPE truncation levels from $(6,3)$ and $(3,8)$ to $(8,4)$ and $(4,10)$, respectively, changes the relative value of the imaginary parts of the noncompact and
thermal amplitudes by $0.0021\%$ and $0.00069\%$.

At $R=4$, increasing the momentum nodes from $32$ to $48$ changes
the thermal amplitude by  $0.0034\%$, while reducing the collision radius from $0.5$
to $0.1$ instead changes it by $0.234\%$.

\subsection{Genus one \texorpdfstring{$1\to 2$}{1 to 2} amplitude}\label{App:Genus11to2Numerics}

\subsubsection*{Conformal Block}

We place the incoming operator at the origin and the outgoing operators
at $w_1=z_2-z_1$ and $w_2=-z_1$. 

For the genus one 3-point block, there are three topologically distinct conformal block decomposition channels: the necklace channel, the pair OPE channel when two operators fuse, and the comb channel when all three fuse (see figure~\ref{fig:torus-3-point-necklace-block} for diagrammatic representations of the three conformal block decomposition). 

In the necklace channel, the plumbing parameters are related with moduli parameters by
\ie\label{eq:g1_necklace_parameters}
q_1=e^{iw_{1}},\qquad
q_2=e^{i(w_{2}-w_{1})},\qquad
q_3=e^{i(2\pi\tau-w_{2})},\qquad
q_1q_2q_3=e^{2\pi i\tau}
\fe
at $0<\Im w_1<\Im w_2<2\pi\tau_2$. We truncate the necklace conformal block as
\ie\label{eq:g1_block_truncation}
F_\Gamma^{(N_1,N_2,N_3)}(q_1,q_2,q_3)
=\sum_{n_1=0}^{N_1} \sum_{n_2=0}^{N_2}\sum_{n_3=0}^{N_3}
F_{\Gamma;n_1n_2n_3}\,q_1^{n_1}q_2^{n_2}q_3^{n_3},
\fe
and use $(N_1,N_2,N_3)=(4,4,4)$ in the practical computation. The coefficients $F_{n_1,n_2,n_3}$ are computed by the $c$-recursion of~\cite{Cho-Collier-Yin}. A similar truncation in the plumbing parameters is conducted for the pair OPE channel and the comb channel.

\subsubsection*{Liouville Momentum Integration}

The torus three point block involves three Liouville momentum integrations. We treat it with the same strategy as the sphere 6-point block computation (see section \ref{App:Genus01to5Numerics}), by splitting each integral into two parts $P_a\in (0,s_a)$ and $P_a\in (s_a,\infty)$ and changing the variables to
\ie\label{eq:g1_momentum_panels}
P_a=s_a u,\qquad P_a=s_a-2w_a\log(1-u),\qquad 0<u<1.
\fe
We then apply Gauss--Legendre quadrature to compute the integration over $u$ numerically. We use five integration points to cover $P_a\in (0,s_a)$ and three integration points to cover $P_a\in (s_a,\infty)$ for a generic point on the moduli space.

Analogously to the 6-point case, the parameters $s_a,w_a$ are chosen adaptively to ensure convergence of the integral. We adopt a prescription that balances the contributions of both the long plumbing tube region, which tends to suppress large momenta, and of the DOZZ structure constants, which tend to suppress at small momentum. 

Denote the product of DOZZ structure constants in a given channel as $C_\Gamma(\mathbf{P})$. We estimate the behavior of the DOZZ structure constants near the origin by the metric $\chi_a$, defined as
\ie
\chi_a&\equiv\frac{H_\Gamma(\mathbf{P}^{(0)}+\epsilon \mathbf{e}_a)-H_\Gamma(\mathbf{P}^{(0)})}{3\epsilon^2}, ~~\mathbf{P}^{(0)}=(\epsilon,\epsilon,\epsilon),~~\epsilon=0.005,\\
H_\Gamma(\mathbf{P})&\equiv \log|C_\Gamma(\mathbf{P})|-2\sum_{b=1}^3\log P_b.
\fe
The leading order behavior of the DOZZ structure constants $C_\Gamma(\mathbf{P})$ near $P=0$ is $P^2$, so the metric $\chi_a$ captures degree of suppression near the origin for the structure constants beyond the leading order behavior. Defining $D_a\equiv \text{max}\left\{-\frac{3}{2}\log|q_a|,-2\log|q_a|-\chi_a\right\}$ and $\sigma_a\equiv D_a^{-1/2}$, our choice of $s_a, w_a$ is
\ie
s_a=\frac{3}{2}\sigma_a,~~w_a=\frac{1}{4}\sigma_a.
\fe

In the degeneration region of the necklace channel, i.e. when a certain plumbing tube becomes long, we increase the number of integration points when another one of the tubes is short. Specifically, when any of the $q_a$ have $-\log |q_a|\leq \frac{1}{2}$ in the degeneration region, we increase the number of integration points over the tube momentum in the region $(0,s_a)$ from five to six. We also change the prescription for choosing $s_a$ and $w_a$. We choose $s_a=\frac{3}{2}$ for $D_a<1$, $s_a=\frac{3}{2}\sigma_a,1\leq D_a\leq 8$, and $s_a=\frac{9}{4}\sigma_a,D_a> 8$. For $w_a$, we choose $w_a=\frac{1}{4}\sigma_a$ for $D_a<8$ and $-\log|q_a|\geq \frac{1}{2}$, and $w_a=\frac{11}{40}\sigma_a$ otherwise.

\subsubsection*{Moduli Integration}

Evaluating the genus one 3-point correlator at each moduli
point is substantially more expensive than in the $1\to1$
calculation. We therefore use degeneration approximations
to cover as much of moduli space as possible, and use Monte Carlo sampling to compensate the contribution from the region that is not covered by the degeneration approximations. We use degeneration approximations to cover $98.5\%$ of the moduli space.

In the degeneration regions, we retain only primary intermediate states propagating through the long tube, integrate the
plumbing parameter corresponding to the long tube analytically, and
evaluate the remaining integrals numerically. The rest of the moduli space that cannot be covered by this method is computed by Monte Carlo evaluation. For each external energy, the compact contribution uses sixteen
independent MC sampling sets of four points, giving
$64$ moduli points in total. Unlike the $1\to1$ calculation, we restrict the MC sampling to the region that is not covered by the long tube approximation.

To be specific, define $d_{ij}$, the distance between local operators, as
\ie
d_{ij}=\min_{m,n\in\mathbb Z}
|z_i-z_j+2\pi(m+n\tau)|,\qquad
d_{\min}=\min_{i<j}d_{ij}.
\fe
We use an analytic approximation to cover the moduli space when $d_{ij}< 1/8$. If the remaining insertion $k$ satisfies
$d_{jk}<r_{\mathrm t}=1/2$, we use the comb channel conformal block decomposition and approximate the longest tube; otherwise, we use the OPE channel. The region with  $d_{ij}>1/8$ is covered by necklace channel.

When the tube of the torus is long, and the torus 3-point function is computed by the necklace channel decomposition, we use a separate prescription. Define
\ie\label{eq:g1_moduli_partition}
\ell_a=-\log|q_a|,\qquad
L=\max_a\ell_a,\qquad
\ell_0=-\log(0.05).
\fe
The region with $L\geq\ell_0$ is treated by approximating the torus 3-point block by the sphere 5-point block, and the 5-point block is computed by elliptic $h$ recursion of \cite[Sec.~5.1]{ArtemevKhromov:2026}.

The contribution from the remaining part of the moduli space is computed by Monte Carlo sampling. This corresponds to a very small piece of the moduli space with $\tau_2<3\ell_0/(2\pi)\simeq1.43036$, which justifies our use of just $64$ Monte Carlo samples to cover this region.

\subsubsection*{Numerical Stability}

We assess numerical stability by comparing conformal block truncations, varying number of momentum integration points at fixed moduli and integration grids in selected nonseparating regions, and estimating the sampling error from independent Monte Carlo sample sets. Over the twenty energies, lowering the native $c$-recursion orders in the collision contributions, changes
the full amplitude by less than $0.007\%$. Lowering the order of elliptic 5-point block when tube of the torus is long changes the full amplitude by
$0.032\%$--$0.154\%$. A variation of the integration points in Liouville momentum integration
changes the amplitude by $0.28\%$--$0.83\%$.

\subsection{Genus 2 thermal free energy}\label{App:Genus2Numerics}

\subsubsection*{Plumbing Parameters and the Period Matrix}

In the main text, we presented the relation between the plumbing parameters and the period matrix
using Schottky product in
\eqref{Period_Matrix_from_Schottsky}, under the choice of the homology cycle chosen in figure \ref{fig:genus-two-plumbing-channels}. When the plumbing parameters are small, we compute the period matrix numerically by restricting the Schottky product to group elements
whose word length does not exceed a chosen cutoff.  In the bulk of moduli space, however, the plumbing parameters can
be larger, so convergence may require a higher word length cutoff,
making the evaluation computationally expensive. In this region, we instead compute
the period matrix by constructing normalized holomorphic one forms
directly on the plumbing fixture.

Let $z$ and $w$ denote the coordinates on the two copies of the
three-punctured sphere. We write the restrictions of the holomorphic one-forms
to these copies as $\omega_I=f_I(z)\,dz$ and
$\omega_I=\widetilde f_I(w)\,dw$, respectively, with $I=1,2$. 
To define a one form on the sewn surface, we require its local
expressions to agree on each pair of annuli identified by the
plumbing relations in Section~\ref{sec:genus2liouville}.
For annuli on different copies of the three punctured sphere,
this requires
\[
f_I(z)\,dz=\widetilde f_I(w)\,dw.
\]
For self sewn annuli, the same matching condition uses the
coefficient function on the corresponding sphere on both sides.

In the numerical computation, we approximate the holomorphic one form via a truncated Laurent series at each puncture at order $M$
\ie
f_I(z)=
\sum_{n=1}^{M}
\left(
\frac{a_{I,n}}{z^n}
+\frac{b_{I,n}}{(z-1)^n}
\right)
-\sum_{n=2}^{M}c_{I,n}z^{n-2},
\fe
together with an analogous expansion for $\widetilde f_I(w)$ with independent coefficients. We choose equally spaced angular points on each sewing circle,
using $4M$ points per circle in the theta channel and $6M$ in the
glasses channel, and determine their images under the plumbing map. With these points fixed, we approximate the sewing condition by minimizing
\[
\sum_{\text{points}}
\left|f_I(z)-\widetilde f_I(w)\frac{dw}{dz}\right|^2,
\]
subject to 
\ie
\int_{A_J}\omega_I=\delta_{IJ}.
\fe
We then compute the period matrix by numerically integrating
\ie
\int_{B_J}\omega_I=\Omega_{IJ}.
\fe

\subsubsection*{Conformal Block}

As in Appendix~\ref{App:Genus01to4Numerics}, we choose the conformal block decomposition separately at each sampled point to improve numerical convergence. For each period matrix, we search over theta and glasses plumbing representations, requiring that the annuli used for plumbing are non-overlapping on a given sphere and that the period matrix is stable, i.e. does not change by more than $10^{-5}$, as the truncation order $M$ is increased. If both channels are acceptable, we use the channel with smaller values of $\max_i|q_i|$.

In the selected channel, we compute the genus two Virasoro block numerically at $c=25$ using the $c$-recursion of~\cite{Cho-Collier-Yin}. For the large $c$ block, we perform a resummation to achieve better numerical convergence. The large $c$ limit of the conformal block factorizes as~\cite{Cho-Collier-Yin}
\[
F_\Gamma^{(\infty)}(\boldsymbol h;\boldsymbol q)
=\mathcal V_\Gamma(\boldsymbol q)\,
G_\Gamma(\boldsymbol h;\boldsymbol q),
\qquad
\Gamma\in\{\mathrm{Theta},\mathrm{Glasses}\},
\]
where $\mathcal V_\Gamma$ is the large $c$ vacuum block
\[
\mathcal V_\Gamma(\boldsymbol q)
=
\prod_{[\gamma]\in\mathcal P_\Gamma}
\prod_{\ell=2}^{\infty}
\left(1-k_\gamma(\boldsymbol q)^\ell\right)^{-1}
\]
and $G_\Gamma(\boldsymbol h;\boldsymbol q)$ is the global $\operatorname{SL}(2)$ block, which admits an explicit series expansion formula given in \cite{Cho-Collier-Yin}.
In the theta channel, we resum the global block as
\[
G_{\mathrm{Theta}}
=
\sum_{m,n=0}^{\infty}
\frac{q_\infty^m q_0^n\,s_{mn}^{\,2}}
     {m!\,n!\,(2h_\infty)_m(2h_0)_n}
\,{}_2F_1\!\left(a_{mn},a_{mn};2h_1;q_1\right),
\qquad
a_{mn}=h_1+h_0+n-h_\infty-m,
\]
where
\[
\begin{aligned}
s_{mn}
={}&
\sum_{p=0}^{\min(m,n)}
\binom{m}{p}\frac{n!}{(n-p)!}
(2h_0+n-p)_p
\\
&\qquad\times
(h_0+h_1-h_\infty)_{n-p}
(h_\infty+h_1-h_0+p-n)_{m-p},
\end{aligned}
\]
and $(a)_n$ denotes the rising Pochhammer symbol. In the glasses channel, the $\operatorname{SL}(2)$ block can be resummed into
\ie
G_{\mathrm{Glasses}}
= {}_2F_1(h_B,h_B;2h_B;q_B)
\prod_{a=L,R}
(1-q_a)^{h_B-1}
{}_2F_1\!\left(
h_B,2h_a+h_B-1;2h_a;q_a
\right).
\fe
We evaluate these expressions and the vacuum product separately from the finite-$c$ recursion, so their numerical cutoffs are independent of the recursion order $N$.

We truncate the conformal blocks at level $N$ using $c$-recursion. Most bulk evaluations use $N=8$, with higher orders up to $N=22$ at selected points. Following the checks described in Appendix~\ref{App:Genus01to4Numerics}, we assess truncation effects by increasing $N$ and comparing $\mathcal Q_L$ between the theta and glasses channels on their overlap.

\subsubsection*{Liouville Momentum Integration}

We integrate the three internal Liouville momenta numerically over
$[0,\infty)$. For each plumbing tube $e$, the conformal block decomposition contains a Gaussian factor coming from the conformal weight of the Liouville primary operator
\ie
|q_e|^{2P_e^2}=e^{-\kappa_e P_e^2},
\qquad
\kappa_e=-2\log|q_e|>0.
\fe
We first change variables to $u_e=\kappa_e P_e^2$, then approximate
the integral over $u_e\in[0,\infty)$ as a weighted sum using
generalized Gauss--Laguerre quadrature with weight
$u_e^{-1/2}e^{-u_e}$. Denoting the integration points and weights by
$(u_j,v_j)$, the precise approximation we use is
\ie
\int_0^\infty\frac{dP_e}{\pi}\,f(P_e)
\simeq
\sum_j
\frac{v_j e^{u_j}}{2\pi\sqrt{\kappa_e}}\,
f\!\left(\sqrt{\frac{u_j}{\kappa_e}}\right),
\fe
where $f$ is the complete momentum integrand, including the Gaussian
factor, the DOZZ structure constants and conformal block. For a fixed quadrature order, the momentum quadrature nodes $u_j$ are fixed,
whereas the corresponding momenta
$P_{e,j}=\sqrt{u_j/\kappa_e}$ depend on the plumbing parameter.
The quadrature therefore adapts to each tube without introducing a finite momentum cutoff.

We apply this quadrature rule to three Liouville momentum integrals in the genus 2 conformal block decomposition. The number of quadrature nodes we use to approximate the integral may vary across the moduli space. Most bulk evaluations use eight nodes per integral, with higher orders at selected points. We evaluate the DOZZ factors and conformal blocks on the resulting momentum
grid to obtain the Liouville partition function. Combining it
with the free-boson partition function in the same plumbing frame
then gives the Liouville factor $\mathcal Q_L$.

\subsubsection*{Moduli Integration}

We conduct the moduli integration using Monte Carlo methods along with a prescription for the degeneration region where $|\Omega_{12}|\to 0$. We integrate over the genus two moduli space by sampling period matrices. We now explain a finite set of inequalities that can be checked to determine if a genus two period matrix lies in the Siegel-Gottschling fundamental domain.

We denote the genus two period matrix as
\ie
\Omega=X+iY=
\begin{pmatrix}
\Omega_{11}&\Omega_{12}\\
\Omega_{12}&\Omega_{22}
\end{pmatrix}\in\mathbb H_2,
\fe
with $X$ and $Y$ real matrices. The Siegel–Gottschling fundamental domain is specified by
\ie\label{eq:basicConstants}
-\frac12\leq X_{11},X_{12},X_{22}\leq\frac12,
\qquad
0\leq2Y_{12}\leq Y_{11}\leq Y_{22},
\fe
together with the determinant condition
\ie
\label{eq:determinantCondition}
|\det(C\Omega+D)|\geq1
\quad \forall
\begin{pmatrix}A&B\\ C&D\end{pmatrix}
\in\operatorname{Sp}(4,\mathbb Z).
\fe
After imposing the conditions in equation \eqref{eq:basicConstants}, the determinant condition in equation  \ref{eq:determinantCondition} is shown in \cite{Gottschling:1959} to be equivalent to the following 19 constraints:
\ie
|\Omega_{11}|\geq1,\qquad
|\Omega_{22}|\geq1,
\fe
\ie
|\Omega_{11}+\Omega_{22}-2\Omega_{12}+1|\geq1,
\qquad
|\Omega_{11}+\Omega_{22}-2\Omega_{12}-1|\geq1,
\fe
as well as
\ie
|\det(\Omega+S)|\geq1
\fe
for the 15 matrices
\ie
S\in\Biggl\{&
\begin{pmatrix}0&0\\0&0\end{pmatrix},
\ \pm\begin{pmatrix}1&0\\0&0\end{pmatrix},
\ \pm\begin{pmatrix}0&0\\0&1\end{pmatrix},
\ \pm\begin{pmatrix}1&0\\0&1\end{pmatrix},
\pm\begin{pmatrix}1&0\\0&-1\end{pmatrix},
\ \pm\begin{pmatrix}0&1\\1&0\end{pmatrix},
\ \pm\begin{pmatrix}1&1\\1&0\end{pmatrix},
\ \pm\begin{pmatrix}0&1\\1&1\end{pmatrix}
\Biggr\}.
\fe
To conduct Monte Carlo sampling of the moduli space, we first satisfy the constraints in equation \eqref{eq:basicConstants} by changing the variable
\ie
Y_{11}=a e^u,\qquad
Y_{22}=a e^{u+v},\qquad
Y_{12}=\frac{s}{2}Y_{11},
\qquad
a=\frac{\sqrt3}{2},
\fe
where $u,v\geq0$ and $0\leq s\leq1$. The Jacobian is
\ie
J(u,v)=
\left|
\det\frac{\partial(Y_{11},Y_{12},Y_{22})}
{\partial(u,v,s)}
\right|
=
\frac{a^3}{2}e^{3u+v}.
\fe
We sample the three independent entries of $X$ uniformly on
$[-1/2,1/2]$ and $s$ uniformly on $[0,1]$, while sampling
$(u,v)\in[0,\infty)^2$ with the weight function
\ie
p(u,v)=\frac14
\sum_{\substack{\lambda\in\{3,3/4\}\\
                \eta\in\{2,1/2\}}}
\lambda\eta\,e^{-\lambda u-\eta v}.
\fe
For each Monte Carlo sampling point, we evaluate the weighted integrand
\ie
\label{eq:weightedGenus2}
W_R(\Omega)=
\mathbf1_{\mathcal F_2}(\Omega)\,
\mathcal I_R(\Omega)\frac{J(u,v)}{p(u,v)},
\fe
where $\mathcal I_R$ is the moduli integrand for $F^{(2)}(R)$ in
equation \eqref{eq:genus2-moduli-integral}. The indicator function
$\mathbf1_{\mathcal F_2}(\Omega)$ returns $1$ when all 19 Gottschling
inequalities are satisfied and $0$ otherwise.

To improve numerical accuracy, we treat the region $|\Omega_{12}|<\rho_0,\rho_0:=0.08$ separately.
As $|\Omega_{12}|\to0$, the genus two surface develops an increasingly long tube connecting two tori, which is the limit discussed in Appendix~\ref{App:Factorialization}. The Monte Carlo sampling procedure above does not adequately resolve the contribution from this region. We take into account the contribution from this region by working with the plumbing parameters in the glasses coordinate, approximating the conformal block by retaining only primary operators propagating in the long tube and integrating the plumbing parameter describing the long tube connecting the two tori, denoted as $q_B$, analytically. Following the derivation in Appendix.~\ref{App:Factorialization}, we approximate the genus 2 moduli integrand by equation \eqref{eq:genus2-separating-moduli-integrand}, where the normalization constant $\mathcal{N}_{2}$ is given by equation~\eqref{eq:NormalizationConstant}.

After integrating out $q_B$, the remaining moduli integrand in the degeneration region of the moduli space becomes
\ie\label{eq:Approximated Integrand}
\mathcal G_R(\rho_0)
&={}
-\frac{1}{2\pi}
\int_0^\infty dP_B\,
\frac{\pi e^{-2L_0P_B^2}}{P_B^2}
\\
&\quad\times
\left[
\int_{\mathcal F_1}d^2\tau\,
\frac{\Theta_R^{(1)}(\tau)|\eta(\tau)|^2}
{\sqrt{\operatorname{Im}\tau}}
\int_0^\infty dP\,
\mathcal C(P,P_B,P)
\left|
q^{h_P-25/24}
F_{\mathrm{tor}}(h_P,h_B;q)
\right|^2
\right]^2,
\fe
where $h_P=1+P^2$, $h_B=1+P_B^2$, $L_0=-\log(2\pi\rho_0)$. The remaining momentum and torus moduli integrals are performed numerically.

For Monte Carlo sampling points that have $|\Omega_{12}|<0.08$, we subtract from the weighted integrand in equation \eqref{eq:weightedGenus2} the leading long tube approximation. We then add the analytically evaluated $\mathcal{G}_{R}(\rho_0)$ given by equation \eqref{eq:Approximated Integrand} separately to avoid double counting. Below $|\Omega_{12}|<0.04$, we further sample the difference in the separate intervals
\[
\begin{alignedat}{3}
&[10^{-5},0.001], &\qquad &[0.001,0.002], &\qquad &[0.002,0.005],\\
&[0.005,0.01],    &\qquad &[0.01,0.02],   &\qquad &[0.02,0.04],
\end{alignedat}
\]
with sampling in each interval uniform in $\log|\Omega_{12}|$. Each interval uses 8 samples, except for the interval $[10^{-5},0.001]$, which uses 16 samples.

With these ingredients, we generate sampling points throughout the Siegel–Gottschling fundamental domain and evaluate the Liouville factor and sampling factor $J(u,v)/p(u,v)$ appearing in equation \ref{eq:weightedGenus2}. Since these quantities are independent of the compactification radius \(R\), we reuse them to compute the free energy at different radii.

\subsubsection*{Numerical Stability}

We assess the numerical stability of our result by varying the number of integration points for the Liouville momentum integration, the level of the conformal block truncation, comparing the Liouville factor $\mathcal{Q}_L$ computed in theta and glasses channels, and estimating fluctuations between independent sets of moduli samples used in the moduli space integration.

The accuracy of the conformal block truncation is supported by the agreement of the Liouville factor $\mathcal{Q}_L$ on the domain of the moduli space accessed by both the theta channel and glasses channel. When the conformal block is truncated at level $12$ and momentum integration uses $12$ integration points, the two channels agree in $\mathcal{Q}_L$ to within $2\times10^{-5}$ at five sample period matrices, as shown in Table~\ref{tab:genus-two-liouville-channel-agreement}.

For the contribution from the degenerate region at $R=1$ and $\rho_0=0.08$, increasing the number of torus and long tube momentum integration points from $(24,32)$ to $(32,48)$ and then $(40,64)$ changes the relative free energy by approximately $5.9\times10^{-8}$ and $2.3\times10^{-10}$, respectively. Increasing the torus conformal block truncation level from $2$ to $4$ changes the relative free energy by $1.2\times10^{-11}$. These comparisons use identical moduli samples at each order.

For the moduli integration, we estimate the sampling uncertainty from independent MC sampling sets, retaining correlations between different radii. At $R=1$, the relative sampling standard error is approximately $0.60\%$.

\printbibliography
\end{document}